\documentclass[prx,twocolumn,floatfix,notitlepage,superscriptaddress,longtable]{revtex4-2}
\usepackage[caption=false]{subfig}
\usepackage{amsmath}
\usepackage{lipsum} 
\usepackage{verbatim}
\usepackage{epsfig}
\usepackage{graphicx}
\usepackage{amsfonts}
\usepackage[figuresright]{rotating}
\usepackage{amssymb}
\usepackage{array}
\usepackage{graphicx}
\usepackage{booktabs}
\usepackage{adjustbox}
\usepackage{adjustbox}
\usepackage{graphicx}
\usepackage{booktabs}
\usepackage{dsfont}

\usepackage{psfrag}
\usepackage{esint} 
\usepackage{bm}
\usepackage[colorlinks,linkcolor=blue,anchorcolor=blue,citecolor=blue,urlcolor=blue]{hyperref}
\usepackage{float}
\usepackage[version=4]{mhchem}
\usepackage[svgnames]{xcolor}

\def\be{\begin{equation}} \def\ee{\end{equation}}
\def\bea{\begin{eqnarray}} \def\eea{\end{eqnarray}}

\def\bk{{\bf k}}
\def\bp{{\bf p}}
\def\bq{{\bf q}}

\renewcommand{\vec}[1]{\mathbf{#1}}

\newcommand{\ket}[1]{| #1 \rangle}
\newcommand{\bra}[1]{\langle #1 |}
\newcommand{\braket}[2]{\langle #1 |#2\rangle}

\newcommand{\up}{\uparrow}
\newcommand{\down}{\downarrow}

\def\bpm{\begin{pmatrix}} \def\epm{\end{pmatrix}}

\DeclareMathOperator{\sgn}{sgn}

\definecolor{Qicolor}{RGB}{3, 136, 252}

\usepackage{framed}
\usepackage{listings}
\usepackage{color}
\usepackage{tikz}
\usetikzlibrary{shapes.arrows}
\usetikzlibrary{backgrounds}
\usetikzlibrary{positioning}
\usetikzlibrary{arrows}
\definecolor{gray}{gray}{0.3}
\definecolor{darkgreen}{rgb}{0,0.55,0}
\definecolor{purple}{rgb}{0.5,0,1}

\makeatletter
\newcommand*{\balancecolsandclearpage}{%
  \close@column@grid
  \clearpage
}
\makeatother

\begin{document}

\author{Omid Tavakol}
\affiliation{Department of Physics and Astronomy, University of California, Irvine, California 92697, USA}

\author{Thomas Scaffidi}
\email{tscaffid@uci.edu}
\affiliation{Department of Physics and Astronomy, University of California, Irvine, California 92697, USA}

\title{Pairing around a Single Dirac Point: A Unifying View of Kohn-Luttinger Superconductivity in Chern Bands, Quarter Metals, and Topological Surface States}

\begin{abstract}
Superconductivity of a single two-dimensional Dirac fermion offers a natural route to topological superconductivity. While usually considered extrinsic---arising from proximity to a conventional superconductor---we investigate when a doped Dirac cone can \emph{spontaneously} develop superconductivity from a short-range repulsive interaction $U$ via the Kohn--Luttinger mechanism.  
We show that an ideal, linear Dirac cone is immune to pairing at leading order in $U^2$. Superconductivity instead emerges only through higher-order in $k$ corrections to the dispersion, which are unavoidable in any lattice realization and crucially dictate the pairing symmetry. The form of the pairing thus reflects how the well-known obstruction to realizing a single Dirac cone on a lattice is circumvented.
When a Dirac cone arises from broken time-reversal symmetry---for instance, at a transition between Chern insulators or in a valley-polarized phase---we find a topological $p - ip$ state whose chirality is opposite to that of the parent chiral metal above $T_c$. By contrast, for a surface Dirac cone of a 3D topological insulator, superconductivity is stabilized by anisotropies in the dispersion. For $C_{3v}$-symmetric warping, as in \ce{Bi2Te3}, pairing is strongest when the Fermi surface becomes hexagonal, leading to order in the $(d \pm id)\times(p+ip)$ channel with accidental near-nodes. In the highly anisotropic limit $v_x \gg v_y$, relevant to side surfaces of layered materials, the Fermi surface splits into two branches, and nesting favors a pairing symmetry $\Delta \sim \mathrm{sgn}(k_x)\cos(k_y)$ reminiscent of organic superconductors.  
\end{abstract}

\date{\today}

\maketitle


\section{Introduction}

Realizing superconductivity in a single two-dimensional Dirac cone has long been proposed as a promising route to topological superconductivity~\cite{Fu_Kane_2008,Shou-Cheng_2010}.
The underlying idea is that the non-trivial winding of the Dirac spinor around the Dirac point imparts a non-trivial winding to the superconducting gap, even for a short-range attractive potential that would otherwise yield a topologically trivial $s$-wave pairing in a conventional metal.
In this context, superconductivity is traditionally assumed to be extrinsic to the Dirac fermions, namely (1) induced via the proximity effect in a heterostructure~\cite{Fu_Kane_2008,Shou-Cheng_2010,Fu_Kane_2009,PhysRevLett.105.077001,Potter,BlackSchaffer}, (2) from attractive interactions generated for example by the electron-phonon pairing mechanism~\cite{Sarma_2013,Sarma_2014, PhysRevB.81.184502,PhysRevB.99.184514}, or (3) from bulk superconductivity in the case of a Dirac cone appearing as a topological surface state~\cite{PhysRevLett.104.057001,PhysRevLett.105.097001,doi:10.1126/science.aan4596,doi:10.1126/science.aao1797,PhysRevLett.117.047001,PhysRevResearch.2.022021,PhysRevX.8.041056}.

In this work, we study the possibility of \emph{intrinsic} superconductivity in a single 2D doped Dirac cone arising from the Dirac fermions themselves. Specifically, we investigate the Kohn-Luttinger mechanism, in which the electronic overscreening of a bare repulsive interaction generates an effective attraction in higher angular-momentum channels~\cite{KL_1965}. This mechanism is known to produce a rich variety of superconducting orders dictated by the interplay between band structure and interactions~\cite{Kagan_1992,Chubukov_1992,Chubukov_1993,Richard_1999,González_2008,PhysRevB.81.224505,Raghu2010,Raghu_2012,Thomale2013,Chubukov2012,ScaffidiEA14,Chubukov_2014,Kagan_2014,PhysRevB.94.085106,Scaffidi2017,PhysRevLett.115.087003,Wolf_2018,WolfRashba,PhysRevB.99.140504,Thomas_2016_inter,ScienceSRO,Thomas_2018_inter,Thomas_2019_inter,ScaffidiReview,Jerzembeck2022,PhysRevB.107.014505,ThirdOrderTMD}. More recently, the interplay between non-trivial band topology and superconductivity in the presence of repulsive interactions has been a subject of intense research~\cite{PhysRevB.100.041117,Devakul_2025,jahin2025enhancedkohnluttingertopologicalsuperconductivity,PhysRevB.110.195126,Santos,2024arXiv241018175D,BitanRoy2025,2025arXiv250616508L}. 
The discovery of superconductivity in multilayer graphene systems, where band structure, topology and strong correlations are exquisitely tunable, has provided new impetus to investigate purely electronic mechanisms for superconductivity in such contexts (See \cite{González_2019,Chubukov_2020,You2022,Cea_2022,Christos_2025,Ghazaryan_2021,Ghazaryan_2023,PhysRevB.111.174523,PNAS_KLforTwisted_Review} and references therein).

While such material-specific studies are crucial, their complex Fermiology can make it difficult to extract general principles which could guide the search for exotic superconducting phases. In that context, we turn to the paradigmatic model of Dirac fermions, which has become a cornerstone of condensed matter physics~\cite{ARCMPDirac}.
Single-layer graphene, with its two Dirac cones of opposite chirality, is the canonical two-dimensional Dirac material, and the prospects for superconductivity there have already been studied extensively~\cite{González_2008,ThomaleGraphene2012,Chubukov2012,Chubukov_2014,Kagan_2014}. Our focus, in contrast, is on a \emph{single} Dirac cone, where the intrinsic chirality of the band structure inherently biases superconductivity toward a topological phase~\cite{Fu_Kane_2008,Shou-Cheng_2010}.

A single Dirac cone cannot be realized on a lattice without breaking fundamental symmetries, a constraint established by the ``fermion doubling'' theorem~\cite{NIELSEN1981219}. However, this theorem can be circumvented in several physically relevant scenarios, two of which we investigate here. 
The first scenario occurs by breaking time-reversal symmetry. For instance a single Dirac cone is generically realized at a topological phase transition between two distinct Chern insulating phases whose Chern numbers differ by one~\cite{Shou-Cheng_2010}. 
Broadly in the same category we find the ``quarter metal'' phase of graphene-based materials, in which case electrons form a valley- and spin-polarized metal which spontaneously breaks time-reversal symmetry~\cite{QuarterMetal2021}.
The observation of valley-polarized superconductivity in tetralayer graphene~\cite{ValleyPolarizedSC,PhysRevB.111.174523} thus serves as a motivation for our work as an example of Cooper pairing for electrons around a single Dirac point~\footnote{Note however that for $N>1$ layers of graphene, the Dirac point has higher winding equal to $N$, whereas we will only focus on $N=1$ in this work}.
The second scenario is realized at the surface of a 3D topological insulator~\cite{RevModPhys.83.1057}, for which the ``partner'' Dirac cone is spatially separated on the opposite surface, leaving a single effective cone on the surface of interest.

Our central result is a striking dichotomy: a single ideal Dirac cone with a purely linear dispersion exhibits no superconducting instability at the leading weak-coupling order ($\mathcal{O}(U^2)$) but is, however, robustly stabilized as soon as higher-order in momentum terms—those very terms required to regularize the Dirac cone on a lattice in either of the scenarios just described—are included in the dispersion. Far from being mere corrections, these terms prove decisive in shaping the superconducting state, giving rise to a rich variety of pairing symmetries whose character depends on the specific physical realization of the Dirac cone.

As already mentioned, we will study the emergence of superconductivity from weak, short-range repulsion $U$ within the Kohn-Luttinger (also known as weak-coupling renormalization group) mechanism, for which the effective interaction in the Cooper channels is calculated up to second order in the interaction ($\mathcal{O}(U^2)$).
This calculation is asymptotically exact in the weak coupling limit of $U \ll t$ (with $t$ a measure of the single-particle bandwidth), in which case the only generic instability is superconductivity~\cite{RevModPhys.66.129,PhysRevB.81.224505}. For stronger coupling, a single Dirac cone is known to have a ferromagnetic transition~\cite{Thomale,PhysRevLett.128.225701}.
Note also that we work in the Fermi liquid regime of $k_B T \ll \mu$, as opposed to the Dirac liquid regime found close to charge neutrality~\cite{RevModPhys.84.1067}. Our results will nevertheless reveal important differences with the two-dimensional electron gas, chief among them a dominant role for interband effects, which aligns with recent studies~\cite{Crepel_Fu_2021,Crepel_Fu_2021_2,Crepel_Fu_2022,PhysRevResearch.5.L012009}.

A key challenge for superconductivity in 2D Dirac systems is the vanishing density of states (DOS), which scales linearly with the Fermi momentum ($k_F$) and contrasts sharply with the constant DOS of a two-dimensional electron gas (2DEG). Since the dimensionless pairing strength $\lambda = \rho V_\text{eff}$ is proportional to the DOS $\rho$, one might conclude that Dirac fermions are poor candidates for superconductivity at low carrier density. However, this view is incomplete, as it neglects the crucial scaling of the effective interaction, $V_\text{eff}$, with $k_F$. For example, a 2DEG realized in the low density limit of the square lattice Hubbard model exhibits an unfavorable $V_\text{eff} \sim k_F^4$ scaling~\cite{Chubukov_1992,Kagan_1992}. We will demonstrate a more promising behavior for Dirac fermions. Specifically, we will find $V_\text{eff} \sim k_F^2$ for a single Dirac cone without time-reversal symmetry, and $V_\text{eff} \sim 1$ for two cones of opposite chirality as found in graphene. This favorable scaling, driven by interband effects, can overcome the suppressed DOS, making Dirac fermions surprisingly robust candidates for low-density superconductivity.

In Sec.~\ref{sec:1}, we derive the general weak-coupling gap equation for a two-band system at second order in the interaction, and in Sec.~\ref{sec:Single Dirac cone}, we demonstrate the absence of pairing for an ideal Dirac cone, highlighting the crucial role of interband effects in Sec.~\ref{Sec:interband}. In Sec.~\ref{Multiple}, we contrast these findings with the case of multiple Dirac cones to highlight the unique physics of the single-cone problem.
We then explore three distinct physical scenarios for which superconductivity is stabilized in a single ideal Dirac cone through higher-order in $k$ terms in the single-particle Hamiltonian. In Sec.~\ref{sec:TPT}, for a Dirac cone at a topological phase transition (TPT), we show that a time-reversal symmetry breaking term ($k^2 \sigma^z$) drives superconductivity and stabilizes a topological $p-ip$ state, a mechanism governed purely by the quantum geometry of the bands. In Sec.~\ref{C3}, we turn to the case of a single Dirac cone as a topological insulator surface state (TISS), where higher-order in $k$ terms in the dispersion induce anisotropy. We find that $C_{3v}$-symmetric warping, as seen in \ce{Bi2Te3}, leads to a peak in pairing strength as the Fermi surface goes from convex to concave, and stabilizes a topological but near-nodal $(d \pm id)\times(p+ip)$ state. In Sec.~\ref{1D}, we study the quasi-1D limit of $v_x \gg v_y$, and find superconductivity with a gap $\Delta \sim \sgn(k_x)\cos(k_y)$, driven by a nesting mechanism similar to the organic superconductors.


\section{Weak-coupling superconductivity for a two-band Hamiltonian}
\label{sec:1}

We begin with a general two-band Hamiltonian featuring a single-particle and an interaction term $H = H_0 + H_\text{int}$ and perform a transformation to diagonalize $H_0$ in the band basis. We then calculate the effective interaction in the Cooper channel (or ``pairing kernel'') up to second order in the interaction strength $U$. Finally, we write the linearized gap equation which corresponds to diagonalizing the pairing kernel. Throughout, we work in the limit of $U/t \ll 1$ and $k_B T \ll \mu$.

\begin{figure*}[t!]
    \centering
    \includegraphics[width=0.9\linewidth]{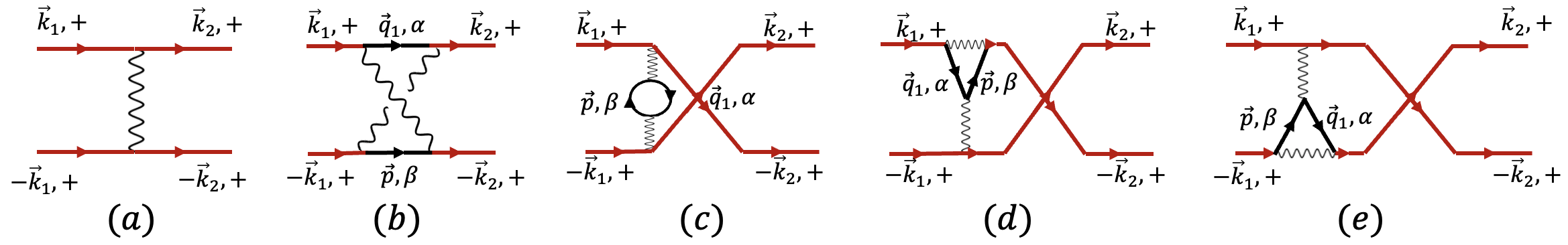}
    \caption{(a) The first-order contribution to the particle-particle (also known as Cooper channel) interaction. 
    (b–e) The one-loop Kohn-Luttinger diagrams contributing to the irreducible particle-particle interaction. Red lines indicate conduction band states, while black lines represent internal propagators, which must be summed over both conduction and valence bands. Note that each diagram $\Gamma_X(k_1,k_2)$ with $X \in \{b,c,d,e\}$ has a partner diagram where the external legs $k_2$ and $-k_2$ are crossed, corresponding to $-\Gamma_X(k_1,-k_2)$. As we are primarily interested in the odd-parity sector, these partner diagrams simply introduce an overall factor of 2, which we omit for brevity in the text but include in our numerical calculations.}
    \label{fig:KL_diagrams_long_range}
\end{figure*}

\subsection{Hamiltonian in the band basis}
The non-interacting Bloch Hamiltonian can be written as 
\begin{equation}
    H_0 = \sum_{\vec{k}} \boldsymbol{\psi}^{\dagger}(\vec{k}) h(\vec{k}) \boldsymbol{\psi}(\vec{k}),
    \label{eq:hofk}
\end{equation}
where the spinor is \( \boldsymbol{\psi}(\vec{k}) = \big(c_{\uparrow}(\vec{k}),\, c_{\downarrow}(\vec{k})\big)^{T} \), and \( c_{\sigma}(\vec{k}) \) is the annihilation operator for an electron with pseudospin \( \sigma \in \{\uparrow,\downarrow\}\). The matrix $h(\vec{k})$ is given by
\begin{equation}
    h(\vec{k}) = \vec{d}(\vec{k})\cdot \boldsymbol{\sigma}
\end{equation}
where $\boldsymbol{\sigma}$ is the vector of Pauli matrices acting in pseudospin space.
Additionally, we include a density-density interaction term with a Fourier-transformed potential $V(q)$:
\bea
H_\text{int} = \frac1{2N} \sum_{\vec{q}} V(q) \rho_\vec{q} \rho_{-\vec{q}}
\label{eq:interaction_H}
\eea
with $N$ being the number of unit cells and
\bea
\rho_{\vec{q}} = \sum_{\vec{k},\alpha_1,\alpha_2} \left\langle \vec{k},\alpha_1|\vec{k}+\vec{q},\alpha_2 \right\rangle c^\dagger_{\vec{k},\alpha_1} c_{\vec{k}+\vec{q},\alpha_2}
\eea
where $\ket{\vec{k},\alpha}$ is the Bloch eigenvector at momentum $\vec{k}$ for band $\alpha$. The Bloch eigenvectors are the columns of the unitary matrix $u(k)$ that diagonalizes the Hamiltonian (with the first column for the upper band, $\alpha=+1$, and the second for the lower band, $\alpha=-1$):
\begin{equation}
    u(k) = 
    \begin{pmatrix}
        \cos(\phi/2) &  \sin(\phi/2)\\
         \sin(\phi/2) e^{+i\Theta} & -\cos(\phi/2) e^{+i\Theta}
    \end{pmatrix},
    \label{eq:Bogoliubov}
\end{equation}
where \( \phi = \arccos\left(\frac{d_z(\vec{k})}{|\vec{d}(\vec{k})|}\right)\) and \(\Theta = \arctan\left(\frac{d_y(\vec{k})}{d_x(\vec{k})}\right)\).
After diagonalization, the free Hamiltonian $H_0$ reads
\begin{equation}
    H_0 = \sum_{\alpha, \vec{k}} \epsilon_{\alpha}(\vec{k}) c_{\vec{k},\alpha}^{\dagger} c_{\vec{k},\alpha},
\end{equation}
where \(\epsilon_\alpha (\vec{k}) = \alpha|\vec{d}(\vec{k})|\) with \(\alpha = \pm\) is the band index, and the new fermionic operators are defined by
\begin{equation}
    c_{\vec{k},\alpha}^\dagger = u_{\alpha,\uparrow}(\vec{k}) c^\dagger_{\vec{k},\uparrow} + u_{\alpha,\downarrow}(\vec{k}) c^\dagger_{\vec{k},\downarrow}\,.
\end{equation}
In the band basis, the interaction Hamiltonian becomes
\bea 
    H_{\text{int}} =&  \frac1{2N} \sum_{\alpha_i\beta_i} \sum_{\vec{k},\vec{p},\vec{q},\vec{s}}  \delta(\vec{k} + \vec{p} - \vec{q} - \vec{s}) \notag \\
    &V(\vec{k},\alpha_1 ; \vec{p},\alpha_2| \vec{q},\beta_1;\vec{s},\beta_2) 
    c_{\alpha_1\vec{k}}^{\dagger} c_{\alpha_2\vec{p}}^{\dagger} 
    c_{\beta_1 \vec{q} } c_{\beta_2\vec{s}} 
\eea
with the vertex function
\bea
    V(\vec{k},\alpha_1 ; \vec{p},\alpha_2| \vec{q},\beta_1;\vec{s},\beta_2) = V_0(\vec{p}-\vec{q}) \notag\\ 
    \braket{u_{\beta_2}(\vec{s})}{u_{\alpha_1}(\vec{k})}\braket{u_{\beta_1}(\vec{q})}{u_{\alpha_2}(\vec{p})} 
    \label{vertexfactor}
\eea
defining the interaction in this basis. This vertex function carries the information about the non-trivial band geometry of the normal state encoded in the overlaps $\braket{u_{\beta}(\vec{q})}{u_{\alpha}(\vec{p})}$.
We will mostly focus on the short-range density-density interaction, $V(q) = U$, for which this form factor can be simplified to
\begin{align}
    &V(\vec{k},\alpha_1 ; \vec{p},\alpha_2| \vec{q},\beta_1;\vec{s},\beta_2) =\\\nonumber & U \left( 
    u_{\alpha_1\uparrow}(\vec{k}) u_{\alpha_2\downarrow}(\vec{p}) 
    u^*_{\beta_1\downarrow}(\vec{q}) u^*_{\beta_2\uparrow}(\vec{s}) 
    + (\uparrow \leftrightarrow \downarrow)
    \right).
\end{align}
For a single Dirac cone this is the only short-range (i.e. intra-unit cell) interaction term possible.
In Section \ref{Multiple}, when we consider multiple Dirac cones in the presence of an additional flavor degree of freedom, there will be more options for the choice of a short-range interaction. In that case, we will also consider a purely intraorbital interaction which, in the context of the honeycomb lattice corresponds to an on-site Hubbard term.

\subsection{Effective interaction in the Cooper channel and the linearized gap equation}

We now calculate the effective interaction in the Cooper channel up to second order in the interaction $U$ (see Fig.~\ref{fig:KL_diagrams_long_range}): $\Gamma(\vec{k}_1,\vec{k}_2) = \Gamma^{(1)}(\vec{k}_1,\vec{k}_2) + \Gamma^{(2)}(\vec{k}_1,\vec{k}_2)$.

The first-order contribution, \( \Gamma^{(1)}(\vec{k}_1, \vec{k}_2) \), is determined by the bare repulsive interaction, as shown in Fig.~\ref{fig:KL_diagrams_long_range}(a), and is given by
\bea
\Gamma^{(1)}(\vec{k}_1,\vec{k}_2) = V(\vec{k}_1,+;-\vec{k}_1,+ | -\vec{k}_2,+;\vec{k}_2,+).
\eea

The second-order contribution, corresponding to diagrams (b)–(e) in Fig.~\ref{fig:KL_diagrams_long_range}, is expressed as:
\begin{align}
\Gamma^{(2)}(\vec{k}_1,\vec{k}_2) = -\frac{1}{N} \sum_{\vec{p},\alpha,\beta}  \chi_{\alpha\beta}(\vec{q}_1, \vec{p}) F_{\alpha\beta}(\vec{k}_1, \vec{k}_2; \vec{q}_1, \vec{p}),
\label{eq:Gamma(2)}
\end{align}
where \( \vec{q}_1 = \vec{p} + \vec{k}_1 + \vec{k}_2 \), and where
\begin{align}
\chi_{\alpha\beta}(\vec{q}_1,\vec{p}) \equiv \frac{n\big(\epsilon_\alpha(\vec{q}_1)\big) - n\big(\epsilon_\beta(\vec{p})\big)}{\epsilon_\alpha(\vec{q}_1) - \epsilon_\beta(\vec{p})}
\end{align}
describes the susceptibility to particle-hole excitations, with $n(\epsilon) = 1/(1+e^{\beta(\epsilon-\mu)})$ the Fermi-Dirac distribution.
The form factor \( F_{\alpha\beta}(\vec{k}_1, \vec{k}_2; \vec{q}_1, \vec{p}) \) is the sum of contributions from each diagram in (b)–(e), where each contribution is a product of two vertex functions $V$ (see Appendix~\ref{App1} for explicit expressions). 

In the limit $\mu \gg k_B T$, the linearized gap equation can be projected onto the Fermi surface, yielding:
\begin{align}
\frac{1}{\Omega} \oint_{\text{FS}} \frac{d\hat{k}_2}{|v_F(\hat{k}_2)|} \Gamma(\hat{k}_1, \hat{k}_2) f(\hat{k}_2) = \lambda f(\hat{k}_1),
\label{eq:graphene_gap}
\end{align}
where $\Omega$ is the Brillouin zone area and where $\hat{k}_{1,2}$ are on the Fermi surface. Here, \( \Gamma(\hat{k}_1, \hat{k}_2) \) is the pairing interaction kernel, and \( f(\vec{k}) \) is the gap function in the band basis, related to the Cooper pair amplitude by $f(\vec{k}) \propto \left\langle c^\dagger_{\vec{k},+} c^\dagger_{-\vec{k},+} \right\rangle$. (We will always work with models where only the conduction band ($\alpha=+1$) crosses the Fermi level, so we only consider pairing on that band).
To make the results more transparent, we will often adopt a gauge $\tilde{c}_\vec{k,\alpha} = c_\vec{k,\alpha} e^{i \Theta_\vec{k}}$ in which the gap reads
\bea
\Delta(\vec{k}) \equiv \left\langle \tilde{c}_{\vec{k},+} \tilde{c}_{-\vec{k},+} \right\rangle = f^*(\vec{k}) e^{2 i \Theta_k}.
\eea
The eigenvalue $\lambda$ determines the critical temperature via $T_c \sim t e^{-1/|\lambda|}$ for an attractive channel ($\lambda < 0$), with $t$ an energy scale on the order of the single-particle bandwidth.
It is often useful to factor out the density of states $\rho \equiv \frac{1}{\Omega} \oint_{\text{FS}} \frac{d\hat{k}}{|v_F(\hat{k})|}$ from $\lambda$ to define an effective interaction strength $V_\text{eff} \equiv \lambda / \rho$.

\section{Superconductivity of Ideal Dirac Cones}

\subsection{Single Ideal Dirac Cone: A Null Result}
\label{sec:Single Dirac cone}

\begin{figure}[t!]
    \includegraphics[scale=0.5]{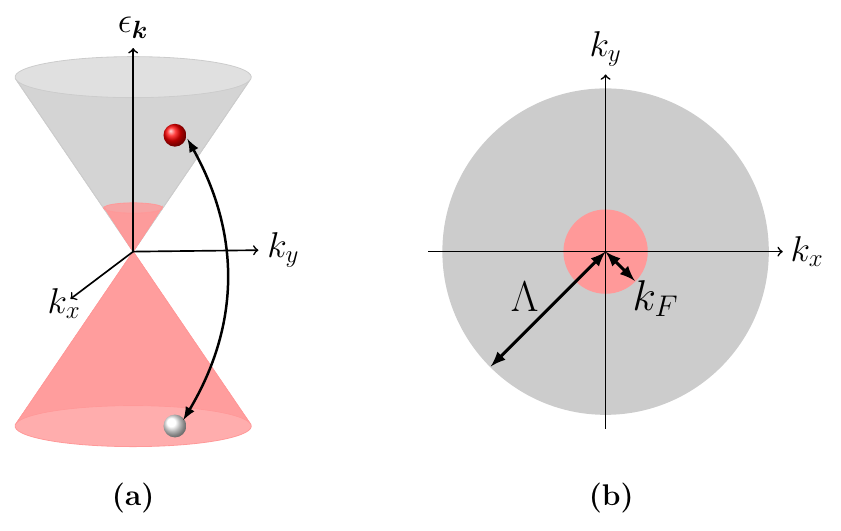}
    \caption{(a) Schematic of the massless Dirac dispersion. Occupied states are shown in red, for the case of  $\mu>0$. A virtual process which creates an interband electron-hole pair far away from the Fermi surface is also shown. As discussed in the main text, these processes dominate the effective interaction in the Cooper channel in many instances. (b) Schematic of the electronic occupation of states in the conduction band, with a Fermi surface of radius $k_F$ and the inverse lattice spacing scale $\Lambda \sim a^{-1}$. We focus on the limit $k_F \ll \Lambda$.}
    \label{fig:interband}
\end{figure}

We are now equipped to calculate the kernel $\Gamma$ and solve the linearized gap equation for a single, ideal Dirac cone, described by the Hamiltonian:
\begin{equation}
    H_0 =  v_F \sum_{|\vec{k}|<\Lambda}   \boldsymbol{\psi}^{\dagger}(\vec{k}) (\vec{k} \times \boldsymbol{\sigma} )  \boldsymbol{\psi}(\vec{k})  ,
    \label{IsotropicDiracH}
\end{equation}
where $\vec{k}\times\boldsymbol{\sigma} \equiv k_x\sigma^y - k_y\sigma^x$, $\sigma^\mu$ denotes the Pauli matrices acting in pseudospin space, and $\Lambda$ is the momentum ``cutoff'' (see Fig.~\ref{fig:interband}). Although we refer to $\Lambda$ as a cutoff, our model is only a low-energy description of an underlying lattice system, so $\Lambda$ is a \emph{physical} ultraviolet scale (of the order of the inverse lattice spacing $a^{-1}$) and is therefore never taken to infinity. For a TISS, $\Lambda$ corresponds to the momentum at which the surface state merges into the bulk bands and can thus be substantially smaller than $a^{-1}$, making it a useful parameter to keep track of~\cite{Thomale}.

The order-\( U \) contribution to the effective interaction, shown in Fig.~\ref{fig:KL_diagrams_long_range}(a), is given by
\begin{equation}
    \Gamma^{(1)}(\hat{k}_1,\hat{k}_2) = \frac{U}{2}e^{i(\theta_1 - \theta_2)},
\end{equation}
where \( \theta_i = \arctan(\hat{k}_{i,y}/\hat{k}_{i,x}) \) is the polar angle of $\hat{k}_i$ on the circular Fermi surface. The intrinsic winding of the Bloch wavefunctions around the Dirac cone thus imparts a corresponding phase winding to the interaction vertex \( \Gamma^{(1)} \). This implies that an attractive interaction ($U<0$) would favor a superconducting gap $f(k) \sim e^{i \theta_k}$, which resembles a spinless $p+ip$ superconductor, as discussed in Ref.~\cite{Fu_Kane_2008}.

In this work, however, we are interested in repulsive interactions ($U>0$). It is convenient to remove the intrinsic winding via a gauge transformation, which allows us to define a real-valued effective interaction \( \tilde{\Gamma} \) in the rotated basis:
\begin{equation}
    \tilde{\Gamma}(\hat{k}_1,\hat{k}_2) \equiv \Gamma(\hat{k}_1,\hat{k}_2) e^{-i(\theta_1 - \theta_2)}.
\end{equation}
In this rotated basis, the gap functions obtained as eigenvectors of $\tilde\Gamma$ are related to the original basis as $\tilde{f}(k) = f(k) e^{-i\theta}$ where we dropped the $k$ index for $\theta$ for brevity.
Using rotational invariance, we can decompose $\tilde\Gamma$ into its Fourier components:
\bea
\tilde\Gamma(\hat{k}_1,\hat{k}_2) =  V_{\text{eff},0} + 2 \sum_{l > 0} V_{\text{eff},l}  \cos( l (\theta_2-\theta_1) ).
\eea
In this notation, the $l=0$ channel corresponds to a gap with $\tilde{f}(k) \sim 1$ in the rotated basis ($f(k) \sim e^{i\theta}$ in the original). For $l=1$, $\tilde{f}(k) \sim e^{\pm i\theta}$ maps to two distinct forms in the original basis: $f(k) \sim e^{2i\theta}$ for the plus sign and $f(k) \sim 1$ for the minus sign. Higher-order channels follow the same structure.
For a single Dirac cone, the Pauli principle allows only odd-parity superconducting states (i.e., $f(k) = -f(-k)$), which correspond to even values of $l=0,2,4,\dots$. Thus, we only need to consider even values of $l$ here.
(Adding an additional flavor degeneracy to the Dirac cone allows for even-parity order parameters, which will be discussed separately in Section \ref{Multiple}).

After a calculation which is too long to reproduce here (see Appendix~\ref{App1}), we have calculated the second-order diagrams analytically:
\bea
\tilde{\Gamma}^{(2)}(\theta_{1},\theta_{2}) = \frac{U^2}{ 8 \pi  v_F}  \left(  \Lambda + \frac{2k_F + \Lambda}{2}  \cos(\theta_{1} - \theta_{2}) \right)
\label{eq:G_func}
\eea
which leads to this final expression for $V_{\text{eff},l}$:
\bea
\label{VeffIsotropic}
V_{\text{eff},0} &=& \frac{U}{2} +  \frac{U^2}{8 \pi   v_F}  \Lambda + \mathcal{O}(U^3) \\
V_{\text{eff},1} &=& \frac{U^2}{32 \pi v_F}    (2k_F + \Lambda) + \mathcal{O}(U^3) \\
V_{\text{eff},l \geq 2} &=& \mathcal{O}(U^3).
\eea
We conclude that the effective interaction is repulsive for $l=0,1$ and zero for $l\geq2$.
We thus find no pairing instability at order $U^2$ for a single, ideal Dirac cone, a result analogous to that for the two-dimensional electron gas (2DEG)~\cite{Chubukov_1993}.
The underlying physics, however, is quite different for Dirac fermions compared to the 2DEG. For the 2DEG, a single-band system, the effective interaction is simply $\Gamma(\bk_1,\bk_2) = U + U^2 \Pi(|\bk_1 + \bk_2|)$, where $\Pi(q)$ is the polarization bubble. Since $\Pi(q)$ is constant for $q \leq 2 k_F$, $\Gamma(\bk_1,\bk_2)$ is a positive constant for all momenta on the Fermi surface. This leads to a repulsive contribution in the $l=0$ channel and zero contribution to all higher channels, explaining the absence of pairing at $\mathcal{O}(U^2)$~\cite{Kagan_1992,Chubukov_1993}.

The situation for Dirac fermions is more subtle. Although the polarization bubble (corresponding to the diagram in Fig ~\ref{fig:KL_diagrams_long_range} c) is also independent of $q$ for $q \leq 2k_F$~\cite{DasSarmaDielectricFunction}, the pairing kernel receives contributions from additional diagrams in this case (Fig.~\ref{fig:KL_diagrams_long_range} b, d, and e). Due to momentum-dependent form factors and interband contributions for each of these diagrams, it was not a priori obvious that the Dirac cone would also exhibit no superconducting instability at this order.

\subsection{On the importance of interband excitations}
\label{Sec:interband}

A particularly important difference between the 2DEG and Dirac fermions is the presence of interband electron-hole excitations for the latter, which corresponds to the two internal legs being on different bands in the diagrams of Fig.~\ref{fig:KL_diagrams_long_range} (b-e).
We find these contributions to be crucial, leading to terms proportional to $\Lambda \sim a^{-1}$ in $\Gamma^{(2)}$ (See Eq.~\ref{eq:G_func}), in contrast to intraband contributions which give contributions proportional to $k_F$.

 The dominance of these interband contributions in the Fermi liquid regime ($k_B T \ll \mu$), for which the valence band is fully occupied at equilibrium, is somewhat surprising. In fact, it is usually assumed that the valence band can be safely disregarded in that regime~\cite{RevModPhys.84.1067}.
 In our case, the dominance of interband effects relies on the dispersion $\epsilon_\bp$ being linear as we now show. (Note however that recent works have also highlighted the importance of interband effects in models without linear dispersion~\cite{Crepel_Fu_2021,Crepel_Fu_2021_2,Crepel_Fu_2022,PhysRevResearch.5.L012009}).
 Unlike intraband electron-hole excitations which are constrained to be close to the Fermi surface, interband electron-hole excitations are summed over the entire area $k_F < |\vec{k}| < \Lambda$  (see Fig.~\ref{fig:interband}).
 These excitations give a contribution to the pairing kernel of the form:
\bea
\Gamma^{(2)}_{+-}(\bk_1,\bk_2) \sim \int d\theta \int_{k_F}^{\Lambda} dp \frac{p}{|\epsilon_\vec{p}| + |\epsilon_{\vec{p}+\vec{q}}|} F(\vec{p},\vec{q})
\eea
with $\bp = p (\cos(\theta),\sin(\theta))$, $\bq=\bk_1+\bk_2$ and $F(\bp,\bq)$ the relevant form factor involving inner products between Bloch eigenstates. 
The energy of interband particle-hole excitations appearing in the denominator increases with $p$, but only linearly so in the case of Dirac electrons with $|\epsilon_\bp| = v_F p$, so this energy penalty is compensated by the added phase space which also grows linearly with $p$.
One might worry that the form factor does decay with $p$, but this does not happen for the double exchange diagram in Fig.~\ref{fig:KL_diagrams_long_range}b for which the form factor $F(\bp,\bq)$ remains finite in the limit of $p \gg |\bq|$.
In the limit of $\Lambda \gg k_F$, the integral is thus dominated by the behavior at large $p$, and gives a contribution proportional to $\Lambda \sim a^{-1}$ (in contrast to contributions proportional to $k_F$ for intraband processes).
The question of whether that contribution proportional to $\Lambda$ is repulsive or attractive, and in which channels, is determined by the precise form of the form factor $F$.
In the case of a single ideal Dirac cone we have just discussed, the contribution proportional to $\Lambda$ is repulsive in the $l=0$ and $l=1$ channels and zero in other channels (see Eq.~\ref{eq:G_func}).
However, as we will show in Section \ref{graphene}, for the case of two Dirac cones of opposite chirality, as in graphene, this contribution is attractive and is responsible for enhanced superconductivity in the limit of small $k_F$ in the $d+id$ and $f$-wave channels.

\subsection{Multiple ideal Dirac cones}
\label{Multiple}

Before moving on to a discussion of how to stabilize superconductivity in a single non-ideal Dirac cone, we first discuss the case of multiple ideal Dirac cones.
There are two distinct ways to increase the number of Dirac cones: one can either increase the number $n_D$ of Dirac spinor components, and/or introduce additional $N_f$ flavors of fermions (following notation in Ref.~\cite{PhysRevB.103.125128}).
In this language, the single Dirac cone we studied in Section \ref{sec:Single Dirac cone} corresponds to $n_D = 2$ and $N_f =1$.

We consider two scenarios. First, we keep $n_D = 2$ and introduce an additional flavor index with $N_f=2$, which can also be described as two Dirac cones with the same chirality.
 Second, we discuss the case of graphene, with two Dirac cones of opposite chirality centered at two different valleys $K$ and $K'$ (which gives $n_D = 4$), and with a spin degree of freedom which gives $N_f=2$.

We will see that these two scenarios have drastically different behavior: for two Dirac cones of the same chirality ($n_D=2, N_f=2$), superconductivity is not generated at order $U^2$ for an ideal linear dispersion, just like the case of $N_f=1$.
By contrast, in the case of two Dirac cones of opposite chirality, superconductivity appears at order $U^2$ even for the ideal Dirac dispersion, and has a finite effective interaction in the Cooper channel $V_\text{eff}$ even as $k_F \to 0$ thanks to interband fluctuations. Having two cones with \emph{opposite} chiralities thus seems crucial to realize superconductivity in the weak coupling limit for the case of an ideal Dirac dispersion.

\subsubsection{Two Dirac cones of the same chirality ($n_D=2, N_f=2$)}

We consider a single Dirac cone with an additional twofold flavor degeneracy ($s=\pm 1$), which can also be regarded as two Dirac cones with identical chirality:
\begin{equation}
    H_0 = v_F \sum_{|\mathbf{k}|<\Lambda } \sum_{s =\pm1} \boldsymbol{\psi}^{\dagger}_{\mathbf{k},s} ( k_x \sigma^x + k_y \sigma^y)  \boldsymbol{\psi}_{\mathbf{k},s} ,
    \label{eq:TwoDiracConesSameChirality}
\end{equation}
where $s$ is the spin index and $\boldsymbol{\psi}_{\mathbf{k},s} = (c_{\mathbf{k},s,a},\, c_{\mathbf{k},s,b})^{T}$ with $a,b$ the two orbitals. Such a model can describe, for example, a spinful Qi-Wu-Zhang model~\cite{Qi2006} or a graphene-like system which is valley-polarized but not spin-polarized. 

As detailed in Appendix~\ref{App:SameChirality}, our analysis of this system with either intra-orbital or density-density interactions reveals no attractive channels at order $U^2$. 
Adding an additional flavor degree of freedom to a single Dirac cone does not, therefore, generate a pairing instability.

\subsubsection{Two Dirac cones of opposite chirality ($n_D=4$, $N_f=2$)}
\label{graphene}

Motivated by graphene, we now consider the case of two isotropic Dirac cones of opposite chirality (each with an additional degeneracy due to spin):
\bea
    H_0  =  v_F \sum_{|\vec{k}|<\Lambda }  \sum_{\nu, s =\pm1} \boldsymbol{\psi}^{\dagger}_{\bk,\nu,s} ( \nu k_x \sigma^x + k_y \sigma^y)    \boldsymbol{\psi}_{\bk,\nu,s}  ,
\eea
where $\nu$ is the valley index, $s$ is the spin index, and \( \boldsymbol{\psi}_{\bk,\nu,s} = \big(c_{\bk,\nu,s,a},\, c_{\bk,\nu,s,b}\big)^{T} \) with $a,b$ the two orbitals in the unit cell.
We consider the case of an intraorbital Hubbard interaction, which corresponds to an on-site Hubbard term on the honeycomb lattice.

As shown in Appendix~\ref{AppGraphene}, we were able to calculate the pairing kernel and solve the linearized gap equation analytically for this model.
We find that both $d\pm id$ and $f$-wave superconducting orders are favored, with a spin-singlet order parameter $\Delta_\nu(\theta) = \nu e^{\pm i \theta}$ for $d\pm id$ and a spin-triplet order parameter $\Delta_\nu(\theta) = \nu $ for $f$-wave, with $\theta$ the angle around the Fermi pockets and $\nu = \pm 1$ the pocket index. (The Fermi surface is composed of two pockets, one in each valley). On the honeycomb lattice, both pairing symmetries are realized by nearest-neighbor (and thus inter-sublattice) pairing~\cite{PhysRevB.99.165112,Li_2022}.
Both have an effective interaction in the Cooper channel (normalized by $(U/t)^2$) given by:
\bea
|V_{\text{eff},d\pm id}|, |V_{\text{eff},f}| \sim v_F  \Lambda.
\eea
Remarkably, $V_\text{eff}$ does not vanish as $k_F \to 0$.
Our analysis (see Appendix~\ref{AppGraphene}) has revealed that interband effects are crucial to favor this robust pairing in the limit of $k_F \to 0$.
As explained in Section~\ref{Sec:interband}, interband fluctuations dominate in the limit of $k_F \ll \Lambda$ since they can be integrated over a region of area $\sim \Lambda^2$, with a contribution of the form:
\bea
\Gamma^{(2)}_{+-}(\bk_1,\bk_2) \sim \int d\theta \int_{k_F}^{\Lambda} dp \frac{p}{|\epsilon_\vec{p}| + |\epsilon_{\vec{p}+\vec{q}}|} F(\vec{p},\vec{q})
\eea
The energy penalty in the denominator increases with $p$, but only linearly so in the case of Dirac electrons, and is thus compensated by the added phase space for larger $p$.
Notably, for the single Dirac cone, interband fluctuations gave a repulsive contribution, whereas here it is attractive.
This difference can be traced back to the sign change in $\Delta(k)$ between the two different valleys, which is present for both $d\pm i d$ and $f$-wave: this sign change makes it possible to turn the $\propto \Lambda$ contribution into an attractive one, which is not possible for a single Dirac cone.

For Dirac fermions, the density of states scales as $\rho \sim k_F$, and therefore the pairing eigenvalue $\lambda = \rho V_\text{eff}$ (such that $T_c \propto \exp[-1/|\lambda|]$) scales as $\lambda \sim k_F$ in this case, or equivalently $\lambda \sim \sqrt{n}$ where $n \sim k_F^2$ is the carrier density.  This finding is consistent with previous numerical results for Kohn-Luttinger SC in graphene~\cite{Kagan_2014,Chubukov_2014} which we have reproduced, see Appendix~\ref{AppGraphene}. (The scaling of $T_c$ with $k_F$ is also reminiscent of the Kondo problem in doped graphene\footnote{This is similar to the Kondo problem for graphene at finite doping, for which the Kondo temperature $T_K$ is given by $\log(T_K) \propto -1/k_F$.
(See discussion below Eq 17 in \cite{Fritz_2013}).}).
Refs~\cite{Kagan_2014,Chubukov_2014} have identified both $d\pm id$ and $f$-wave pairing as the dominant orders in the small $k_F$ limit. Although both corresponding eigenvalues scale as $\sqrt{n}$, the $\lambda_{d\pm id}$ eigenvalue has a larger numerical prefactor and is thus the predicted pairing symmetry for graphene at small carrier density(see also Ref.~\cite{PhysRevB.84.121410} for a Monte Carlo study which shows $d \pm i d$ pairing).
Our analysis shows that while the $\lambda \sim \sqrt{n}$ scaling is a universal feature of $n_D=4$ Dirac systems, the numerical prefactors are not. These prefactors are sensitive to the specific lattice regularization of the Dirac theory, as they are determined by large-momentum excitations near the lattice scale $\Lambda \sim a^{-1}$. Consequently, the relative stability of the $d \pm i d$ and $f$-wave pairing channels may differ in other microscopic models.

\section{Stabilizing superconductivity in a single non-ideal Dirac cone}

\begin{table*}[t!]
\centering
\caption{Summary of superconducting orders realized for a single non-ideal Dirac cone with repulsive interactions. }
\label{tab:pairing_symmetries_summary}
\begin{tabular}{@{} l p{3cm} p{2cm} p{6.5cm} l @{}}
\toprule
 & \textbf{Model} & \textbf{Section} & \textbf{Hamiltonian} & \textbf{Gap symmetry} \\
\midrule
\textbf{(a)} & $k^2$ mass      & \hyperref[sec:TPT]{Sec.~\ref{sec:TPT}} & $H = v_F \vec{k} \times \boldsymbol{\sigma} + (m + B k^2)\sigma_z$             & $\Delta(k) \sim (p-ip)$ \\
\addlinespace
\textbf{(b)} & $C_{3v}$ warping & \hyperref[C3]{Sec.~\ref{C3}}          & $H = v_F \vec{k} \times \boldsymbol{\sigma} + \eta (k_+^3 + k_-^3)\sigma_z$           & $\Delta(k) \sim \text{``}(d \pm id)\text{''}\times(p + ip)$ \\
\addlinespace
\textbf{(c)} & Quasi-1D limit  & \hyperref[1D]{Sec.~\ref{1D}}          & $H = 2 v_x \sin(k_x/2)\sigma^y - 2 t_y \cos(k_y)$ & $\Delta(k) \sim \mathrm{sgn}(k_x)\cos(k_y)$ \\
\bottomrule
\end{tabular}
\label{tab:pairing_symmetries}
\end{table*}

Returning to the case of a single Dirac cone, we recall that it lacks a pairing instability at the leading $\mathcal{O}(U^2)$ perturbative order. To induce superconductivity, one must therefore consider effects beyond this approximation. Two primary pathways exist: extending the perturbative expansion to third order ($\mathcal{O}(U^3)$), or incorporating higher-order momentum ($k$) terms into the single-particle dispersion. Both approaches have precedents, having been successfully applied to find superconductivity in the two-dimensional electron gas (2DEG): see Refs.~\cite{Chubukov_1993,ThirdOrderTMD} and~\cite{Chubukov_1992,Kagan_1992}, respectively. In the weak-coupling regime of interest ($U \ll t$), any $\mathcal{O}(U^2)$ pairing generated by the modified dispersion will necessarily dominate the subleading $\mathcal{O}(U^3)$ contributions. Therefore, we focus this work on the impact of higher-order dispersion terms, deferring the full $\mathcal{O}(U^3)$ analysis to a future study.

\begin{figure}[t!]
\centering
\subfloat[(a) $p-ip$ \label{fig:sub_a}]{%
    \includegraphics[width=2.2cm]{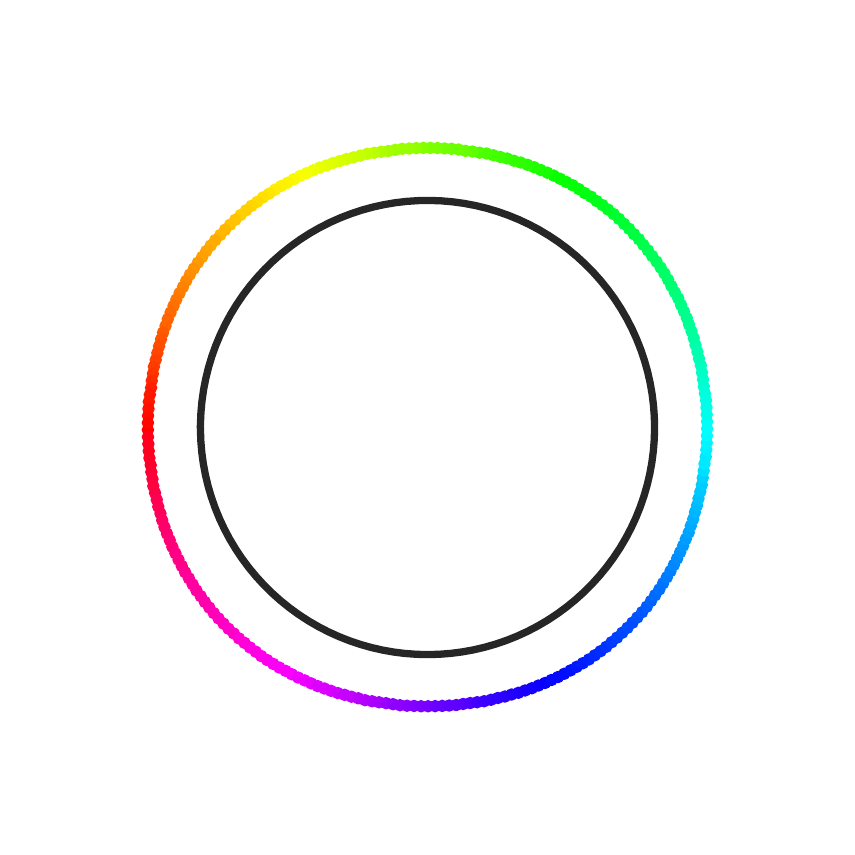}%
}\hspace*{\fill}%
\subfloat[(b) \text{``$(d \pm id)$''}$\times (p + ip)$ \label{fig:sub_b}]{%
    \includegraphics[width=2.2cm]{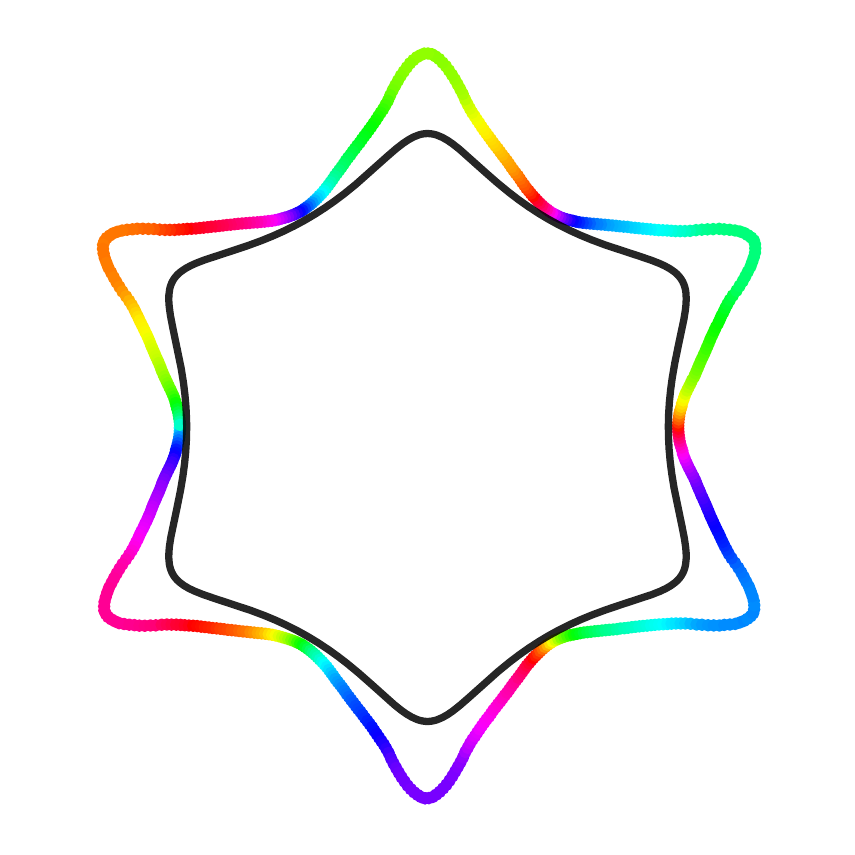}%
    \includegraphics[width=2.2cm]{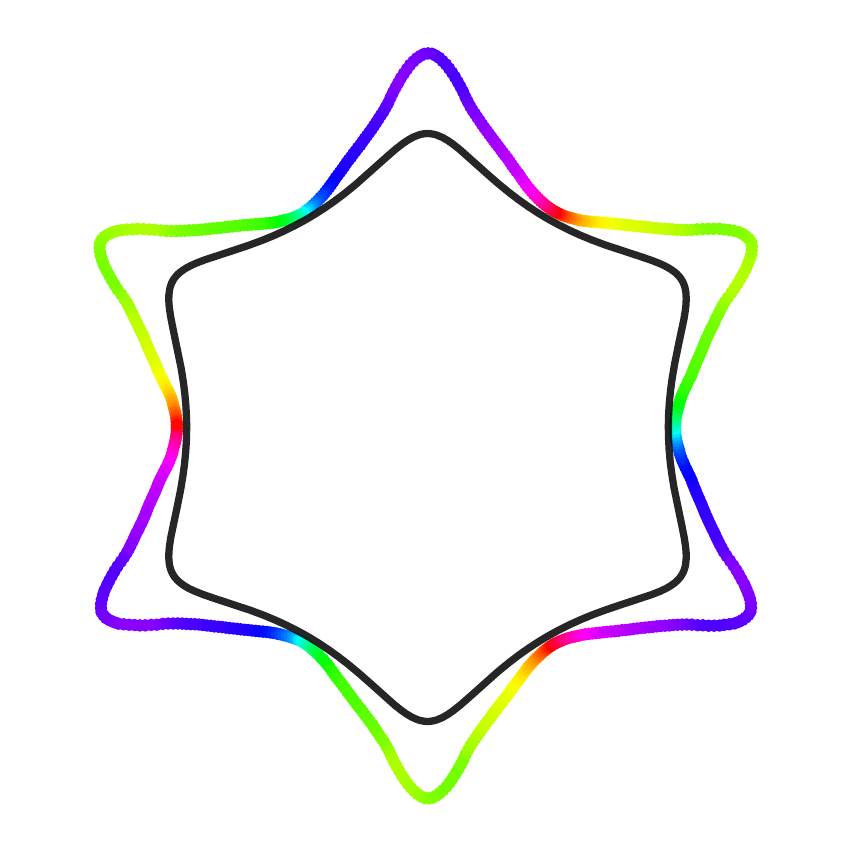}
}\hspace*{\fill}%
\subfloat[(c) $\sgn(k_x)\cos(k_y)$\label{fig:sub_d}]{%
    \includegraphics[width=2.2cm]{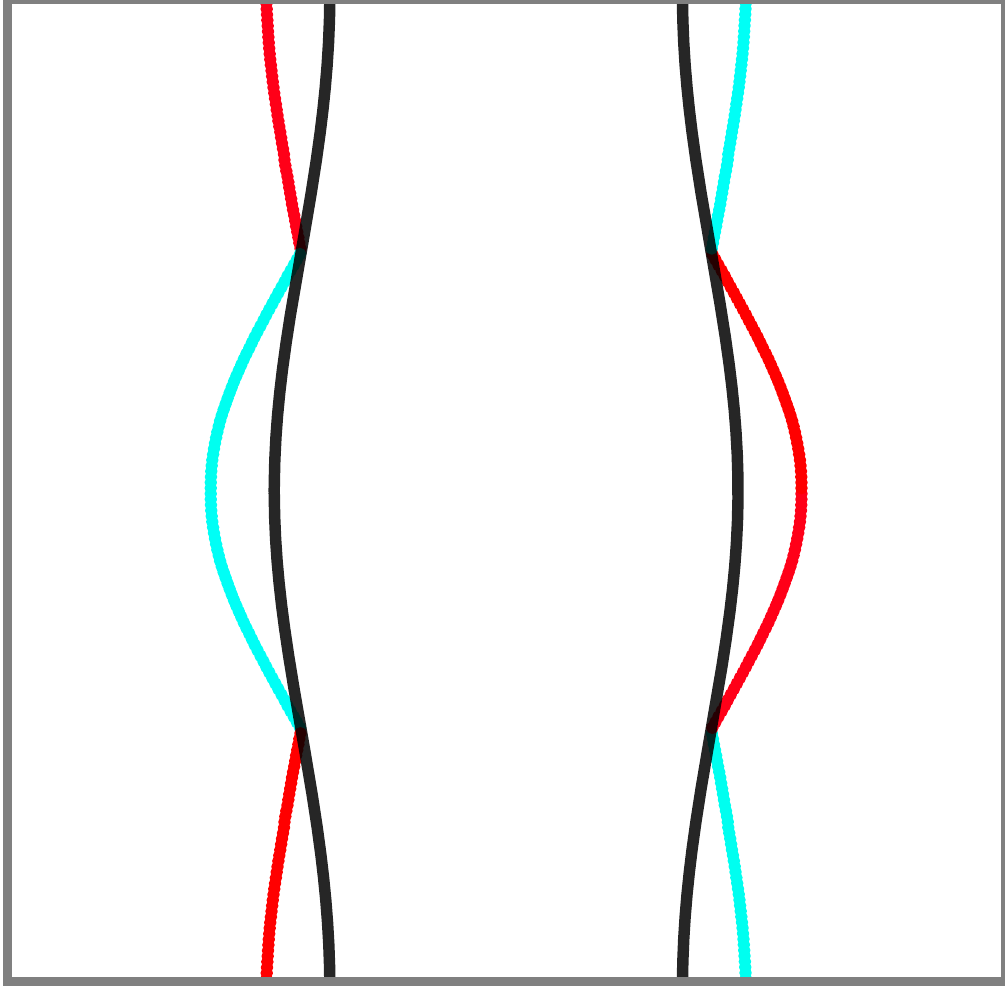}%
}

\subfloat{%
    \includegraphics[width=0.6\linewidth]{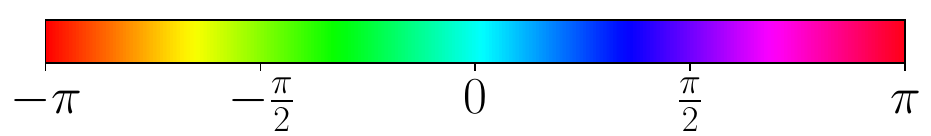}%
}
    
\caption{Visualization of the superconducting gap functions corresponding to the pairing symmetries listed in Table~\ref{tab:pairing_symmetries}. The color encodes the phase of $\Delta(\mathbf{k})$, while the distance from the Fermi surface represents its magnitude. (a) $p-ip$ state with parameters $\mu/t = 0.3$, $B/(ta^2)=-0.25$, $\Lambda a=\pi$, and $\hbar v_F/(ta)=0.5$. (b) Degenerate $(d \pm i d)(p+ip)$ states with $\mu/t = 0.8$, $\eta/(ta^3)=1/8$, $\Lambda a=2\pi$, and $\hbar v_F/(ta)=0.5$. Left: \text{``$(d - id)$''}$\times (p + ip)$; Right: \text{``$(d + id)$''}$\times (p + ip)$. (c) Nodal state $\sgn(k_x)\cos(k_y)$ with $v_x=0.5$, $\mu = 0.6$ and $t_y=0.04$.}
\label{fig:pairing_figures}
\end{figure}

\begin{figure}[t!]
    \centering
    \includegraphics[width=0.35\textwidth]{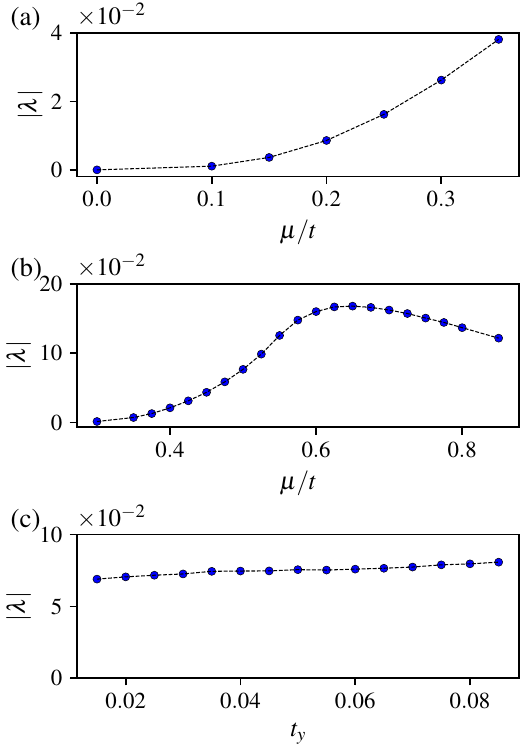}
    \caption{Comparison of the leading attractive eigenvalue $|\lambda|$ (normalized by $(U/t)^2$) for the three models in Table~\ref{tab:pairing_symmetries}. (a) Model with $B k^2 \sigma^z$ term: $v_F / (t a / \hbar)=0.5 $, $B /(t a^2) = -0.25 $, $\Lambda a = \pi $. (b) Model with $C_{3v}$ warping: $v_F / (t a / \hbar)=0.5 $, $\eta / (t a^3)= \frac{1}{8} $, $\Lambda  a = 2\pi $. (c) Quasi-1D model: $v_x=0.5$, $\mu = 0.6$. Results are in good agreement with the prediction from Eq.~\ref{1DAnalytic} which, for these parameters,  gives $|\lambda| \simeq 7.9 \times 10^{-2}$.}
    \label{fig:lambda-panel-combined}
\end{figure}

The three primary cases we study are summarized in Table~\ref{tab:pairing_symmetries} and Figs~\ref{fig:pairing_figures} and \ref{fig:lambda-panel-combined}. 
In each case, we employ a combination of numerical and analytical methods to calculate the effective interaction $\Gamma$ and solve the gap equation to find the leading SC gap symmetry $\Delta(k)$ (see Fig.~\ref{fig:pairing_figures}) and pairing strength $\lambda$ (see Fig.~\ref{fig:lambda-panel-combined}). For more details on this approach, see Refs.~\cite{Scaffidi2017,ScaffidiEA14}.

It is useful to distinguish between two types of systems that realize a single Dirac cone, as they permit different sets of symmetry-allowed higher-order terms in the dispersion: (1) a topological phase transition (TPT) between two different Chern insulators, and (2) a topological insulator surface state (TISS). We will show that superconductivity emerges in both cases, but through distinct mechanisms.

For the TPT, we will find that the breaking of time-reversal symmetry acts as the pairing glue. The quantum geometry of the bands, entering the pairing kernel through the overlaps $\braket{\bk_1,\alpha_1}{\bk_2,\alpha_2}$, generates an isotropic, fully gapped $p-ip$ superconductor.

For the TISS, the addition of higher-order terms generates an anisotropic dispersion, which generically stabilizes nodal or near-nodal superconductivity. We consider two prominent cases: (1) $C_{3v}$-symmetric warping, relevant to the surface of \ce{Bi2Te3}, which leads to a $(d \pm i d)(p+ip)$ pairing state with an enhanced critical temperature as the Fermi surface approaches a hexagonal shape; and (2) the quasi-1D limit of $v_x \gg v_y$, where the Fermi surface consists of two disconnected branches, leading to a gap of the form $\Delta \sim \sgn(k_x) \cos(k_y)$, analogous to pairing in quasi-1D organic superconductors.

\subsection{Superconductivity stabilized by time-reversal symmetry breaking at a topological phase transition}
\label{sec:TPT}

\begin{figure}[t!]
\centering
\includegraphics[width=0.9\linewidth]{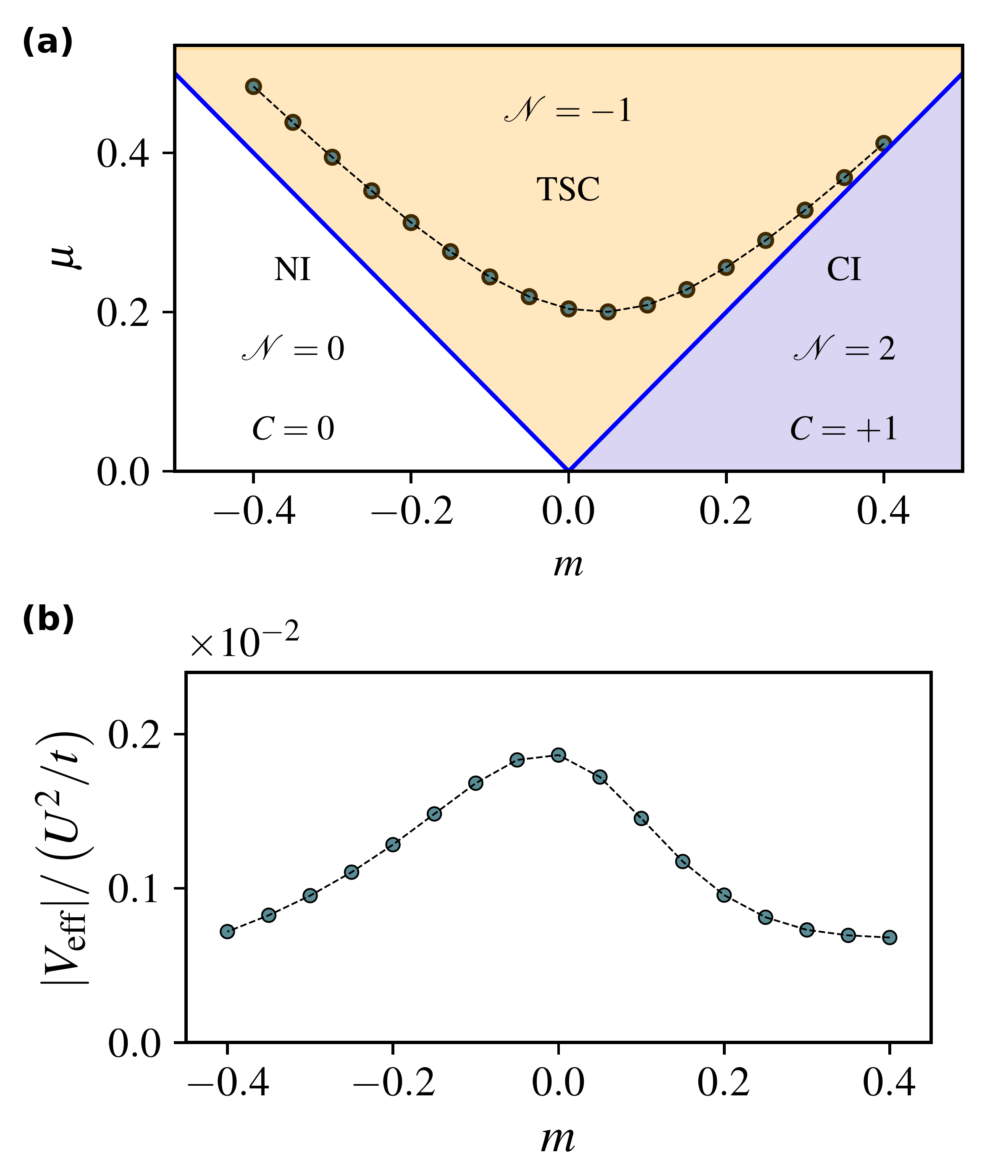}
\caption{(a) Phase diagram for the Hamiltonian in Eq.~\ref{HwithBterm} with $B<0$ and in the presence of repulsive interactions. NI denotes a normal (or trivial) insulator, CI a Chern insulator, and TSC a topological superconductor. The case of attractive interactions, discussed in Ref.~\cite{Shou-Cheng_2010}, has a similar structure, but the TSC phase has a BdG Chern number of $\mathcal{N}=+1$ instead of $-1$. The scatter points trace a contour of constant Fermi momentum, \( k_F a = 0.4 \). (b) Effective interaction \( V_{\mathrm{eff}} \) in the $p-ip$ channel as a function of the mass parameter \( m \), evaluated at \( k_F a = 0.4 \) and \( B/(t a^2) = -0.25 \). A peak is observed at the topological phase transition ($m=0$), driven by enhanced interband particle-hole excitations.}
\label{fig:combinedphasediag}
\end{figure}

A single Dirac cone can be realized on a 2D lattice if one allows for the breaking of time-reversal symmetry, in which case they are referred to as ``Wilson fermions''~\cite{PhysRevD.10.2445}.
A physical realization of such fermions occurs at the topological phase transition between two Chern insulators whose Chern numbers $C$ differ by one. The paradigmatic model for such a transition is given by~\cite{Qi2006,Shou-Cheng_2010}:
\bea
h(\bk) = v_F \vec{k} \times \boldsymbol{\sigma}  + (m + B k^2) \sigma^z 
\label{HwithBterm}
\eea
with $\vec{k} \times \boldsymbol{\sigma} \equiv k_x \sigma^y - k_y \sigma^x $, and where the $\sigma^z$ term breaks time-reversal symmetry.
Without loss of generality, we will assume $B<0$ throughout\footnote{For $B>0$, the results are modified as follows. The Hamiltonian describes a transition at $m=0$ between a normal insulator for $m>0$ (with $C=0$) and a non-trivial Chern insulator for $m<0$ (with $C=-1$). For repulsive interaction, we find a gap $\Delta(k) \sim e^{+3 i \theta}$ with BdG Chern number $\mathcal{N}=+1$. Note that the BdG Chern number does not simply correspond to the winding of $\Delta(k)$ in this case, see Appendix \ref{App:Chern} for more details on the calculation of $\mathcal{N}$. Attractive interaction would give a superconducting gap with $\mathcal{N}=-1$.}.
In that case, at zero chemical potential, this Hamiltonian describes a transition at $m=0$ between a normal insulator for $m<0$ (with $C=0$) and a non-trivial Chern insulator for $m>0$ (with $C=1$), as depicted in Fig.~\ref{fig:combinedphasediag}. 
 Here, $C \equiv \frac1{2\pi} \int d^2\bk  (\Omega_{\bk,+} n_{\bk,+} + \Omega_{\bk,-}n_{\bk,-})$ with $n_{\bk,\pm}$ the occupation number of band $\pm|\epsilon_\bk|$ and $\Omega_{\bk,\pm}$ the Berry curvature of each band, given by 
\bea
\Omega_{\bk,\pm} = \mp \frac{m - B k^2}{2 (k^2 + (m+B k^2)^2)^{3/2}}.
\eea
Upon doping, for $|\mu|>|m|$, this model describes a chiral metal with positive chirality, i.e. with $C>0$.

Topological superconductivity at a TPT described by Eq.~\ref{HwithBterm} was previously studied, but assuming an attractive interaction (or a proximity effect)~\cite{Shou-Cheng_2010,PhysRevB.92.064520,PhysRevA.111.043310}. In that scenario, Ref.~\cite{Shou-Cheng_2010} showed the emergence of an intermediate state with a single ($\mathcal{N}=1$) Majorana edge mode, with $\mathcal{N}$ the Chern number of the Bogolyubov-de Gennes (BdG) Hamiltonian (see Appendix~\ref{App:Chern}).
This state appears during the transition between a trivial Chern insulator, which has no edge modes ($\mathcal{N}=0$), and a topological Chern insulator with a chiral Dirac edge mode, which can be viewed as two chiral Majorana edge modes ($\mathcal{N}=2C=2$)~\cite{Shou-Cheng_2010}. In that context, the intermediate topological superconductor phase can be seen as an intermediate phase of $\mathcal{N}=1$ when transitioning between $\mathcal{N}=0$ (normal insulator) and $\mathcal{N}=2$ (Chern insulator).

Remarkably, we find that superconductivity appears at such a TPT even in the presence of purely \emph{repulsive} interactions, which is experimentally advantageous as it circumvents the challenges of proximity-effect engineering. 
Since we work in the weak coupling limit, we study the metallic range of $|\mu|>|m|$, in which case the normal state is a chiral metal with positive chirality ($C>0$), as mentioned earlier.
By numerically solving the gap equation, we find that the $k$-dependent mass term $B k^2 \sigma^z$ term is responsible for an attractive interaction at order $U^2$ in the $p-ip$ channel. The resulting superconductor has a gap function $\Delta(k) \sim e^{-i \theta}$ (see Fig.~\ref{fig:pairing_figures}(a)) and a superconducting Chern number $\mathcal{N} = -1$, in contrast to the $\mathcal{N}=+1$ state favored by attractive interactions. This means that, for repulsive interactions, the chirality of the superconducting phase is \emph{opposite} to the chirality of the chiral metal in the normal state above $T_c$. This finding is consistent with recent work on valley-polarized superconductors which also found $p-ip$ pairing~\cite{k8s3-dgfs,Devakul_2025}, and provides an experimental signature for repulsion-driven SC in a chiral metal, since it means that the Hall conductivity $\sigma^{xy}$ (or its thermal equivalent $\kappa^{xy}$) should flip sign between the normal state above $T_c$ and the superconducting state below it.
 We also note that $p$-wave superconductivity was predicted in a repulsive doped Dirac insulator at stronger coupling using $\epsilon$-expansion methods~\cite{BitanRoy2025}.

Treating $B$ as a small perturbation (which is justified for $B  \ll v_F \Lambda^{-1}$) and setting $m=0$ for now, we find numerically that the SC pairing strength scales as:
\bea
|V_\text{eff}|/(U^2/t^2) \sim |B| k_F^2
\eea
and thus vanishes for $B \to 0$, as expected from Section~\ref{sec:Single Dirac cone}.
It is noteworthy that for small $B$, the dispersion only deviates from the ideal dispersion at order $B^2$:
\bea
|\epsilon_\bk| = \sqrt{v_F^2 k^2 + (B k^2)^2} = v_F k + \mathcal{O}(B^2),
\eea
whereas the Bloch eigenvectors $u(\vec{k})$ are modified already at first order in $B$: 
\begin{equation}
    u(\bk) = 
    \frac{1}{\sqrt{2}}\begin{pmatrix}
        1+\frac{B |k|}{2} &  1-\frac{B |k|}{2}\\
         \left(1-\frac{B |k|}{2}\right) e^{+i\Theta} & -\left(1+\frac{B |k|}{2}\right) e^{+i\Theta}
    \end{pmatrix} + \mathcal{O}(B^2)
\end{equation}
Since we find $V_\text{eff} \propto B$, this means superconductivity is driven purely by the change in the eigenvectors $u(\bk)$ rather than a change in the dispersion $\epsilon_\bk$.
These eigenvectors enter the pairing kernel through the overlaps $\braket{\bk_1,\alpha_1}{\bk_2,\alpha_2}$ contained in the form factor $F_{\alpha,\beta}$ (see Eq.~\ref{eq:Gamma(2)}).
This provides a clear example of how the quantum geometry~\cite{Geometry} of the underlying bands (as encoded in the overlaps) can induce superconductivity in a repulsive system.

Another key feature of this model is that its superconductivity originates from the term $B k^2 \sigma^z$, which explicitly breaks time-reversal symmetry (TRS). This is somewhat unusual, as TRS breaking is traditionally thought to be detrimental to superconductivity. However, this finding is consistent with a growing number of theoretical works finding TRS-breaking terms that generate SC~\cite{PhysRevLett.121.157003,Santos,2025arXiv250616508L}, including cases that emerge at topological phase transitions~\cite{2024arXiv241018175D}. 
On the experimental front, magnetic fields have been reported to stabilize superconductivity in Bernal bilayer graphene~\cite{doi:10.1126/science.abm8386}. However, the superconductivity there appears to rely on Fermi surface anisotropy and proximity to a Zeeman-tunable van Hove singularity~\cite{PhysRevB.110.214517}, and is therefore qualitatively different from our isotropic model.




We now consider the effect of the mass term $m \sigma^z$ (while staying in the metallic phase of $|\mu|>|m|$). We find that this term does not generate superconductivity on its own; the $k^2 \sigma^z$ term is essential. The inability of the $m \sigma^z$ term to generate superconductivity is not unexpected, as in the limit of large $m$, the system reduces to a two-dimensional electron gas with quadratic dispersion, for which no pairing occurs at order $U^2$~\cite{Chubukov_1993}. In fact, we find a maximum for the pairing strength $V_\text{eff}$ centered at $m=0$ (see Fig.~\ref{fig:combinedphasediag}b), thereby producing a ``superconducting dome'' around the topological phase transition. This behavior can be attributed to the enhancement of interband electron-hole excitations, which are responsible for the pairing attraction and become strongest as the band gap closes. We have checked numerically that interband effects are indeed primarily responsible for pairing, see Appendix~\ref{BtermApp}. This model thus provides a well-controlled example where quantum critical fluctuations mediate the pairing for superconductivity, and where these fluctuations arise intrinsically from the electrons rather than being externally introduced.

We now discuss the prospects for experimental realizations. The Hamiltonian in Eq.~\ref{HwithBterm} describes the low-energy physics of lattice models such as the Qi-Wu-Zhang model~\cite{Qi2006} (see Appendix~\ref{QWZ}), which was recently realized on optical lattices~\cite{PhysRevResearch.5.L012006}.
This model also represents one spin sector of the Bernevig-Hughes-Zhang model for the Quantum Spin Hall effect~\cite{bernevig2006quantum}. In fact, a topological phase transition between different Chern insulators can be realized by magnetically doping a quantum spin Hall insulator~\cite{doi:10.1126/science.1187485,doi:10.1126/science.1234414,PhysRevB.92.064520}. Such transitions have also been observed in multilayer graphene systems, and are sometimes adjacent to superconducting domes ~\cite{PhysRevLett.127.197701,PhysRevX.15.011045,Young2025}. 

As mentioned in the introduction, particularly relevant is the ``quarter metal'' phase of graphene-based materials, in which electrons form a valley- and spin-polarized metal which spontaneously breaks time-reversal symmetry~\cite{QuarterMetal2021}.
The intra-valley superconductivity observed in rombohedral tetralayer graphene~\cite{ValleyPolarizedSC} could thus be driven by a similar mechanism (see Refs.~\cite{PhysRevB.111.174523,k8s3-dgfs,Devakul_2025,2025arXiv250700158R} for related theoretical work), especially since a $k$-dependent mass term $k^2 \sigma^z$ occurs in an effective two-band low-energy description of this system~\cite{Ghazaryan_2023}. 
The quarter metal phase in tetralayer graphene has however important differences with the model studied here: it has higher winding ($N=4$) around the Dirac point, and it has no inversion symmetry since it sits at the $K$ point.
In the absence of inversion symmetry, another quadratic term is allowed which is responsible for trigonal warping ($k_+^2 \sigma^+ + k_-^2 \sigma^-$). This term spoils the condition $\epsilon_\vec{k} = \epsilon_{-\vec{k}}$ and one would thus need to consider beyond weak-coupling methods to study pairing in such systems. An additional complication when studying intravalley pairing at the K point is the fact the finite momentum of the Cooper pairs leads to a pair density wave phase~\cite{k8s3-dgfs}.

\subsection{Superconductivity stabilized by anisotropy at the surface of a topological insulator}
\label{sec:Single_Dirac_at_the_surface_of_TI}

We now study how to stabilize superconductivity in a single Dirac cone realized as a topological insulator surface state (TISS), in the presence of weak, repulsive interactions. Again, the idea is to include higher-order in $k$ terms in the single particle Hamiltonian to see if they stabilize superconductivity at order $U^2$ in the interaction.
Based on the previous section, one might consider applying a magnetic field to break time-reversal symmetry. However, a Zeeman field only introduces a mass term $m \sigma^z$ which does not by itself stabilize superconductivity, as we have shown in the previous section. 
We therefore focus on time-reversal-symmetric higher-order momentum terms in this section, which inevitably appear due to lattice effects.
We find that these terms are not merely a correction but are in fact the primary driver of the emergent superconducting order, with the anisotropy they introduce serving as the stabilizing mechanism.

Our weak-coupling approach complements earlier studies using exact diagonalization on finite-size systems with strong interactions, which found only Fermi liquid and ferromagnetic phases for repulsive $U$~\cite{Thomale}. We surmise that the absence of superconductivity in those studies could be attributed to their use of an ideal Dirac dispersion, which we have shown has no pairing instability at order $U^2$.

We also note earlier work on interacting TISS which found gapped, symmetry-respecting surfaces, which is made possible due to the appearance of anomalous surface topological order~\cite{PhysRevX.5.011011,PhysRevB.92.125111}.
By contrast, the superconducting phases we consider here are conventional in the sense that they break the $U(1)$ symmetry, and thus need not be anomalous.

\subsubsection{$C_{3v}$ warping}
\label{C3}
\begin{figure}[t!]
    \centering
    \includegraphics[width=0.95\linewidth]{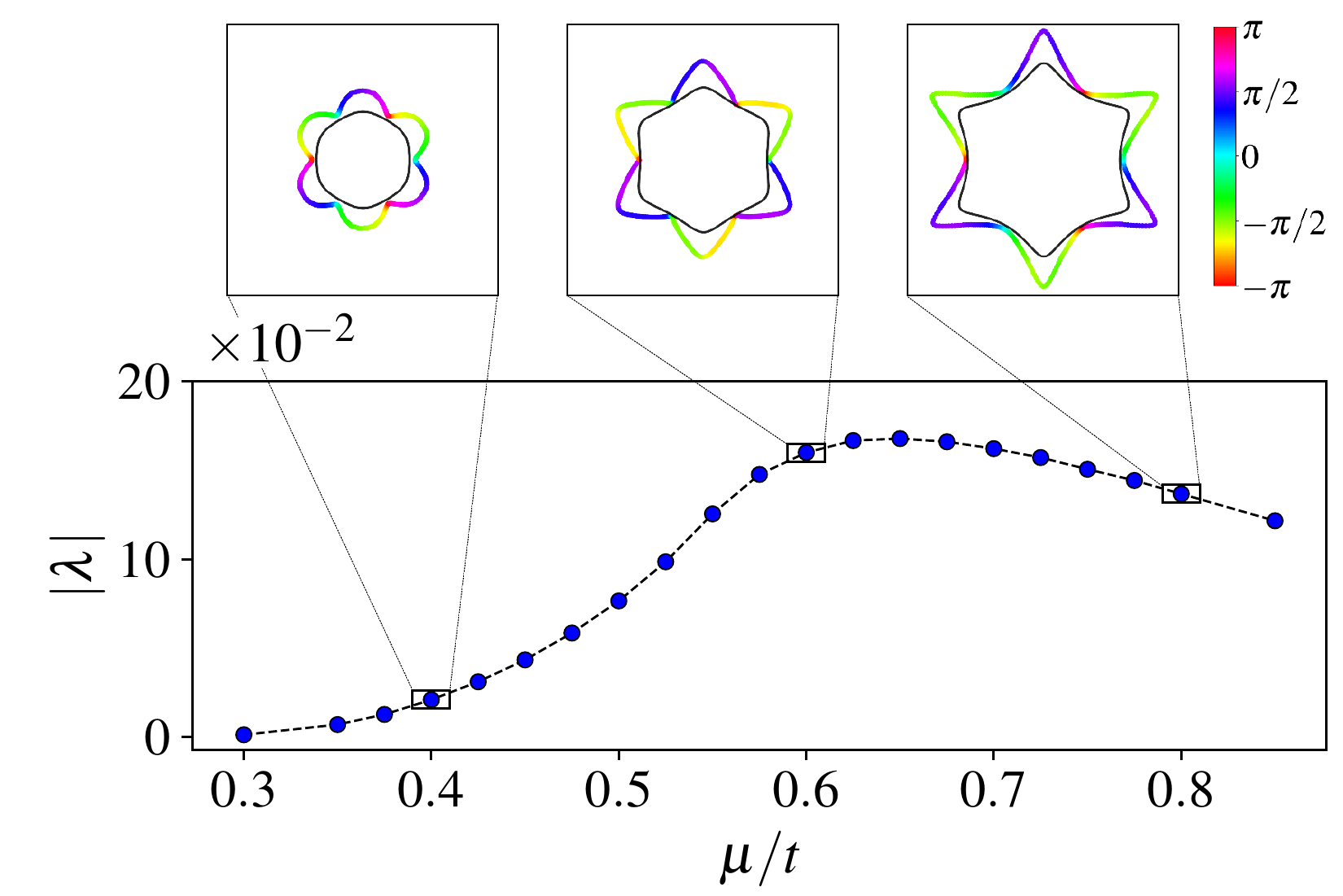}
    \caption{Pairing eigenvalue for a $C_{3v}$-warped TISS (Eq.~\ref{Hwithetaterm}) as a function of chemical potential $\mu$. The three insets at the top show the superconducting order parameter $\Delta(k)$ in the $(d+id)(p+ip)$ sector for $\mu \in \{ 0.4, 0.6, 0.8 \}$. The color encodes the phase of $\Delta(\mathbf{k})$, while the radial distance from the Fermi surface represents its magnitude. The gap magnitude has six minima which are deep but still finite, see Fig.~\ref{fig:C3-gap}. Calculations were done with $\eta/(ta^3)=1/8$, $\Lambda a=2\pi$, and $\hbar v_F/(ta)=0.5$.  }
    \label{fig:C3v_lambda}
\end{figure}

It was recognized early on that warping effects can be significant on the surfaces of TIs~\cite{ARPESBi2Te3,PhysRevLett.103.266801,PhysRevB.82.045122}. (Highly anisotropic Dirac cones can also be engineered using Moiré superlattices~\cite{PhysRevB.103.155157}). For the case of $C_{3v}$-symmetric systems, the leading-order single-particle Hamiltonian is given by
\bea
H_0 = v_F \vec{k} \times \boldsymbol{\sigma} + \eta (k_+^3 + k_-^3) \sigma^z,
\label{Hwithetaterm}
\eea
where the $\eta$ term is responsible for a hexagonal warping of the dispersion~\cite{PhysRevLett.103.266801,PhysRevB.82.045122}. As $\mu$ increases, the Fermi surface transitions from a circle, to a hexagonal shape, and eventually to a concave snowflake-like shape (see Fig.~\ref{fig:C3v_lambda}).
This model provides a good fit to angle-resolved photoemission spectroscopy (ARPES) experiments on \ce{Bi2Te3}~\cite{ARPESBi2Te3}.

We find that the inclusion of the $\eta$ term induces a superconducting instability. The pairing strength, $\lambda$, reaches its maximum in the range of $\mu$ corresponding to the transition from a convex to a concave Fermi surface (see Fig.~\ref{fig:C3v_lambda}). This enhancement is driven by nesting, as the Fermi surface in this region is approximately hexagonal. This mechanism is analogous to that discussed for graphene doped to its van Hove singularity, which also features a hexagonal Fermi surface and nesting-enhanced superconductivity~\cite{Chubukov2012}.
(Unlike in graphene, however, the density of states here does not diverge.) The presence of nesting does imply a competition with charge or spin density waves~\cite{PhysRevLett.103.266801}, which would require a parquet RG analysis for a complete description at the nesting point~\cite{Chubukov2012}. Such a calculation is beyond the scope of this work. However, for any chemical potential away from perfect nesting, the logarithmic divergence in the particle-hole channel is cut off, and a sufficiently weak interaction~$U$ will always favor the superconducting instability first. Our weak-coupling results should therefore accurately describe the system, provided it is at a sufficient distance from perfect nesting.

\begin{figure}[t!]
    \centering
    \includegraphics[width=0.45\textwidth]{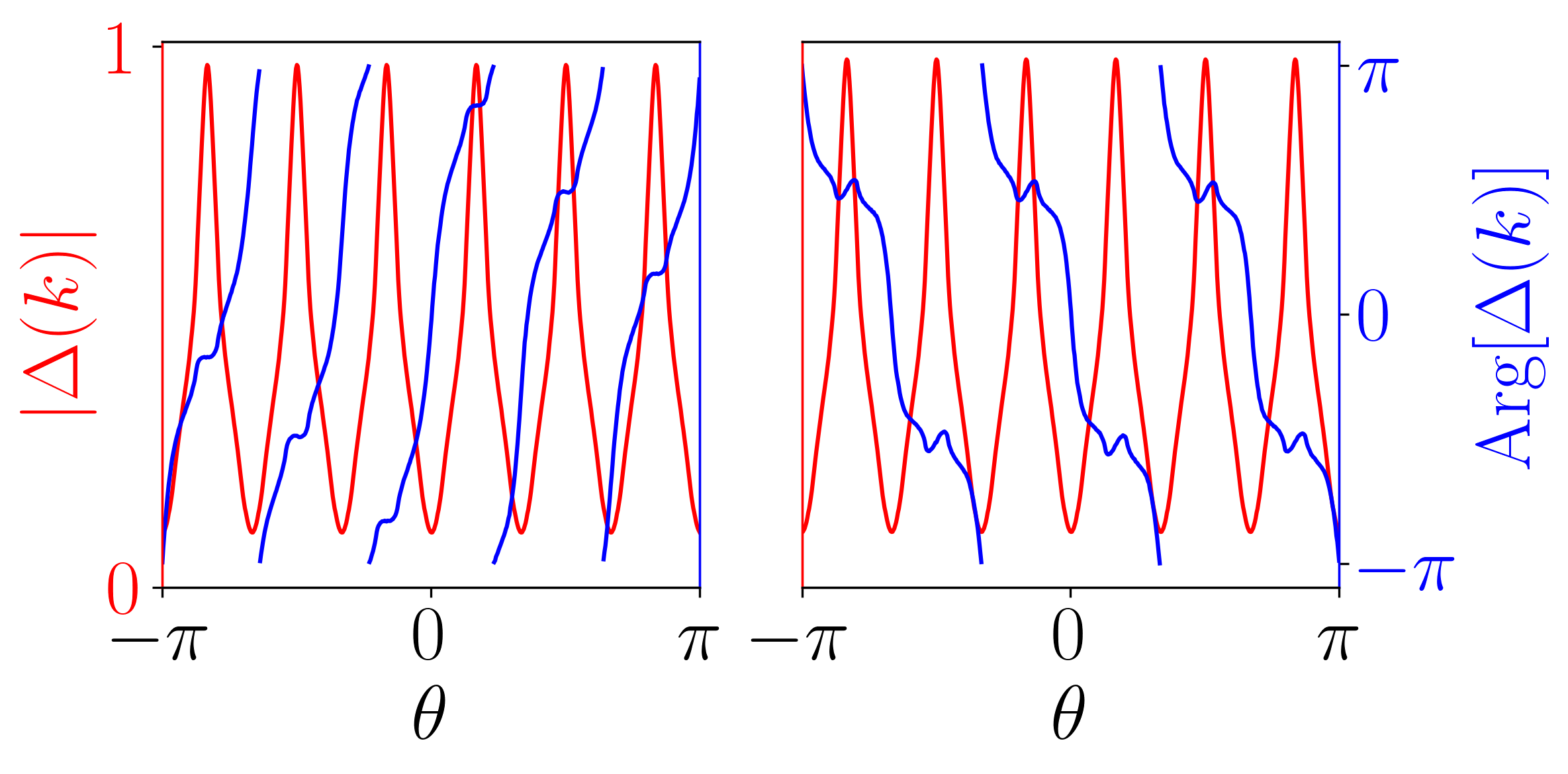}
    \caption{The magnitude and phase of the two degenerate gap functions, $\Delta(k)$, for the $C_{3v}$-warped model at $\mu/t=0.8$. Left: $(d-id)(p+ip)$. Right: $(d+id)(p+ip)$.}
    \label{fig:C3-gap}
\end{figure}

The resulting superconducting order parameter belongs to the $E$ representation of the $C_{3v}$ point group (See Appendix ~\ref{appsym}), with a gap of the form $\Delta(k) \sim \text{``}(d \pm i d)\text{''}(p+ip)$, which spontaneously breaks time-reversal symmetry (see Fig.~\ref{fig:C3v_lambda} and Fig.~\ref{fig:C3-gap}). 
This order parameter can be regarded as an analog of the $d\pm id$ state predicted for the hexagonal FS of graphene doped to its van Hove singularity~\cite{Chubukov2012}, the main difference being that the hexagonal FS in our case encloses a Dirac point, hence the extra factor of $(p+ip)$ in $\Delta(k)$.

Note however that, although the gap is in the $E$ representation, it exhibits deep minima for $|\Delta(k)|$ which are accidental, i.e. not enforced by symmetry. This is the reason why we used quotes around $d \pm i d$.
As shown in Fig.~\ref{fig:C3v_lambda}, $|\Delta(k)|$ becomes very small in the middle of the edges of the hexagonal Fermi surface. This is due to a close competition between two harmonics in the same representation: for the $(d-id)(p+ip)$ state, the gap has the form $\Delta(k) \sim e^{i \theta} (A e^{- 2i \theta} + B e^{4 i \theta})$ with $|B| \gtrsim |A|$. For the $(d+id)(p+ip)$ state, the gap has the form $\Delta(k) \sim e^{i \theta} (A^* e^{+ 2i \theta} + B^* e^{-4 i \theta})$.
Figure~\ref{fig:C3-gap} shows that the gap remains finite at its minima and displays a total phase winding of $+5$ for $(d-id)(p+ip)$ and $-3$ for $(d+id)(p+ip)$.

Even with deep minima, the gap is fully open, allowing an analysis of its topological properties. In the case of attractive interactions (which is equivalent to the state proposed by Fu and Kane~\cite{Fu_Kane_2008}), the state with $\Delta(\vec{k}) \sim e^{i\theta}$ state is adiabatically connected to a $p+ip$ superconductor of spin-down electrons and hosts a Majorana zero mode in vortices~\cite{RevModPhys.83.1057}. Similarly, our state can be adiabatically connected to a superconductor of only spin-down electrons by applying a magnetic field $m \sigma^z$ with $m < 0$ (for $\mu > 0$). By tuning $\mu$ just above $|m|$, we are left with effectively spin-polarized quadratic fermions. Since $\Delta(\vec{k})$ is given in a gauge where the spin-down electron Bloch wavefunction has no momentum-dependent phase winding, one can directly read off the superconducting Chern number $\mathcal{N}$ of the resulting state from the winding of $\Delta(\vec{k})$ around the Fermi surface.
We thus find that the $(d-id)(p+ip)$ state is adiabatically connected to a superconductor for spin-down electrons with $\mathcal{N}=+5$, while the $(d+id)(p+ip)$ state is connected to a state with $\mathcal{N}=-3$. In either case, these states are adiabatically connected to superconductors with an odd Chern number and thus host a single Majorana zero mode in vortices, just like the state in the original Fu-Kane proposal~\cite{Fu_Kane_2008}.


A chiral superconducting gap with accidental deep minima seems to be a common feature of the Kohn-Luttinger mechanism in 2D. For instance, it also occurs in the square lattice Hubbard model, where the order parameter stabilized at weak coupling in the $E_u$ irreducible representation is often dominated by higher lattice harmonic (e.g. $\sin(3 k_x) \pm i \sin(3 k_y)$) with accidental (near-)nodes~\cite{ScaffidiEA14,PhysRevB.94.085106,Thomas_2018_inter}. This can be understood by noting that, unlike in 3D, rotationally-symmetric models in 2D tend to have no superconducting instability at order $U^2$: this was shown for the 2DEG in Ref.~\cite{Chubukov_1993}, and we have shown it to be true as well for Dirac fermions in Section~\ref{sec:Single Dirac cone}. Consequently, anisotropic terms in the single-particle Hamiltonian are responsible for the non-zero pairing eigenvalues. The favored gap structure is thus dictated by the specific warping of the Fermi surface, which promotes sign changes between different regions of the Fermi surface, which in turn generates accidental near-nodes in the gap.

In this context, the $H_0 = v_F \vec{k} \times \boldsymbol{\sigma} + B k^2 \sigma^z$ model we studied in Sec.~\ref{sec:TPT} is remarkable because it produces an isotropic chiral superconducting gap from repulsion---a seemingly rare outcome in 2D.
This suggests that explicitly breaking time-reversal symmetry in the Hamiltonian is a promising route to stabilizing fully gapped topological superconductors in 2D, a somewhat counterintuitive conclusion since TRSB is usually considered detrimental to superconductivity.

\subsubsection{Quasi-1D limit}
\label{1D}

\begin{figure}[t!]
    \centering
    \includegraphics[width=0.95\linewidth]{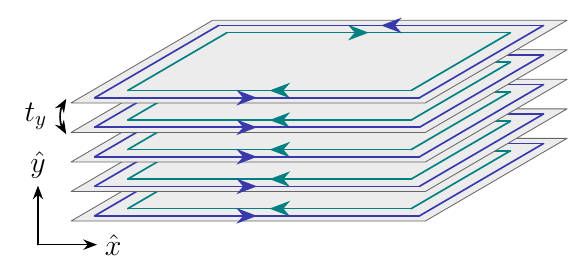}
    \caption{Schematic of a layered topological insulator with helical edge modes (in blue and teal, corresponding to different spin orientations) in the limit of small interlayer coupling. We are interested in the anisotropic topological surface state appearing on one of the side surfaces (shown as the $x,y$ plane here).}
    \label{fig:quasi1dschematic}
\end{figure}

For surfaces with low symmetry, the single-particle Hamiltonian $H = v_x k_x \sigma^y - v_y k_y \sigma^y$ can already be anisotropic at linear order in $k$ by having $v_x \neq v_y$.
We focus on the quasi-1D limit of $v_x \gg v_y$, which is experimentally motivated by the side surfaces of layered topological insulators for which inter-layer hopping is comparatively small, like \ce{ZrTe5}~\cite{ZrTe5} and \ce{TaSe3}~\cite{TaSe3}. In that case, the $x$ direction is taken along the layer, and the $y$ direction is taken perpendicular to the layers (See Fig.~\ref{fig:quasi1dschematic}).

\begin{figure}[t!]
    \centering
    \includegraphics[width=0.55\linewidth]{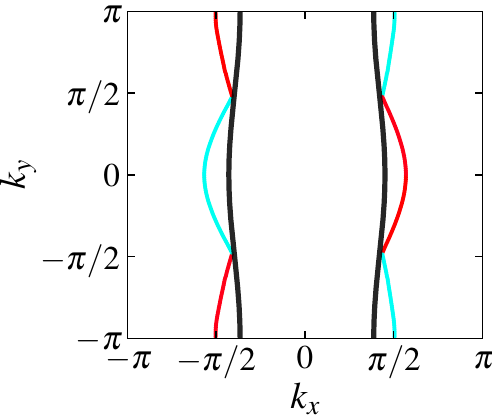}
    \caption{Anisotropic surface state appearing on the $x, y$ side surface of a layered topological insulator (see Fig.~\ref{fig:quasi1dschematic}), calculated based on Eq.~\ref{1DLattice} for parameters $v_x=0.5$, $\mu = 0.6$ and $t_y=0.04$. The Fermi surface is shown in black and is made of two nearly straight branches. The gap $\Delta(k) \sim \sgn(k_x) \cos(k_y)$ is shown in color, with the distance from the Fermi surface giving $|\Delta(k)|$ and the color giving the sign. }
    \label{fig:quasi1dfs}
\end{figure}

In the very anisotropic limit, the Fermi surface splits into disconnected branches (see Fig.~\ref{fig:quasi1dfs}) which wrap around $k_y$.
To capture the periodicity of the Brillouin zone in such systems, it is necessary to place the model on a lattice along the $y$-direction. A natural starting point is the coupled-wire construction for an anisotropic Dirac cone from Ref.~\cite{PhysRevX.5.011011}, which gives the Hamiltonian $H(k) = v_x k_x \,\sigma^y - v_y \sin\!\left(k_y/2\right) \sigma^x$ with $k_y \in [-\pi, \pi[$. This model produces a single Dirac cone at $\mathbf{k} = 0$.  

In the strongly anisotropic limit $\mu, v_x \gg v_y$, the Fermi surface consists of two nearly straight, disconnected branches at $k_x = \pm k_F(k_y)$, with only a weak corrugation in the $k_y$-direction (see Fig.~\ref{fig:quasi1dfs}). From earlier work on the quasi-1D limit of the square-lattice Hubbard model~\cite{PhysRevB.107.014505}, such a geometry is known to favor superconductivity, driven by the nesting between these two branches. Because the pairing kernel $\Gamma$ in this regime is dominated by intraband scattering processes with momenta close to the Fermi surface, the precise band structure away from the Fermi surface is unimportant.

This motivates two simplifications of the Hamiltonian. First, instead of $v_x k_x \sigma^y$, we use the lattice-regularized form $2 v_x \sin(k_x/2)\,\sigma^y$ with $k_x \in [-\pi,\pi)$. Second, the main role of the $\sin(k_y/2)\,\sigma^x$ term is to induce a corrugation $k_F(k_y) = k_{F,0} + \delta k_F \cos(k_y)$ in the Fermi surface. It also produces a small $k$-dependent rotation of the spin away from the $y$-axis, but this effect is negligible for $v_x \gg v_y$. We therefore replace $\sin(k_y/2)\,\sigma^x$ by a spin-conserving hopping term $-2 t_y \cos(k_y)\,\sigma^0$, which produces the same Fermi-surface corrugation without altering the spin texture.  

With these simplifications, the effective model becomes
\begin{equation}
\label{1DLattice}
H_0 = 2 v_x \sin\!\left(\frac{k_x}{2}\right)\sigma^y - 2 t_y \cos(k_y) \sigma^0,
\end{equation}
which we henceforth refer to as the quasi-1D TISS model. This should accurately describe a quasi-1D topological-insulator surface state when $v_x, \mu \gg t_y$ for momenta close to the Fermi surface. We have verified that similar results are obtained if we keep the original $k_y$-term, i.e. in the model $H = 2 v_x \sin(k_x/2) \sigma^y - 2 v_y \sin(k_y/2) \sigma^x$ with $v_y \ll v_x$.

We now apply the two-band Kohn--Luttinger formalism developed in Sec.~\ref{sec:1} to the quasi-1D TISS model in Eq.~\ref{1DLattice} with a repulsive short-range interaction \(U\). We construct the effective interaction in the Cooper channel numerically from Eq.~\ref{eq:Gamma(2)} and solve the linearized gap equation in Eq.~\ref{eq:graphene_gap}. We find that the leading superconducting instability has the form
\begin{equation}
    \Delta(\bk)\propto \mathrm{sgn}(k_x)\cos(k_y),
\end{equation}
as illustrated in Fig.~\ref{fig:quasi1dfs}. The corresponding pairing eigenvalue as a function of \(t_y\) is shown in Fig.~\ref{fig:lambda-panel-combined}(c).

Numerically, we find that the pairing kernel is dominated by a logarithmic enhancement of the intraband contribution near the nesting wavevector \(\mathbf{Q}=(2k_F,\pi)\) connecting the two Fermi branches. Microscopically, \(\Delta(\bk)\propto \mathrm{sgn}(k_x)\cos(k_y)\) corresponds to \(p\)-wave pairing along \(x\) together with pairing between neighboring chains along \(y\). This nesting-driven interchain pairing is reminiscent of the quasi-1D organic superconductors~\cite{lebed2008physics,CRPHYS_2024__25_G1_17_0}.

A closely related mechanism appears in the quasi-1D limit of the square-lattice Hubbard model with dispersion \(\epsilon(\bk)=-2t_x\cos k_x-2t_y\cos k_y\) and \(t_x\gg t_y\). In that setting, Ref.~\cite{PhysRevB.107.014505} analytically identified a logarithmic divergence arising from nesting between two disconnected Fermi branches. While both models feature nested quasi-1D Fermi surfaces, their internal structure differs in an important way: the quasi-1D TISS is a two-band, effectively spinless system (each Fermi branch is fully pseudospin-polarized), whereas the quasi-1D Hubbard model is single-band with a twofold spin degeneracy. Consequently, the TISS supports only odd-parity superconducting order parameters, while the Hubbard model also admits an even-parity spin-singlet state \(\Delta(\bk)\propto \cos(k_y)\), which forms an accidental singlet--triplet degeneracy with \(\Delta(\bk)\propto \mathrm{sgn}(k_x)\cos(k_y)\) in the regime discussed in Ref.~\cite{PhysRevB.107.014505}.

That said, within the odd-parity sector our results are broadly consistent with Ref.~\cite{PhysRevB.107.014505}. In particular, we find a pairing eigenvalue that depends only weakly on \(t_y\) and approaches a finite value as \(t_y\to 0\), close to the analytic estimate of Ref.~\cite{PhysRevB.107.014505} (see Fig.~\ref{fig:lambda-panel-combined}(c)):
\begin{equation}
\label{1DAnalytic}
|\lambda|=\frac{U^2}{2(2\pi)^2 v_F^2},
\end{equation}
where \(v_F\) is the Fermi velocity averaged over the Fermi surface.

This similarity can be understood as follows. For the quasi-1D TISS model of Eq.~\ref{1DLattice}, the conduction-band pseudospin is fully polarized along \(+\hat y\) for \(k_x>0\) and along \(-\hat y\) for \(k_x<0\), so that the Bloch spinors reduce to the form
\(u_{+,\sigma^y}(\bk)=\delta_{\sigma^y,\mathrm{sgn}(k_x)}\) with \(\sigma^y=\pm1\).
(Since our numerics indicate that intraband processes dominate, we focus on the conduction band only.) Using this simple form for the pseudospin texture, the second-order pairing kernel in Eq.~\ref{eq:Gamma(2)} simplifies dramatically: as shown explicitly in Appendix~\ref{app:quasi1D_formfactors}, the form-factor contributions from diagrams (c), (d), and (e) in Fig.~\ref{fig:KL_diagrams_long_range} cancel identically, leaving diagram (b) as the only nonvanishing second-order contribution. As a result, the intraband part of the second-order kernel reduces to
\begin{equation}
\Gamma^{(2)}_{\mathrm{intra}}(\bk_1,\bk_2)
= U^2 \!\!\int_{\substack{p_x<0\\ q_{1x}>0}}\frac{d^2\bp}{(2\pi)^2}\,
\frac{n\!\left(\epsilon(\bp)\right)-n\!\left(\epsilon(\bq_1)\right)}
{\epsilon(\bq_1)-\epsilon(\bp)},
\label{eq:Gamma(2)bis}
\end{equation}
with \(\bq_1\equiv \bp+\bk_1+\bk_2\). Here we take \(\bk_1\) on the \(k_x>0\) Fermi branch without loss of generality.

Equation~\eqref{eq:Gamma(2)bis} is closely analogous to the susceptibility evaluated in Ref.~\cite{PhysRevB.107.014505} for the quasi-1D Hubbard model, with two differences: (i) the dispersion here is \(\epsilon(\bk)=2v_x\big|\sin(k_x/2)\big|-2t_y\cos k_y\) rather than \(\epsilon(\bk)=-2t_x\cos k_x-2t_y\cos k_y\), and (ii) the Hubbard-model calculation does not impose the restriction \(p_x<0\), \(q_{1x}>0\). Neither difference is expected to affect the qualitative behavior. Indeed, Ref.~\cite{PhysRevB.107.014505} showed that the dominant contribution arises from momenta with \(p_x\simeq -k_F\) and \(q_{1x}\simeq +k_F\), where the integrand is sharply peaked. This justifies linearizing the dispersion near the Fermi surface; after linearization, the difference in \(\epsilon(\bk)\) enters only through the Fermi velocity \(v_F\). Moreover, since the peak lies well within the domain \(p_x<0\), \(q_{1x}>0\) for any finite \(k_F\), the additional restriction in Eq.~\eqref{eq:Gamma(2)bis} is essentially immaterial.

\section{Experimental outlook}
\label{sec:exp}

Although our analysis relied on idealized Dirac cone models, our work still provides testable consequences for experiments.
The first example arises in multilayer graphene. The valley and spin-polarized ``quarter-metal'' phase has been observed experimentally~\cite{QuarterMetal2021} and was shown to realize superconductivity \cite{ValleyPolarizedSC}. Our analysis predicts that in such systems, repulsive interactions stabilize a $p-ip$ state whose chirality is opposite to that of the parent chiral metal. This prediction leads to a falsifiable consequence: the Hall response (electrical or thermal) should flip between the normal state and the superconducting state, which could be probed using optical probes (Kerr effect)~\cite{Charbonneau2009}, magnetic imaging~\cite{Bert2011,Uri2019,Tschirhart2021}, or thermal transport~\cite{majorana}.


Our work also provides falsifiable predictions on the emergence of intrinsic superconductivity 
at the surface of a topological insulator, showing that repulsion-driven pairing in such 
systems is controlled by the anisotropy of the band structure and providing a correspondence between the anisotropy of the band structure and the anisotropy of the preferred gap. 
Because the anisotropy of the band structure can be routinely measured with ARPES~\cite{ARPESBi2Te3}, our work provides a direct way to deduce a candidate gap structure based on such measurements. 
For instance, in the case of $C_{3v}$-symmetric warping (as in \ce{Bi2Te3}), we predict a 
$(d \pm id)\times(p+ip)$ order parameter with deep, accidental near-nodes at the centers of the 
hexagonal Fermi-surface edges, 
while in layered topological insulators such as \ce{ZrTe5} and \ce{TaSe3}~\cite{ZrTe5,TaSe3}, 
corresponding to the quasi-one-dimensional limit $v_x \gg v_y$ of the Dirac cone, 
Fermi-surface nesting instead favors a gap of the form 
$\Delta(\vec{k}) \sim \mathrm{sgn}(k_x)\cos(k_y)$ with line nodes at $k_y = \pm \pi/2$. 
In both cases, the resulting nodal structure could be observed directly through ARPES~\cite{Damascelli2003}
or quasiparticle interference~\cite{Hoffman2002}, and would manifest in characteristic power-law scaling of 
thermodynamic and transport properties such as specific heat, penetration depth, and thermal conductivity~\cite{Sigrist1991}.

\section{Conclusion}
\label{sec:conclusion}

In this work, we addressed the question of when a single Dirac cone develops superconductivity from purely repulsive interactions. Our central finding is that an ideal, perfectly linear Dirac cone does not exhibit a superconducting instability at the leading weak-coupling order ($\mathcal{O}(U^2)$). 
However, we have shown that superconductivity is robustly restored once higher-order in momentum terms are included in the dispersion relation. These terms, which are unavoidable in any lattice realization of a Dirac cone, play a decisive role in determining the fate of the system and the symmetry of the resulting superconducting state. We explored three physically distinct scenarios, each yielding a unique pairing mechanism and gap structure.

First, for a single Dirac cone realized thanks to a breaking of time-reversal symmetry---which occurs at a topological phase transition between different Chern insulators, or in valley-polarized phases--- the term $k^2 \sigma^z$ stabilizes a fully gapped, topological $p-ip$ superconductor. This pairing is driven by the non-trivial quantum geometry and Berry curvature of the bands, with interband scattering providing the essential attractive glue. Notably, the chirality of the resulting superconductor is opposite to that of the chiral metal above $T_c$, highlighting a distinctive feature of repulsion-driven pairing in topological bands.

Second, for a Dirac cone on the surface of a 3D topological insulator, time-reversal-symmetric warping terms induce anisotropy, which in turn drives superconductivity. For the case of $C_{3v}$ warping, relevant to materials like \ce{Bi2Te3}, we found that the pairing strength is maximized when the Fermi surface becomes hexagonal. This leads to a time-reversal-symmetry-breaking state of the form $(d \pm id)(p+ip)$, characterized by a topological gap with accidental near-nodes.

Third, in the quasi-1D limit ($v_x \gg v_y)$ relevant to the side surfaces of layered topological insulators, the Fermi surface splits into two disconnected, quasi-one-dimensional sheets. Here, superconductivity is driven by a nesting mechanism analogous to that in organic superconductors, stabilizing a nodal gap with $\sin(k_x)\cos(k_y)$ symmetry.

We also reviewed the case of multiple ideal Dirac cones, revealing a starkly different picture. Specifically, a system with two Dirac cones of opposite chirality, such as graphene, exhibits a robust superconducting instability at order $\mathcal{O}(U^2)$ even without higher-order band structure terms. The pairing mechanism in this case is driven by interband electron-hole excitations, which, in contrast to the single-cone case, generates a strong attraction. This is facilitated by a gap function that changes sign between the two valleys, such as in the favored $d\pm id$ or $f$-wave channels, allowing the attraction to persist down to the $k_F \to 0$ limit. By contrast, a system of two Dirac cones with the same chirality shows no such instability.

Our results demonstrate that the details of the band structure beyond the linear Dirac approximation are not merely minor corrections but are, in fact, the primary drivers of intrinsic superconductivity from repulsive interactions. The diverse array of pairing symmetries we predict—from fully gapped chiral p-wave to nodal and near-nodal states—could be tested experimentally. More generally, our work underscores the crucial importance of including interband electron-hole excitations within the Kohn-Luttinger formalism, even when only a single band crosses the Fermi level.

\begin{acknowledgments}
This work was supported by the U.S. Department of Energy, Office of Science, Office of Basic Energy Sciences under Early Career Research Program Award Number DE-SC0025568. OT gratefully acknowledges the support of the Eddleman Quantum Institute (EQI) graduate fellowship. We gratefully acknowledge discussions with Luis Jauregui and Javier Sanchez-Yamagishi.
\end{acknowledgments}

\bibliography{KL_SC_fixed}

\newpage\clearpage
\appendix
\onecolumngrid
\setlength{\belowcaptionskip}{0pt}

\begin{center}
    \large\textbf{Appendices}
\end{center}


\section{Kohn-Luttinger for a single ideal Dirac cone}
\label{App1}


We consider an ``ideal'' Dirac cone with a purely linear dispersion: 
\begin{equation}  
H_0 = v (-k_y \sigma_x + k_x \sigma_y),  
\label{eq:H_TI}  
\end{equation}  
where \( \sigma \) represents the Pauli matrices acting in the pseudospin $\{\uparrow,\downarrow\}$ space and $v$ is the Fermi velocity.
\begin{figure}[h!]
    \centering
    \includegraphics[width=0.2\linewidth]{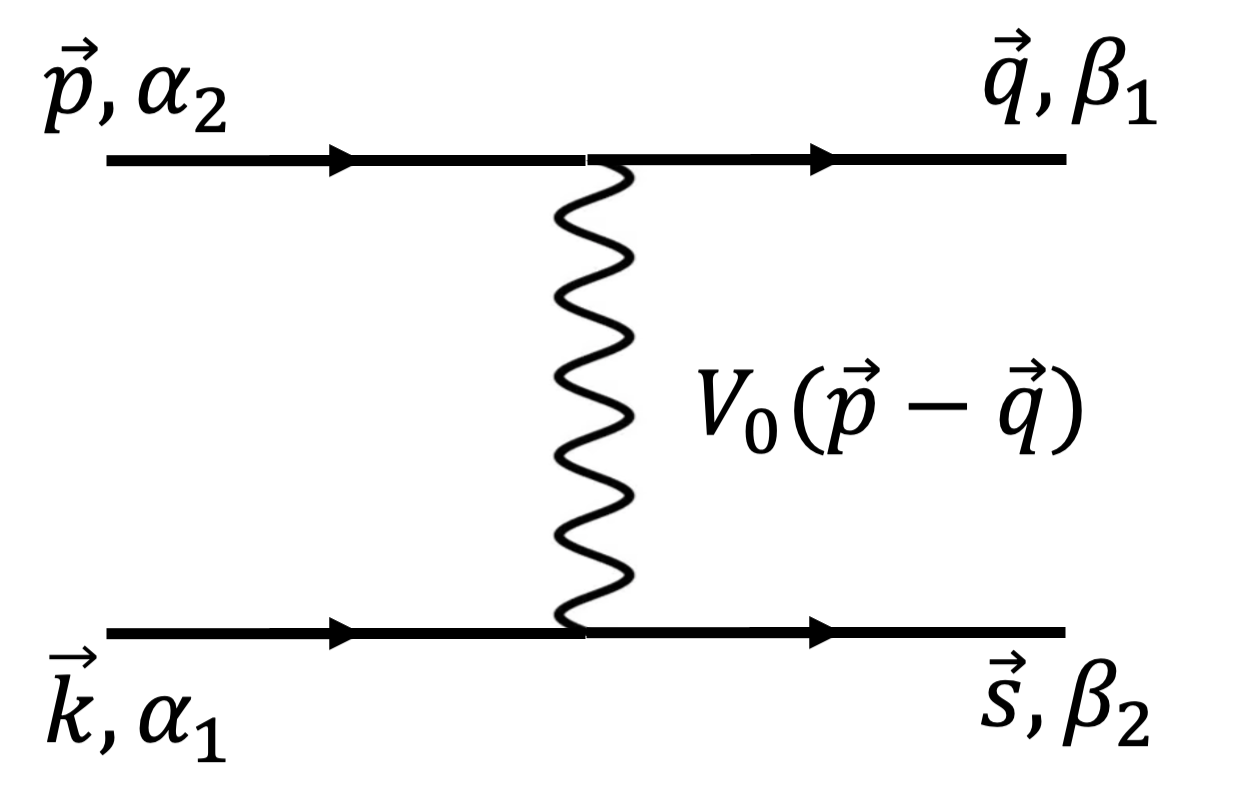}
    \caption{The diagram representation of the interaction vertex $V(\vec{k},\alpha_1 ; \vec{p},\alpha_2| \vec{q},\beta_1;\vec{s},\beta_2)$}
    \label{fig:Long_range_vertex}
\end{figure}

The interaction vertex shown in Fig.~\ref{fig:Long_range_vertex} can be written as follows 
\begin{equation}
    V(\vec{k},\alpha_1 ; \vec{p},\alpha_2| \vec{q},\beta_1;\vec{s},\beta_2) = V_0(\vec{p}-\vec{q})\braket{u_{\beta_2}(\vec{s})}{u_{\alpha_1}(\vec{k})}\braket{u_{\beta_1}(\vec{q})}{u_{\alpha_2}(\vec{p})} 
\end{equation}
with $V(q)$ the Fourier transform of the interaction potential, and where the overlap matrix elements are given by
\bea
\braket{u_{\alpha}(\vec{k})}{u_{\beta}(\vec{p})} = u_{\alpha{\uparrow}}^*(\vec{k})u_{\beta{\uparrow}}(\vec{p})  + u_{\alpha{\downarrow}}^*(\vec{k})u_{\beta{\downarrow}}(\vec{p}).
\eea
The $u(\vec{k})$ matrix can be found by diagonalizing the single particle Hamiltonian \ref{eq:H_TI}, and $\alpha,\beta $ are the band indices running over conduction (with value $+1$) and valence (with value $-1$) band.
Here
\bea
\label{eq:Blochev}
u (\vec{k})= 
\begin{pmatrix}
u(\vec{k})_{+,\uparrow} & u(\vec{k})_{-,\uparrow} \\
u(\vec{k})_{+,\downarrow} & u(\vec{k})_{-,\downarrow}
\end{pmatrix} =
\begin{pmatrix}
\frac{1}{\sqrt{2}} & \frac{1}{\sqrt{2}} \\
\frac{e^{i \Theta_\vec{k}}}{\sqrt{2}} & -\frac{e^{i \Theta_\vec{k}}}{\sqrt{2}}
\end{pmatrix}
\label{app:TI_us}
\eea
with $e^{i\Theta_\vec{k}} = \frac{1}{\sqrt{(k_x)^2+(k_y)^2}}(-k_y+i k_x)$.

%

\begin{figure}
    \centering
    \includegraphics[width=0.9\linewidth]{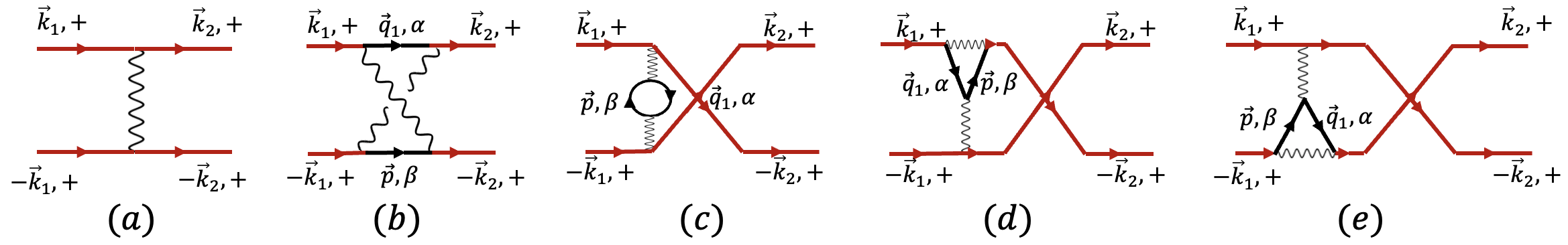}
    \caption{(a) The first-order contribution to particle-particle interactions. 
    (b–e) Kohn-Luttinger diagrams representing particle-particle interactions up to one-loop order, contributing to the irreducible interaction. The red lines indicate conduction band states, while the black lines represent internal propagators, which must be summed over both conduction and valence bands.}
    \label{fig_app:KL_diagrams_long_range}
\end{figure}

The first order particle-particle interaction is given by Fig.~\ref{fig_app:KL_diagrams_long_range}(a) and can be written as follows 
\begin{equation}
    \Gamma^{(1)}(-\vec{k}_1;\vec{k}_1|\vec{k}_2;-\vec{k}_2) = V_0(\vec{k}_1 - \vec{k}_2)\braket{u_{+}(-\vec{k}_2)}{u_{+}(-\vec{k}_1)}\braket{u_{+}(\vec{k}_2)}{u_{+}(\vec{k}_1)}
    \label{eq:Gamma1}
\end{equation}

For the second-order Kohn-Luttinger diagrams, we utilize diagrams (b-e) in Fig.~\ref{fig_app:KL_diagrams_long_range} to express:
\begin{equation}
    F_{b,\alpha\beta}(\vec{k_1},\vec{k_2},\vec{q}_1,\vec{p}) = V(-\vec{k}_1,+ ; \vec{q}_1,\alpha| \vec{k}_2,+;\vec{p},\beta)V(\vec{p},\beta ; \vec{k}_1,+| \vec{q}_1,\alpha;-\vec{k}_2,+)
    \label{eq_app:Fb}
\end{equation}
\begin{equation}
    F_{c,\alpha\beta}(\vec{k_1},\vec{k_2},\vec{q}_1,\vec{p}) = V(-\vec{k}_1,+ ; \vec{q}_1,\alpha| \vec{p},\beta;\vec{k}_2,+)V(\vec{k}_1,+ ; \vec{p},\beta| \vec{q}_1,\alpha;-\vec{k}_2,+)
\end{equation}
\begin{equation}
    F_{d,\alpha\beta}(\vec{k_1},\vec{k_2},\vec{q}_1,\vec{p}) = -V(-\vec{k}_1,+ ; \vec{q}_1,\alpha| \vec{p},\beta;\vec{k}_2,+)V(\vec{k}_1,+ ; \vec{p},\beta| -\vec{k}_2,+;\vec{q}_1,\alpha)
\end{equation}
\begin{equation}
    F_{e,\alpha\beta}(\vec{k_1},\vec{k_2},\vec{q}_1,\vec{p}) =- V(\vec{q}_1,\alpha ; -\vec{k}_1,+| \vec{p},\beta;\vec{k}_2,+)V(\vec{p},\beta ; \vec{k}_1,+| -\vec{k}_2,+;\vec{q}_1,\alpha)
    \label{eq_app:Fe}
\end{equation}
and
\begin{equation}
    \Gamma_{X,\alpha\beta}(\vec{k}_1,\vec{k}_2) = -\frac{1}{N}\sum_{\vec{p},\vec{q}_1} \chi_{\alpha\beta}(\vec{q}_1,\vec{p})F_{X,\alpha\beta}(\vec{k_1},\vec{k}_2,\vec{q}_1,\vec{p})\delta_{\vec{q}_1,\vec{p}+\vec{k}_1+\vec{k}_2}.
    \label{eq_app:Gamma_X}
\end{equation}
where, from here onward, sums over momenta are implied to extend up to the cutoff: $|\bp|, |\bq_1|<\Lambda$, and where
\begin{equation}
    \chi_{\alpha\beta}(\vec{q}_1,\vec{p}) = \frac{n(\epsilon_\alpha\big(\vec{q}_1)\big)-n(\epsilon_\beta\big(\vec{p})\big)}{\epsilon_\alpha\big(\vec{q}_1)-\epsilon_\beta\big(\vec{p})}.
\end{equation}

Since we are interested in Cooper pairs at the Fermi surface, we take  
\[
\vec{k}_1 = \left(k_F \cos\theta_1, k_F \sin\theta_1\right), \quad
\vec{k}_2 = \left(k_F \cos\theta_2, k_F \sin\theta_2\right).
\]
The total contribution at second order is obtained by summing over diagrams (b-e) in Fig.~\ref{fig:KL_diagrams_long_range} and taking $\bk_1,\bk_2$ on the Fermi surface, giving:
\begin{equation}
    \Gamma_{\alpha\beta}^{(2)}(\theta_1,\theta_2) = \sum_{X \in \{b,c,d,e\}} \Gamma_{X,\alpha\beta}(\vec{k}_1, \vec{k}_2),
\end{equation}
and finally by summing over internal band indices:
\bea
\Gamma^{(2)}(\theta_1,\theta_2) = \sum_{\alpha\beta} \Gamma_{\alpha\beta}^{(2)}(\theta_1,\theta_2).
\eea

\subsection{Short-range interaction}
\label{app_sec:short_range}
We now focus on the case of a short-range Hubbard interaction, corresponding to $V(q)=U$. In this case, using Eq.~\ref{eq:Gamma1}, the first-order contribution becomes
\begin{equation}
    \Gamma^{(1)}(\theta_1,\theta_2) = \frac{U}{2} e^{i(\theta_1-\theta_2)}.
\end{equation}

At second order, it can be checked by direct calculation that there is a cancellation between the last three diagrams (c), (d) and (e), leaving only the contribution from the diagram (b), whose matrix elements read:
\begin{equation}
    F_{\alpha\beta} \equiv F_{b,\alpha\beta} = U^2\braket{u_\alpha(\vec{p}+\vec{k}_1+\vec{k_2})}{u_+(\vec{k}_1)}\braket{u_+(-\vec{k}_2)}{u_\beta(\vec{p})}\braket{u_{+}(\vec{k}_2)}{u_\alpha(\vec{p}+\vec{k}_1+\vec{k_2})}\braket{u_\beta(\vec{p})}{u_+(-\vec{k}_1)} 
    \label{MatrixElementsShortRange}
\end{equation}
where $\alpha$ and $\beta$ refer to the band indices for the internal propagators.
(Note that the cancellation between diagrams (c), (d) and (e) is only true for the ideal Dirac model and not true in general).
The second-order pairing kernel is then given by: 
\bea
\Gamma^{(2)}(\theta_1;\theta_2) = -\frac{1}{N}\sum_{\alpha\beta} \sum_{\bp}\chi_{\alpha\beta}(\vec{p}+\vec{k}_1+\vec{k_2},\vec{p})F_{\alpha\beta}(\vec{k}_1,\vec{k}_2;\vec{p}+\vec{k}_1+\vec{k_2},\vec{p})
\eea

To simplify the derivation, we find it convenient to apply the transformation $\bk_2 \to - \bk_2$, which is equivalent to calculating the same diagram as Fig.~\ref{fig_app:KL_diagrams_long_range}b but with the two right external legs crossed.
We call this diagram $\Gamma'^{(2)}$ with:
\bea
\Gamma'^{(2)}(\theta_1;\theta_2) =   \frac{1}{N}\sum_{\alpha\beta} \sum_{\bp}\chi_{\alpha\beta}(\vec{p}+\vec{k}_1-\vec{k_2},\vec{p})F_{\alpha\beta}(\vec{k}_1,\vec{k}_2;\vec{p}+\vec{k}_1-\vec{k_2},\vec{p})
\label{Gamma2prime}
\eea
such that $\Gamma'^{(2)}(\theta_1;\theta_2+\pi) = - \Gamma^{(2)}(\theta_1;\theta_2)$.
At the end of the calculation, we will use this property to recover $\Gamma^{(2)}$ from $\Gamma'^{(2)}$.

Using the formulas above for the Bloch eigenvectors (Eq.~\ref{eq:Blochev}), and combining them with Eq.~\ref{MatrixElementsShortRange}, we obtain an analytical expression for the form factors (we omit the factor of $U^2$ for the rest of the derivation): 
\begin{align}
\tilde{F}_{\alpha\beta}(\theta_1,\theta_2;\theta_\vec{p+k_1-k_2},\theta_\vec{p}) & = \frac1{16} \bigg( -2 (1 - \cos(\theta_1 - \theta_2) )
+\big((\alpha e^{-i \theta_\vec{p}}  - \beta e^{-i\theta_\vec{p+k_1-k_2}}) (e^{i\theta_1 }  - e^{i\theta_2})  + \text{h.c.} \big)\\\nonumber
& -2\alpha \beta \cos(\theta_\vec{p} + \theta_\vec{p+k_1-k_2}-\theta_1-\theta_2)+ 2\alpha \beta \cos(-\theta_\vec{p} + \theta_\vec{p+k_1-k_2}) \bigg)
\end{align}
where we defined $\tilde{F} \equiv F e^{-i(\theta_1-\theta_2)}$, and where $\theta_{\bp}, \theta_\vec{p+k_1-k_2}$ are the polar angles of momenta $\bp$ and $\vec{p+k_1-k_2}$, respectively

To further simplify this expression, we exploit the rotational invariance of the dispersion. Let \(\vec{q} = \vec{k}_1 - \vec{k}_2\) and let us express it in terms of a reference direction as  
\begin{equation}  
\vec{q} = R_{\phi} \vec{q}_0,  
\end{equation}  
where \(\vec{q}_0 = q \hat{x}\) and \(R_{\phi}\) represents a rotation by an angle \(\phi\).  
Considering that both \(\vec{k}_1\) and \(\vec{k}_2\) lie on the Fermi surface, the rotation angle is given by 
\begin{equation}
\phi = \frac{\pi}{2} \, \text{sgn} \left( \sin\left(\frac{\theta_1-\theta_2}{2} \right) \right) + \frac{\theta_1 + \theta_2}{2}.
\label{app:phi}
\end{equation}  

We also modify the summation variable from \(\bp\) to \(R_{\phi} \bp\) in Eq.~\ref{Gamma2prime}, leading to:
\begin{align}
& \tilde{\Gamma}'_{\alpha\beta}(\theta_1,\theta_2) = \frac1{16N} \sum_{\bp} 
 \chi_{\alpha\beta}(\vec{p}+\bq_0,\vec{p})\bigg( -2 (1 - \cos(\theta_1 - \theta_2) )\\\nonumber & 
 +\big((\alpha e^{-i \theta_{\bp}}  - \beta e^{-i\theta_{\bp+\vec{q}_0}}) e^{-i\phi}(e^{i\theta_1}  - e^{i\theta_2})  + \text{h.c.}\big) \\\nonumber
& -2\alpha \beta  \cos(\theta_{\bp} + \theta_{\bp+\vec{q}_0}+2\phi-\theta_1-\theta_2)+ 2\alpha \beta \cos(-\theta_{\bp} + \theta_{\bp+\vec{q}_0}) \bigg)
\end{align}
where we defined $\tilde{\Gamma}' \equiv \Gamma' e^{-i(\theta_1-\theta_2)}$.

Using Eq.~\ref{app:phi}, we can separate different terms in the pairing kernel according to their $\theta_1,\theta_2$ dependence:
\bea
& \tilde{\Gamma}'_{\alpha\beta}(\theta_1,\theta_2) = \frac{-1}{8}\Pi_{1}^{\alpha\beta}(q_0) + \frac{1}{8}\Pi_2^{\alpha\beta}(q_0)\cos(\theta_1 - \theta_2) +\frac{1}{4} \left|\sin(\frac{\theta_1 - \theta_2}{2})\right|\Pi_3^{\alpha\beta}(q_0)
\eea
where 
\begin{align}
\Pi_{1}^{\alpha\beta}(\vec{q}_0) &=
\frac{1}{N} \sum_{\vec{p}}
\chi_{\alpha\beta}(\vec{p}+\vec{q}_0,\vec{p})
\Bigl[1 - 2\alpha\beta
\cos\bigl(\theta_{\vec{p}}\bigr)
\cos\bigl(\theta_{\vec{p}+\vec{q}_0}\bigr)\Bigr],
\\
\Pi_{2}^{\alpha\beta}(\vec{q}_0) &=
\frac{1}{N} \sum_{\vec{p}}
\chi_{\alpha\beta}(\vec{p}+\vec{q}_0,\vec{p}),
\\
\Pi_{3}^{\alpha\beta}(\vec{q}_0) &=
\frac{1}{N} \sum_{\vec{p}}
\chi_{\alpha\beta}(\vec{p}+\vec{q}_0,\vec{p})
\Bigl[\alpha \cos\bigl(\theta_{\vec{p}}\bigr)
- \beta \cos\bigl(\theta_{\vec{p}+\vec{q}_0}\bigr)\Bigr].
\end{align}

\subsubsection{\texorpdfstring{Calculation of $\Pi_1$}{Calculation of Pi1}}

To obtain analytical results, we separate the contributions with a factor of \( n\left(\epsilon_{+}(k)\right) \), from the contributions with a factor of \( n\left(\epsilon_{-}(k)\right)  \), with $n(\epsilon)$ the Fermi-Dirac distribution. Note that this decomposition mixes contributions from interband and intraband processes. We write this decomposition as follows:
\begin{equation}
    \sum_{\alpha\beta}\Pi_1^{\alpha\beta}(q) = \Pi_1^{+}(q)+\Pi_1^-(q)
\end{equation}
We begin with the expression for \({\Pi}_1^{+}(q)\),
\begin{align}
\Pi_{1}^{+}(q) = \frac{1}{N} 
\sum_{\vec{p}} \Biggl\{
& \frac{n(\epsilon_{\vec{p},+}) - n(\epsilon_{\vec{p}+\vec{q}_0,+})}
       {\epsilon_{\vec{p},+}-\epsilon_{\vec{p}+\vec{q}_0,+}}
  \Bigl[1 - 2\cos\bigl(\theta_{\vec{p}}\bigr)
           \cos\bigl(\theta_{\vec{p}+\vec{q}_0}\bigr)\Bigr]
\nonumber \\[4pt]
&+ \frac{n(\epsilon_{\vec{p},+})}
        {\epsilon_{\vec{p},+}-\epsilon_{\vec{p}+\vec{q}_0,-}}
  \Bigl[1 + 2\cos\bigl(\theta_{\vec{p}}\bigr)
           \cos\bigl(\theta_{\vec{p}+\vec{q}_0}\bigr)\Bigr]
\nonumber
- \frac{n(\epsilon_{\vec{p}+\vec{q}_0,+})}
        {\epsilon_{\vec{p},-}-\epsilon_{\vec{p}+\vec{q}_0,+}}
  \Bigl[1 + 2\cos\bigl(\theta_{\vec{p}}\bigr)
           \cos\bigl(\theta_{\vec{p}+\vec{q}_0}\bigr)\Bigr]
\Biggr\}.
\end{align}

We now reorganize the sum by separating terms according to whether the Fermi function appears at momentum \(\bp\) or at \(\bp + \bq_0\). This leads to the following decomposition:
\begin{align}
\Pi_{1}^{+}(q) = 
& \frac{1}{N} \sum_{\vec{p}} \Biggl[
   \frac{n(\epsilon_{\vec{p},+})}
        {\epsilon_{\vec{p},+}-\epsilon_{\vec{p}+\vec{q}_0,+}}
   \Bigl(1 - 2\cos\bigl(\theta_{\vec{p}}\bigr)
             \cos\bigl(\theta_{\vec{p}+\vec{q}_0}\bigr)\Bigr)
 + \frac{n(\epsilon_{\vec{p},+})}
        {\epsilon_{\vec{p},+}-\epsilon_{\vec{p}+\vec{q}_0,-}}
   \Bigl(1 + 2\cos\bigl(\theta_{\vec{p}}\bigr)
             \cos\bigl(\theta_{\vec{p}+\vec{q}_0}\bigr)\Bigr)
\Biggr]
\nonumber \\[6pt]
& - \frac{1}{N} \sum_{\vec{p}} \Biggl[
   \frac{n(\epsilon_{\vec{p}+\vec{q}_0,+})}
        {\epsilon_{\vec{p},+}-\epsilon_{\vec{p}+\vec{q}_0,+}}
   \Bigl(1 - 2\cos\bigl(\theta_{\vec{p}}\bigr)
             \cos\bigl(\theta_{\vec{p}+\vec{q}_0}\bigr)\Bigr)
 + \frac{n(\epsilon_{\vec{p}+\vec{q}_0,+})}
        {\epsilon_{\vec{p},-}-\epsilon_{\vec{p}+\vec{q}_0,+}}
   \Bigl(1 + 2\cos\bigl(\theta_{\vec{p}}\bigr)
             \cos\bigl(\theta_{\vec{p}+\vec{q}_0}\bigr)\Bigr)
\Biggr].
\end{align}
To simplify the second sum, we perform a change of variables \(\bp' = \bp + \bq_0\), and relabel the summation accordingly. This yields:
\begin{align}
\Pi_{1}^{+}(q) = 
& \frac{1}{N} \sum_{\vec{p}} \Biggl[
   \frac{n(\epsilon_{\vec{p},+})}
        {\epsilon_{\vec{p},+}-\epsilon_{\vec{p}+\vec{q}_0,+}}
   \Bigl(1 - 2\cos\bigl(\theta_{\vec{p}}\bigr)
             \cos\bigl(\theta_{\vec{p}+\vec{q}_0}\bigr)\Bigr)
 + \frac{n(\epsilon_{\vec{p},+})}
        {\epsilon_{\vec{p},+}-\epsilon_{\vec{p}+\vec{q}_0,-}}
   \Bigl(1 + 2\cos\bigl(\theta_{\vec{p}}\bigr)
             \cos\bigl(\theta_{\vec{p}+\vec{q}_0}\bigr)\Bigr)
\Biggr]
\nonumber \\[6pt]
& + \frac{1}{N} \sum_{\vec{p}} \Biggl[
   \frac{n(\epsilon_{\vec{p},+})}
        {\epsilon_{\vec{p},+}-\epsilon_{\vec{p}-\vec{q}_0,+}}
   \Bigl(1 - 2\cos\bigl(\theta_{\vec{p}-\vec{q}_0}\bigr)
             \cos\bigl(\theta_{\vec{p}}\bigr)\Bigr)
 + \frac{n(\epsilon_{\vec{p},+})}
        {\epsilon_{\vec{p},+}-\epsilon_{\vec{p}-\vec{q}_0,-}}
   \Bigl(1 + 2\cos\bigl(\theta_{\vec{p}-\vec{q}_0}\bigr)
             \cos\bigl(\theta_{\vec{p}}\bigr)\Bigr)
\Biggr].
\end{align}
Now we can define 
\begin{align}
    \Pi_1^{+}(q) = \tilde{\Pi}_1^{+}(q) +\tilde{\Pi}_1^{+}(-q) 
\end{align}
where
\begin{align}
\tilde{\Pi}_{1}^{+}(q) = 
\frac{1}{N} \sum_{\vec{p}} \Biggl[
   \frac{n(\epsilon_{\vec{p},+})}
        {\epsilon_{\vec{p},+}-\epsilon_{\vec{p}+\vec{q}_0,+}}
   \Bigl(1 - 2\cos\bigl(\theta_{\vec{p}}\bigr)
             \cos\bigl(\theta_{\vec{p}+\vec{q}_0}\bigr)\Bigr)
 + \frac{n(\epsilon_{\vec{p},+})}
        {\epsilon_{\vec{p},+}-\epsilon_{\vec{p}+\vec{q}_0,-}}
   \Bigl(1 + 2\cos\bigl(\theta_{\vec{p}}\bigr)
             \cos\bigl(\theta_{\vec{p}+\vec{q}_0}\bigr)\Bigr)
\Biggr].
\end{align}
By using the linear dispersion \(\epsilon_{\bp,\alpha} = \alpha \hbar v_F |\vec{p}|\), we can derive the following equation:
\begin{equation}
\tilde{\Pi}_1^{+}(q) = -\frac{2}{N}\sum_{\bp} n(\epsilon_{\bp,+}) \left( \frac{p - 2|\bp + \vec{q}| \cos(\theta) \cos(\theta_\vec{p+q})}{q^2 + 2pq \cos(\theta)} \right)
\end{equation}
where we have used the shorthand $\theta$ for $\theta_\bp$.
Using the relation \(|\bp + \vec{q}| \cos(\theta_{p+q}) = p\cos(\theta) + q\), we find 
\begin{equation}
\hbar v \tilde{\Pi}_1^{+}(q) = - 2\int_{0}^{\infty} \frac{p \,dp}{(2\pi)^2} n(\epsilon_{\bp,+}) \int_{0}^{2\pi} \left( \frac{p - 2p\cos(\theta)^2-2q\cos(\theta)}{q^2 + 2pq \cos(\theta)} \right) d\theta
\end{equation}
After a suitable rearrangement, we obtain:
\begin{equation}
\hbar v \tilde{\Pi}_1^{+}(q) = - 2\int_{0}^{\infty} \frac{p \,dp}{(2\pi)^2} n(\epsilon_{\bp,+}) \int_{0}^{2\pi} \left( \frac{p - q\cos(\theta)}{q^2 + 2pq \cos(\theta)} - \frac{\cos(\theta)}{q} \right) d\theta
\end{equation}
The second term in the integrand integrates to zero over $\theta$, leaving
\begin{equation}
\hbar v \tilde{\Pi}_1^{+}(q) = - 2\int_{0}^{\infty} \frac{\,dp}{(2\pi)^2} n(\epsilon_{\bp,+}) \int_{0}^{2\pi} \left( \frac{p^2 - pq\cos(\theta)}{q^2 + 2pq \cos(\theta)} \right) d\theta
\end{equation}
From the expression above, we observe that  \(\tilde{{\Pi}}_1^{+}(q)\) is an even function of momentum,
\begin{equation}
\tilde{\Pi}_1^{+}(q) = \tilde{\Pi}_1^{+}(-q),
\end{equation}
as can be verified by performing the change of variables \(\theta \to \theta + \pi\) in the angular integral. Under this shift, \(\cos\theta \to -\cos\theta\), which leaves the integrand invariant due to the combined transformation \(q \to -q\) and \(\cos\theta \to -\cos\theta\).

Thus,
\begin{equation}
\hbar v \Pi_1^{+}(q) = -4 \int_{0}^{\infty} \frac{dp}{(2\pi)^2} n(\epsilon_{\bp,+}) \int_{0}^{2\pi} \left( \frac{p^2 - pq \cos(\theta)}{q^2 + 2pq \cos(\theta)} \right) d\theta
\end{equation}

Focusing solely on the angular component of the integral, we find that 
\[
\int_{0}^{2\pi} \left( \frac{p^2 - pq \cos(\theta)}{q^2 + 2pq \cos(\theta)} \right) d\theta =
\begin{cases}
-\pi, & \text{if }  \frac{q}{2}<p \\[10pt]
\displaystyle
\frac{\pi (2p^2 + q^2)}{q \sqrt{q^2 - 4p^2}}-\pi, 
& \text{if } \frac{q}{2}>p
\end{cases}
\]
Thus, for \( q < 2k_F \), we find
\begin{equation}
\hbar v\, \Pi_1^{+}(q) = -4 \int_0^{q/2} \frac{dp}{(2\pi)^2} 
\left( \frac{\pi (2p^2 + q^2)}{q \sqrt{q^2 - 4p^2}} - \pi \right) 
+ 4 \int_{q/2}^{k_F} \frac{dp}{(2\pi)^2} \pi.
\label{eq:Pi_integral}
\end{equation}
After performing the integration, we obtain
\begin{equation}
\Pi_1^{+}(q) = \frac{1}{\hbar v (2\pi)^2} \left( 4\pi k_F - \frac{5\pi^2}{4} q \right).
\label{eq:Pi_result}
\end{equation}

The function \(\Pi_1^-(q)\) can also be evaluated analytically, and the result is:
\begin{equation}
\Pi_1^-(q) = \frac{-2}{(2\pi)^2} \int_{|\bp|<\Lambda} d^2\bp  \frac{n(\epsilon_{\bp,-})}{|\bp| + |\bp + \bq| } \left( 1 + 2 \cos(\theta_\bp) \cos(\theta_{\bp+\bq}) \right) = -\frac{1}{\hbar v (2\pi)^2} 4 \pi \Lambda + \frac{1}{\hbar v (2\pi)^2} \frac{5 \pi^2}{4} q
\end{equation}

(Strictly speaking, one should also impose the condition $|\bp+\bq| < \Lambda$ when performing the integral over $\bp$. However, one can show that
\begin{equation}
\frac{-2}{(2\pi)^2} \int_{|\bp|,|\bp+\bq|<\Lambda} d^2\bp  \frac{n(\epsilon_{\bp,-})}{|\bp| + |\bp + \bq| } \left( 1 + 2 \cos(\theta_p) \cos(\theta_{p+q}) \right) = -\frac{1}{\hbar v (2\pi)^2} 4 \pi \Lambda + \frac{1}{\hbar v (2\pi)^2} \frac{5 \pi^2}{4} q + \mathcal{O}(q^3\Lambda^{-2})
\end{equation}
We thus recover the formula above to leading order in $q / \Lambda$, which is sufficient for our purpose. The same expansion was used when calculating $\Pi^-_2$ below.)

Finally, we find:
\begin{equation}
\sum_{\alpha\beta}\Pi_1^{\alpha\beta}(q) = \Pi_1^+(q) + \Pi_1^-(q) = \frac{1}{\hbar v (2\pi)^2} \left( 4 \pi k_F - 4 \pi \Lambda \right)
\end{equation}


\subsubsection{\texorpdfstring{Calculation of $\Pi_2$}{Calculation of Pi2}}
The calculation of \(\sum_{\alpha\beta}\Pi^{\alpha\beta}_{2}(q) = {\Pi}_{2}^+(q) + {\Pi}_{2}^-(q)\) follows a similar procedure to that of \(\Pi_1\):
\begin{equation}
{\Pi}_2^{+}(q) = 2\frac{1}{N}\sum_{\bp} \left(\frac{n(\epsilon_{\bp,+})}{\epsilon_{\bp,+}-\epsilon_{\bp+\vec{q}_0,+}} + \frac{n(\epsilon_{\bp,+})}{\epsilon_{\bp,+}-\epsilon_{\bp+\vec{q}_0,-}}\right).
\end{equation}
By using the linear dispersion, we have
\begin{equation}
{\Pi}_2^{+}(q) = -4\frac{1}{N}\sum_{\bp}n(\epsilon_{\bp,+})\frac{p}{q^2+2pq\cos(\theta)}.
\end{equation}
Thus, by performing the integral, we obtain the following result:
\begin{equation}
{\Pi}_2^{+}(q) = -4\int_{0}^{\infty}\frac{p\,dp}{(2\pi)^2}n(\epsilon_{\bp,+})\int_{0}^{2\pi}\frac{p}{q^2+2pq\cos(\theta)}\,d\theta = -\frac{1}{\hbar v(2\pi)^2}\frac{\pi^2}{4}q.
\end{equation}

For the \({\Pi}^-(q)\) contribution:
\begin{equation}
\Pi_2^{-}(q) = 2\frac{1}{N}\sum_{\bp} \frac{n(\epsilon_{\bp,-})}{\epsilon_{\bp,-}-\epsilon_{\bp+\vec{q}_0,+}} = -2\frac{1}{\hbar v(2\pi)^2}\pi \Lambda + \frac{1}{\hbar v(2\pi)^2}\frac{\pi^2}{4}q.
\end{equation}

Finally, we obtain:
\begin{equation}
\sum_{\alpha\beta}\Pi^{\alpha\beta}_2(q) = {\Pi}_2^+(q) + {\Pi}_2^-(q) = -\frac{2}{\hbar v(2\pi)^2}\pi \Lambda.
\end{equation}

\subsubsection{\texorpdfstring{Calculation of $\Pi_3$}{Calculation of Pi3}}
The same procedure applies to \(\Pi_3\). In this case \(\Pi_3^{-}\) vanishes, leaving only \(\Pi_3^{+}\):
\begin{equation}
    \sum_{\alpha\beta}\Pi_{3}^{\alpha\beta}(q) = {\Pi}_3^+(q)
\end{equation}
with
\bea
{\Pi}_3^{+}(q) = 2\frac{1}{N}\sum_{\bp} \bigg(\frac{n(\epsilon_{\bp,+})}{\epsilon_{\bp,+}-\epsilon_{\bp+\vec{q}_0,+}}\big(\cos(\theta_\bp)-\cos(\theta_{\bp+\bq_0})\big) + \frac{n(\epsilon_{\bp,+})}{\epsilon_{\bp,+}-\epsilon_{\bp+\vec{q}_0,-}}\big(\cos(\theta_\bp)+\cos(\theta_{\bp+\bq_0})\big)\bigg)\,.
\eea
Using the linear dispersion we find:
\bea
{\Pi}_3^+(q) = -\frac{4}{\hbar v}\frac{1}{N}\sum_{p}n(\epsilon_{\bp,+}) \frac{p\cos(\theta)-|\vec{p}+\vec{q}|\cos(\theta_{p+q})}{q^2+2 pq\cos(\theta)}\,.
\eea
which after some algebra gives
\bea
{\Pi}^+_3 = \frac{4}{(2\pi)^2\hbar v }\int_{0}^{\infty}n(\epsilon_{\bp,+})\,dp \int_{0}^{2\pi}\frac{pq}{q^2+2 pq\cos(\theta)}\,d\theta = 2\pi q\,.
\eea
Thus
\bea
\sum_{\alpha\beta}\Pi^{\alpha\beta}_3(q) = \frac{1}{(2\pi)^2\hbar v}2\pi q\,.
\eea

\subsubsection{\texorpdfstring{Calculation of $\tilde{\Gamma}^{(2)}(\theta_1,\theta_2)$}{Calculation of G}}\label{app:sec_cal_gamma_TI}
As defined earlier, the second order contributions to the pairing kernel can be decomposed as follows:
\bea
\tilde{\Gamma}'^{(2)}(\theta_1,\theta_2) = &\sum_{\alpha\beta} \tilde{\Gamma}'_{\alpha\beta}(\theta_1,\theta_2) =\sum_{\alpha\beta}\bigg( \frac{-1}{8}\Pi_{1}^{\alpha\beta}(q_0) + \frac{1}{8}\Pi_2^{\alpha\beta}(q_0)\cos(\theta_1 - \theta_2) +\frac{1}{4} \left|\sin(\frac{\theta_1 - \theta_2}{2})\right|\Pi_3^{\alpha\beta}(q_0)\bigg)
\eea
Using the results from the previous sections, we find:
\bea
\tilde{\Gamma}'^{(2)}(\theta_1,\theta_2) =\frac{1}{\hbar v(2\pi)^2}\bigg( -\frac{\pi}{2}k_F + \frac{\pi}{2}\Lambda - \frac{\pi}{4} \Lambda\cos(\theta_1-\theta_2)+ \frac{\pi}{2} q \left|\sin\left(\frac{\theta_1-\theta_2}{2}\right)\right|\bigg)
\eea
By using $q= 2k_F \left|\sin(\frac{\theta_1-\theta_2}{2})\right|$ we obtain:
\bea
\tilde{\Gamma}'^{(2)}(\theta_1,\theta_2) =\frac{1}{\hbar v(2\pi)^2}\bigg( \frac{\pi}{2}\Lambda - \frac{\pi}{4}(2k_F+\Lambda)\cos(\theta_1-\theta_2)\bigg) 
\eea
Finally, using $\Gamma'^{(2)}(\theta_1;\theta_2+\pi) = - \Gamma^{(2)}(\theta_1;\theta_2)$, we find the formula in the main text:
\bea
\tilde{\Gamma}^{(2)}(\theta_1,\theta_2) =\frac{1}{\hbar v(2\pi)^2}\bigg( \frac{\pi}{2}\Lambda + \frac{\pi}{4}(2k_F+\Lambda)\cos(\theta_1-\theta_2)\bigg) 
\eea

\section{Kohn-Luttinger for two ideal Dirac cones with the same chirality}
\label{App:SameChirality}

Here we analyze Kohn-Luttinger superconductivity of a Dirac cone with an additional twofold degeneracy due to a flavor index $s$ we call spin, i.e.
\begin{equation}
    H_0 = v_F \sum_{|\mathbf{k}|<\Lambda } \sum_{s =\pm1} \boldsymbol{\psi}^{\dagger}_{\mathbf{k},s} ( k_x \sigma_x + k_y \sigma_y) \boldsymbol{\psi}_{\mathbf{k},s} ,
    \label{eq:TwoDiracConesSameChiralityApp}
\end{equation}
with $\boldsymbol{\psi}_{\mathbf{k},s} = (c_{\mathbf{k},a,s},\, c_{\mathbf{k},b,s})^{T}$, with $a,b$ the two orbitals.
(For concreteness, in this section we refer to the space on which the Pauli matrices act non-trivially as the orbital index, and the $s$ index as spin which simply gives an additional degeneracy to the bands).
We consider two different types of interaction.

\subsection{Intra-orbital interaction}
\begin{figure}[h!]
    \centering
    \includegraphics[width=0.35\linewidth]{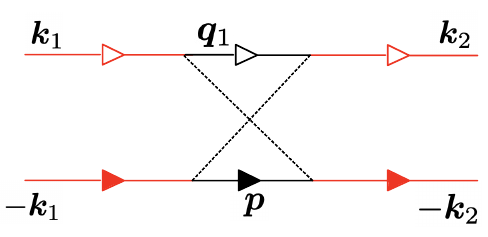}
    \caption{Second-order Kohn-Luttinger diagram in the case of an intra-orbital interaction between electrons of opposite spins.
A filled arrow represents a spin down electron, while an empty arrow represents a spin up electron.}
    \label{fig:TwoConesIntra}
\end{figure}

We start with a purely intra-orbital interaction of the form
\bea
H_\text{int} = U \sum_{\vec{x}}  (n_{\vec{x},a,\uparrow} n_{\vec{x},a,\downarrow} + n_{\vec{x},b,\uparrow} n_{\vec{x},b,\downarrow} ) 
\eea
where $\vec{x}$ is the unit cell position.
Since the only interaction is between opposite spins, it is sufficient to calculate a single diagram given by Fig.~\ref{fig:TwoConesIntra}, and the parity-odd (respectively parity-even) sector gives pairing in the triplet (resp. singlet) channel.
This gives
\bea
\Gamma^{(2)}(\bk_1,\bk_2) = - \sum_{\alpha,\beta} \int \frac{d^2\bp}{(2\pi)^2} \chi_{\alpha\beta}(\bp,\bp+\bk_1+\bk_2) F_{\alpha\beta}(\vec{k_1},\vec{k}_2,\vec{q}_1,\vec{p})
\eea
with $\bq_1 = \bp + \bk_1 + \bk_2$ and $F_{\alpha\beta}(\vec{k_1},\vec{k}_2,\vec{q}_1,\vec{p}) = V_1 V_2$ with
\bea
V_1
=
U \Big(
u_{+, 1}(\vec{k}_1)
u_{\alpha 1}(\vec{p}+\bk_1+\bk_2)
u_{+, 1}^*(-\vec{k}_2)
u_{\beta 1}^*(\vec{p})
+
u_{+, 2}(\vec{k}_1)
u_{\alpha 2}(\vec{p}+\bk_1+\bk_2)
u_{+, 2}^*(-\vec{k}_2)
u_{\beta 2}^*(\vec{p})
\Big)
\eea
and
\bea
V_2
=
U \Big(
u_{\beta 1}(\vec{p})
u_{+, 1}(-\vec{k}_1)
u_{\alpha 1}^*(\vec{p}+\bk_1+\bk_2)
u_{+, 1}^*(\vec{k}_2)
+
u_{\beta 2}(\vec{p})
u_{+ 2}(-\vec{k}_1)
u_{\alpha 2}^*(\vec{p}+\bk_1+\bk_2)
u_{+ 2}^*(\vec{k}_2)
\Big).
\eea
After some algebra, one finds 
\bea
F_{\alpha\beta}(\vec{k_1},\vec{k}_2,\vec{q}_1,\vec{p}) = \frac{U^2}{8} e^{i (\theta_1 - \theta_2)} [\cos(\theta_1 - \theta_2) - \alpha \beta \cos(\theta_{\bp+\bk_1+\bk_2} - \theta_\bp)]
\eea
with $\theta$ the polar angle for each momentum vector.
We did not find a closed form formula for $\Gamma^{(2)}$, but we found numerically that it is of the form
\bea
\tilde\Gamma^{(2)}(\theta_1,\theta_2) \equiv  e^{-i (\theta_1 - \theta_2)}\Gamma^{(2)}(\theta_1,\theta_2) = |A| + |B| \cos(\theta_1 - \theta_2)
\eea
with $|A|,|B| \sim \Lambda$.
The interaction is thus repulsive in $l=0$ and $l=1$ and zero in $l \geq 2$ channels, just like the case of a single Dirac cone without flavor degeneracy.

\subsection{Density-density interaction}
\label{App:TwoDiracInter}
\begin{figure}[t!]
    \centering
    \includegraphics[width=0.75\linewidth]{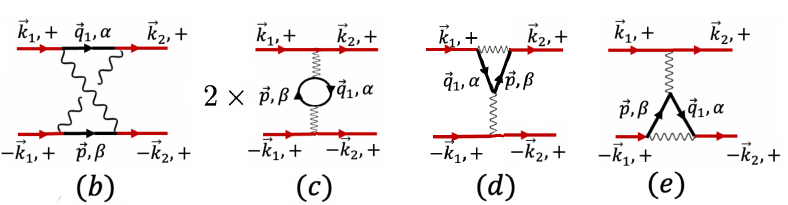}
    \caption{Diagrams for $\Gamma^{(2)}$ in the case of density-density interactions. }
    \label{fig:TwoConesInter}
\end{figure}

We also consider the case of density-density interaction as defined in the main text, which in the limit of short-range interaction $V(q)=U$ gives both intra- and inter-orbital coupling:
\bea
H_\text{int} = U \sum_{\vec{x}} \sum_{s,s'} n_{\vec{x},s} n_{\vec{x},s'}
\eea
with $n_{\vec{x},s} = n_{\vec{x},a,s} + n_{\vec{x},b,s}$, where $\vec{x}$ is the unit cell position.

In this case, the relevant diagrams are shown in Fig.~\ref{fig:TwoConesInter} where it is implicitly assumed that the top set of fermion lines is spin-up and the bottom set of fermion lines is spin-down. In that case, the parity-odd (respectively parity-even) sector gives pairing in the spin triplet (resp. singlet) channel.
The bubble diagram in Fig.~\ref{fig:TwoConesInter}(c) has an additional factor of two coming from the sum over spin species for the internal bubble. 
Overall this gives a form factor:
\bea
F = F_b + 2 F_c + F_d + F_e
\eea
As we showed in Appendix~\ref{App1}, $F_c + F_d + F_e = 0$, leading to $F = F_b + F_c$.

The contribution from $F_b$ is the same as the one calculated in Appendix~\ref{App1}, and is thus repulsive and is dominated by interband contributions proportional to $\Lambda$:
\bea
\tilde{\Gamma}_b^{(2)}(\theta_1,\theta_2) =\frac{1}{\hbar v(2\pi)^2}\bigg( \frac{\pi}{2}\Lambda + \frac{\pi}{4}(2k_F+\Lambda)\cos(\theta_1-\theta_2)\bigg) 
\eea
The contribution from $F_c$ goes as
\bea
\tilde\Gamma^{(2)}_c(\theta_1,\theta_2) = -  \frac12 (1 + \cos(\theta_1 - \theta_2) ) \Pi(\bk_1 - \bk_2)
\eea
with $\Pi(q)$ the polarization function, which was calculated in Ref.~\cite{DasSarmaDielectricFunction} to be constant (and equal to the density of states at the Fermi level) for $q < 2 k_F$, which means $\tilde\Gamma^{(2)}_c \sim \Pi(q \leq 2 k_F) \sim k_F$.

Both $\Gamma_c$ and $\Gamma_b$ only contribute to the $l=0$ and $l=1$ channels.
In the limit of $k_F \ll \Lambda$, we have $\tilde\Gamma^{(2)}_c \ll \tilde{\Gamma}_b^{(2)}$ and we can thus neglect $\tilde\Gamma^{(2)}_c$. We are therefore back to the case of a single Dirac cone for which there is no superconductivity.


\section{Kohn-Luttinger for two ideal Dirac cones of opposite chirality}
\label{AppGraphene}

In this appendix, we perform the Kohn-Luttinger calculation for two Dirac cones of opposite chirality (each cone having an additional twofold degeneracy due to spin).

\subsection{General formalism and graphene dispersion}
\label{grapheneappendixdispersion}
Since this section is motivated by graphene in the small $k_F$ regime, we start with the Bloch Hamiltonian for electrons on a honeycomb lattice with nearest-neighbor hopping. (This section follows closely Ref.~\cite{Kagan_2014}.) The single-particle Hamiltonian reads:
\begin{equation}
    H_0 = \sum_{\vec{k},s} \psi^{\dagger}_{s}(\vec{k}) h(\vec{k}) \psi_{s}(\vec{k}),
    \label{eq:graphene_free_H}
\end{equation}
where $\psi_{s}(\vec{k}) = (c_{a,s}(k),\, c_{b,s}(\vec{k}))^T$, with $a$ and $b$ denoting orbital indices and $s$ representing spin indices. The matrix $h(\vec{k})$ is given by
\begin{equation}
    h(k) = -t\begin{pmatrix}
    0 & g_\vec{k} \\
    g^*_{\vec{k}} & 0
    \end{pmatrix},
\end{equation}
where $t$ is the nearest neighbour hopping parameter, and $g_k = \sum_{\delta}e^{i\vec{k}\cdot\vec{\delta}}$, where $\vec{\delta}$ are the nearest neighbour vectors on a honeycomb lattice. 

Additionally, the model includes an intra-orbital Hubbard interaction:
\bea
H_\text{int} = U \sum_{\vec{x}}  (n_{\vec{x},a,\uparrow} n_{\vec{x},a,\downarrow} + n_{\vec{x},b,\uparrow} n_{\vec{x},b,\downarrow} ) 
\label{eq:graphene_Hubbard}
\eea
where $U$ denotes the interaction strength, and $n_{\vec{x},a,s}$ and $n_{\vec{x},b,s}$ are the occupation number operators in unit cell $\vec{x}$, with spin $s$, in orbitals $a$ and $b$, respectively. Explicitly, we find:
\begin{equation}
    H_\text{int} = \frac{1}{N}U\sum_{\vec{k}_1,\vec{k}_2,\vec{k}_3,\vec{k}_4}\bigg(c^{\dagger}_{\vec{k}_1a,\up}c^{\dagger}_{\vec{k}_2,a,\down}c_{\vec{k}_3,a,\down}c_{\vec{k}_4,a,\up}+c^{\dagger}_{\vec{k}_1,b,\up}c^{\dagger}_{\vec{k}_2,b,\down}c_{\vec{k}_3,b,\down}c_{\vec{k}_4,b,\up}\bigg)\delta(\vec{k}_1+\vec{k}_2-\vec{k}_3-\vec{k}_4)
\end{equation}

After diagonalizing the single particle Hamiltonian, we can define the band-basis (with band index $\alpha = \pm$ such that $\epsilon_k = \pm t \sqrt{3 + f_k}$) as $c^{\dagger}_{\vec{k},\alpha,s}=w_{\alpha,a}c^{\dagger}_{\vec{k},a,s}+w_{\alpha,b}c^{\dagger}_{\vec{k},b,s}$, where the $w$ matrix reads:
\bea
\begin{pmatrix}
w_{+a} & w_{+b}\\
w_{-a} & w_{-b} 
\end{pmatrix} &=&\begin{pmatrix}
\frac{-r_\bk^*}{\sqrt{2}} & \frac{1}{\sqrt{2}}\\
\frac{-r_\bk^*}{\sqrt{2}} & \frac{-1}{\sqrt{2}} 
\end{pmatrix}
\label{app:usfull}
\eea
with
\bea
r_{k}&=&\frac{e^{-ik_x}+2e^{ikx/2}\cos\left(\frac{\sqrt{3}}{2}k_y\right)}{\sqrt{3+f_{k}}}\\
f_{k} &=& 2\cos\left(\sqrt{3}k_y\right)+4\cos\left(\frac{\sqrt{3}}{2}k_y\right)\cos\left(\frac{3}{2}k_y\right).
\eea

We now rewrite the interaction Hamiltonian in the band basis:
\bea
H_\text{int} = \frac{1}{N}\sum_{\alpha_1,\alpha_2,\beta_1,\beta_2}\sum_{\vec{k}\vec{p}\vec{q}\vec{s}}V_{\alpha_1\alpha_2;\beta_1\beta_2}(\vec{k}\vec{p}|\vec{q}\vec{s})c^{\dagger}_{\vec{k},\alpha_1\up}c^{\dagger}_{\vec{p},\alpha_2\down}c_{\vec{q},\beta_1\down}c_{\vec{s},\beta_2\up}\delta(\vec{k}+\vec{p}-\vec{q}-\vec{s})
\eea
where 
\bea
V_{\alpha_1\alpha_2;\beta_1\beta_2}(\vec{k}\vec{p}|\vec{q}\vec{s}) = U\big(w_{\alpha_1a}(\vec{k})w_{\alpha_2a}(\vec{p})w_{\beta_1a}^*(\vec{q})w^*_{\beta_2a}(\vec{s})+ w_{\alpha_1b}(\vec{k})w_{\alpha_2b}(\vec{p})w_{\beta_1b}^*(\vec{q})w^*_{\beta_2b}(\vec{s})\big).
\label{app:V_vertex}
\eea

Since the only interaction is between opposite spins, it is sufficient to calculate one type of diagram (the ``double exchange diagram'', see Fig.~\ref{fig_app:Scattering_diagrams}), and the parity-odd (respectively parity-even) sector gives pairing in the triplet (resp. singlet) channel.
This gives:
\bea
\Gamma^{(2)}_{\up\down\down\up}(\vec{k}_1,\vec{k}_2) = -\frac{1}{N}\sum_{\alpha\beta}\sum_{\vec{p}_1}\chi_{\alpha\beta}(\vec{q}_2,\vec{p}_1) F_{\alpha\beta}(\vec{k}_1,\vec{k}_2;\vec{p}_1,\vec{q}_2)
\label{app:Gamma_2}
\eea
with $\vec{q}_2 = \vec{p}_1 - \bk_1 - \bk_2$,
\bea
\chi_{\alpha\beta}(\vec{k},\vec{p}) = \frac{n_{F}\big(\epsilon_\alpha(\vec{k})\big)-n_{F}\big(\epsilon_\beta(\vec{p})\big)}{\epsilon_\alpha(\vec{k})-\epsilon_\beta(\vec{p})}
\eea
and
\bea
F_{\alpha\beta}(\vec{k}_1,\vec{k}_2;\vec{p}_1,\vec{q}_2) = V_{+\alpha;+\beta}(\vec{k}_1,\vec{q}_2|-\vec{k}_2,\vec{p}_1)V_{\beta+;\alpha+}(\vec{p}_1,-\vec{k}_1|\vec{q}_2,\vec{k}_2).
\label{FFF}
\eea
\subsection{Appoximate dispersion of two ideal Dirac cones at the two valleys}

 In order to perform the calculation analytically, we approximate the graphene dispersion as two ideal Dirac cones in the \(\vec{K}\) and \(\vec{K}'\) valleys:
 \bea
    H_0  \simeq v_F \sum_{|\vec{k}|<\Lambda }  \sum_{\nu, s =\pm1} \boldsymbol{\psi}^{\dagger}_{\bk,\nu,s} ( \nu k_x \sigma^x + k_y \sigma^y)    \boldsymbol{\psi}_{\bk,\nu,s}  ,
    \label{TwoDiracCones}
\eea
where $\nu = \pm 1$ is the valley index, $s$ is the spin index, and \( \boldsymbol{\psi}_{\bk,\nu,s} = \big(c_{\nu \vec{K}+\bk,a,s},\, c_{\nu \vec{K}+\bk,b,s}\big)^{T} \) with $a,b$ the two orbitals.
The Bloch eigenvectors then read \( w(\vec{K}' + \delta \vec{k}) \simeq u(\delta \vec{k}) \) for the valley we call $B$ (with $\nu=-1$) and \( w(\vec{K} + \delta \vec{k}) \simeq v(\delta \vec{k}) \) for valley we call $A$ (with $\nu=+1$), with
\bea
u(\delta \vec{k})  = \begin{pmatrix}
u_{+a} & u_{+b}\\
u_{-a} & u_{-b} 
\end{pmatrix} &=& \begin{pmatrix}
\frac{e^{i\theta_{\delta \vec{k}}}}{\sqrt{2}} & \frac{1}{\sqrt{2}}\\
\frac{e^{i\theta_{\delta \vec{k}}}}{\sqrt{2}} & \frac{-1}{\sqrt{2}} 
\end{pmatrix}
\\
v(\delta \vec{k})  = \begin{pmatrix}
v_{+a} & v_{+b}\\
v_{-a} & v_{-b} 
\end{pmatrix} &=&\begin{pmatrix}
\frac{e^{-i\theta_{\delta \vec{k}}}}{\sqrt{2}} & \frac{1}{\sqrt{2}}\\
\frac{e^{-i\theta_{\delta \vec{k}}}}{\sqrt{2}} & \frac{-1}{\sqrt{2}} 
\end{pmatrix}
\label{app:us}
\eea
where $\theta_{\delta \bk}$ is the the polar angle of $\delta \bk$.

\subsection{Calculation of pairing kernel}
The pairing always occurs between two electrons of opposite valleys since we only consider pairing between $\bk$ and $-\bk$.
As shown in Fig.~\ref{fig_app:Scattering_diagrams}, we need to calculate separately Cooper pair scattering which conserves the valley index of each electron in the pair (which we call $\Gamma_{ABBA}$), and Cooper pair scattering which flips it (which we call $\Gamma_{ABAB}$). 
We will start with the latter.

\subsubsection{\texorpdfstring{Scattering $(k_1,\up,A)(-k_1,\down,B)\rightarrow (-k_2,\down,A)(k_2,\up,B)$}{scattering ABAB}}
\label{appendix:scattering_1}

\begin{figure*}[t!]
    \centering
    \includegraphics[scale=0.3]{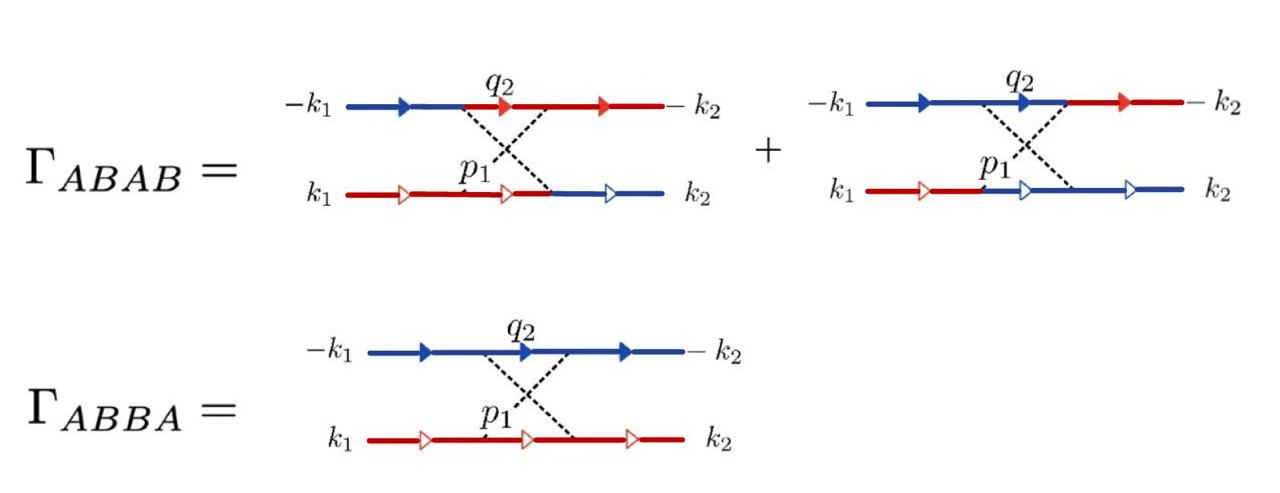}
    \caption{Feynman diagrams for Kohn-Luttinger superconductivity showing Cooper pair scattering in a two-valley system with an intraorbital interaction coupling opposite spins. Empty arrows represent up spins, while filled arrows represent down spins. The A valley is depicted in red, and the B valley is depicted in blue. The dashed lines represent the interaction vertex. The top diagrams corresponds to $\Gamma^{ABAB}_{\up\down\down\up}(k_1,-k_1 | -k_2,k_2)$, for which the valley index of each electron in the pair flips. The bottom diagram corresponds to $\Gamma^{ABBA}_{\up\down\down\up}(k_1,-k_1|-k_2,k_2)$ for which the valley index of each electron in the pair is conserved. Note that this section uses a different labeling of external momenta compared to the rest of the article, whereby the bottom legs have momentum $k_1, k_2$ and the top legs have momentum $-k_1,-k_2$, in order to follow the convention of Ref.~\cite{Kagan_2014}. }
    \label{fig_app:Scattering_diagrams}
\end{figure*}
Following equation~\eqref{FFF} we can write individual $V$ factors as follows:
\begin{align}
V_{\alpha_1\alpha_2;\beta_1\beta_2}
\big(\vec{k}_1,\vec{q}_2\mid -\vec{k}_2,\vec{p}_1\big)
= U \Big(&
w_{\alpha_1, a}(\vec{k}_1)
w_{\alpha_2, a}(\vec{q}_2)
w_{\beta_1, a}^*(-\vec{k}_2)
w_{\beta_2, a}^*(\vec{p}_1)
\nonumber \\
&+
w_{\alpha_1, b}(\vec{k}_1)
w_{\alpha_2, b}(\vec{q}_2)
w_{\beta_1, b}^*(-\vec{k}_2)
w_{\beta_2, b}^*(\vec{p}_1)
\Big)
\end{align}
and
\begin{align}
V_{\alpha_1\alpha_2;\beta_1\beta_2}
\big(\vec{p}_1,-\vec{k}_1 \mid \vec{q}_2,\vec{k}_2\big)
= U \Big(&
w_{\alpha_1, a}(\vec{p}_1)
w_{\alpha_2, a}(-\vec{k}_1)
w_{\beta_1, a}^*(\vec{q}_2)
w_{\beta_2, a}^*(\vec{k}_2)
\nonumber \\
&+
w_{\alpha_1, b}(\vec{p}_1)
w_{\alpha_2, b}(-\vec{k}_1)
w_{\beta_1, b}^*(\vec{q}_2)
w_{\beta_2, b}^*(\vec{k}_2)
\Big)
\end{align}

Now by considering $w(\vec{K}'+\delta \vec{k}) = u(\delta \vec{k})$(for $B$ valley, $\nu=-1$) and $w(\vec{K}+\delta \vec{k}) = v(\delta \vec{k})$(for $A$ valley, $\nu=+1$) we can rewrite the above equations as 
\begin{align}
V_{+\alpha;+\beta}^{A\nu_1 A\nu_1}
\big(\delta\vec{k}_1,\vec{q}_2 \mid -\delta\vec{k}_2,\vec{p}_1\big)
= U \Big(&
v_{+a}(\delta\vec{k}_1)
w_{\alpha, a}(\vec{q}_2)
v_{+a}^*(-\delta\vec{k}_2)
w_{\beta, a}^*(\vec{p}_1)
\nonumber \\
&+
v_{+b}(\delta\vec{k}_1)
w_{\alpha, b}(\vec{q}_2)
v_{+b}^*(-\delta\vec{k}_2)
w_{\beta, b}^*(\vec{p}_1)
\Big) \label{Vs1}
\end{align}
and 
\begin{align}
V_{\beta+;\alpha+}^{\nu_1 B \nu_1 B}
\big(\vec{p}_1,-\delta\vec{k}_1 \mid \vec{q}_2,\delta\vec{k}_2\big)
= U \Big(&
w_{\beta, a}(\vec{p}_1)
u_{+a}(-\delta\vec{k}_1)
w_{\alpha, a}^*(\vec{q}_2)
u_{+a}^*(\delta\vec{k}_2)
\nonumber \\
&+
w_{\beta, b}(\vec{p}_1)
u_{+b}(-\delta\vec{k}_1)
w_{\alpha, b}^*(\vec{q}_2)
u_{+b}^*(\delta\vec{k}_2)
\Big) \label{Vs2}
\end{align}
where the superscripts of $V$ indicate the valley index of each of the four momenta in the argument of $V$. This means both internal legs ($\vec{p}_1$ and $\vec{q}_2$) are in the same valley ($\nu_1$) in this case, as shown in the top two diagrams of Fig.~\ref{fig_app:Scattering_diagrams}. ( Each diagram corresponds to a different value of $\nu_1$).
 The fact that $\vec{p}_1$ and $\vec{q}_2$ are in the same valley is easily shown using $\vec{q}_2 = \vec{p}_1 - \vec{k}_1 - \vec{k}_2 $ and the fact that $\vec{k}_1$ and $\vec{k}_2$ are in opposite valleys by assumption for this $\Gamma^{ABAB}$ process.

Summing over $\nu_1$ gives the total form factor appearing in Eq.~\ref{app:Gamma_2} for the $ABAB$ contribution:
\begin{align}
F^{ABAB}_{\alpha\beta}
\big(
\theta_{\vec{k}_1}, \theta_{\vec{k}_2};
\theta_{\vec{p}_1}, \theta_{\vec{q}_2}
\big)
=
\sum_{\nu_1}
V_{+\alpha;+\beta}^{A\nu_1 A\nu_1}
\big(
\delta\vec{k}_1, \vec{q}_2 \mid -\delta\vec{k}_2, \vec{p}_1
\big)
V_{\beta+;\alpha+}^{\nu_1 B \nu_1 B}
\big(
\vec{p}_1, -\delta\vec{k}_1 \mid \vec{q}_2, \delta\vec{k}_2
\big)
\end{align}
which expands to:
\begin{align}
F^{ABAB}_{\alpha\beta}(\theta_{\vec{k}_1},\theta_{\vec{k}_2};\theta_{\vec{p}_1},\theta_{\vec{q}_2})=U^2\bigg(&v_{+a}(\delta \vec{k}_1)u_{\alpha a}(\vec{q}_2)v_{+a}^*(-\delta \vec{k}_2)u^*_{\beta a}(\vec{p}_1)+ v_{+b}(\delta \vec{k}_1)u_{\alpha b}(\vec{q}_2)v_{+b}^*(-\delta \vec{k}_2)u^*_{\beta b}(\vec{p}_1)\bigg)\\\nonumber&\bigg(u_{\beta a}(\vec{p}_1)u_{+a}(-\delta \vec{k}_1)u_{\alpha a}^*(\vec{q}_2)u^*_{+a}(\delta \vec{k}_2)+ u_{\beta b}(\vec{p}_1)u_{+ b}(-\delta \vec{k}_1)u_{\alpha b}^*(\vec{q}_2)u^*_{+ b}(\delta \vec{k}_2)\bigg)\\\nonumber&
+\bigg(v_{+a}(\delta \vec{k}_1)v_{\alpha a}(\vec{q}_2)v_{+ a}^*(-\delta \vec{k}_2)v^*_{\beta a}(\vec{p}_1)+ v_{+b}(\delta \vec{k}_1)v_{\alpha b}(\vec{q}_2)v_{+b}^*(-\delta \vec{k}_2)v^*_{\beta b}(\vec{p}_1)\bigg)\\\nonumber&
\bigg(v_{\beta a}(\vec{p}_1)u_{+a}(-\delta \vec{k}_1)v_{\alpha a}^*(\vec{q}_2)u^*_{+a}(\delta \vec{k}_2)+ v_{\beta b}(\vec{p}_1)u_{+b}(-\delta \vec{k}_1)v_{\alpha b}^*(\vec{q}_2)u^*_{+b}(\delta \vec{k}_2)\bigg)\,.
\end{align}

By simplifying the above equation using the definitions in equation \eqref{app:us} we find the following expression
\bea
F^{ABAB}_{\alpha\beta}(\theta_{\vec{k}_1},\theta_{\vec{k}_2};\theta_{\vec{p}_1},\theta_{\vec{q}_2})=U^2\bigg(1-\alpha\beta \cos(\theta_{\vec{k}_1}-\theta_{\vec{k}_2}+\theta_{\vec{q}_2}-\theta_{\vec{p}_1})\bigg)\,.
\eea

\emph{Integration}---
We now need to perform the integral in Eq.~\ref{app:Gamma_2} in order calculate the $ABAB$ contribution to the pairing kernel $\Gamma^{ABAB}_{\uparrow \downarrow \downarrow \uparrow}(\vec{k}_1, \vec{k}_2 )$.
For convenience, we integrate over $\vec{q}_2$ instead of $\vec{p}_1$, and we relabel $\vec{q}_2$ as $\vec{q}$, leading to
\bea
\Gamma^{ABAB}_{\up\down\down\up}(\vec{k}_1,\vec{k}_2) = -\frac{1}{N}\sum_{\alpha\beta}\sum_{\vec{q}}\chi_{\alpha\beta}(\vec{q},\vec{q}+\vec{k}_1+\vec{k}_2) F^{ABAB}_{\alpha\beta}(\theta_{\vec{k}_1}, \theta_{\vec{k}_2}; \theta_{\vec{q} + \vec{k}_1 + \vec{k}_2}, \theta_\vec{q}).
\eea
We focus first on the contribution from interband scattering $\alpha \neq \beta$ which dominates in the limit of $k_F \ll \Lambda$ since it scales as $\Lambda$ whereas intraband contributions scale as $k_F$.
Denoting the interband contribution to $\Gamma$ as $\bar\Gamma$, we have:
\begin{equation}
\bar{\Gamma}^{ABAB}_{\uparrow \downarrow \downarrow \uparrow}(\vec{k}_1, \vec{k}_2 ) = -\int_{k_F}^{\Lambda} dq \ q \int d\theta \frac{2}{(2\pi)^2 \hbar v_F}    \frac{1}{\hbar v_F} \frac{-1}{q + |\vec{q} + \vec{k}_1 + \vec{k}_2|} F^{ABAB}_{+-}(\theta_{\vec{k}_1}, \theta_{\vec{k}_2}; \theta_{\vec{q} + \vec{k}_1 + \vec{k}_2}, \theta)
\end{equation}
with $q \equiv |\bq|$ and $\theta \equiv \theta_\bq$.

In the \(k_F \ll \Lambda\) limit, this simplifies to:
\begin{equation}
\begin{aligned}
&\bar{\Gamma}^{ABAB}_{\uparrow \downarrow \downarrow \uparrow}(\vec{k}_1, -\vec{k}_1 | -\vec{k}_2, \vec{k}_2) = -U^2 \int_{k_F}^{\Lambda} dq \ q \int d\theta \frac{2}{(2\pi)^2 \hbar v_F}  \frac{1}{\hbar v_F} \frac{-1}{2 q} (1 + \cos(\theta_{\vec{k}_1} - \theta_{\vec{k}_2})) \\
&\rightarrow \bar{\Gamma}^{ABAB}_{\uparrow \downarrow \downarrow \uparrow}(\vec{k}_1, -\vec{k}_1 | -\vec{k}_2, \vec{k}_2) = U^2 \frac{2 \pi}{(2\pi)^2 \hbar v_F} \Lambda (1 + \cos(\theta_{\vec{k}_1} - \theta_{\vec{k}_2}))
\end{aligned}
\end{equation}


\subsubsection{\texorpdfstring{Scattering $(k_1,\up,A)(-k_1,\down,B)\rightarrow (-k_2,\down,B)(k_2,\up,A)$}{Scattering ABBA}}
\label{appendix:scattering_2}
The second type of scattering that we need to consider is the $ABBA$ type of scattering for which the valley index of electrons is conserved, see the bottom diagram in Fig.~\ref{fig_app:Scattering_diagrams}. 
We follow the same calculation as in the previous section, except that the counterparts of Eqs.~\ref{Vs1} and \ref{Vs2} are now
\begin{align}
V_{+\alpha;+\beta}^{A\nu_2 B \nu_1}
\big(
\delta\vec{k}_1, \vec{q}_2 \mid -\delta\vec{k}_2, \vec{p}_1
\big)
=
U \Big(&
v_{+a}(\delta\vec{k}_1)
w_{\alpha ,a}(\vec{q}_2)
u_{+a}^*(-\delta\vec{k}_2)
w_{\beta ,a}^*(\vec{p}_1)
\nonumber \\
&+
v_{+b}(\delta\vec{k}_1)
w_{\alpha ,b}(\vec{q}_2)
u_{+b}^*(-\delta\vec{k}_2)
w_{\beta ,b}^*(\vec{p}_1)
\Big)
\end{align}
and
\begin{align}
V_{\beta+;\alpha+}^{\nu_1 B \nu_2 A}
\big(
\vec{p}_1, -\delta\vec{k}_1 \mid \vec{q}_2, \delta\vec{k}_2
\big)
=
U \Big(&
w_{\beta ,a}(\vec{p}_1)
u_{+a}(-\delta\vec{k}_1)
w_{\alpha ,a}^*(\vec{q}_2)
v_{+a}^*(\delta\vec{k}_2)
\nonumber \\
&+
w_{\beta ,b}(\vec{p}_1)
u_{+b}(-\delta\vec{k}_1)
w_{\alpha ,b}^*(\vec{q}_2)
v_{+b}^*(\delta\vec{k}_2)
\Big)
\end{align}
where $\nu_1$, $\nu_2$ are the valley indices for $\vec{p}_1$, $\vec{q}_2$.
Given that $\bk_1$ and $\bk_2$ are in valley A, the only acceptable valley indices in this case to ensure both internal momenta are close to a Dirac point are $\nu_1 = 1$ and $\nu_2=-1$.
This is shown as follows.
Since $\vec{q}_2 = \vec{p}_1 - \vec{k}_1 - \vec{k}_2$, we get $\vec{q}_2 = \vec{p}_1 + 2\vec{K}' - \delta \vec{k}_1 - \delta \vec{k}_2$. If $\vec{p}_1$ was in the $K'$ valley, $\vec{q}_2$ would be close to the $\Gamma$ point of the Brillouin zone, which is unacceptable.
Therefore $\vec{p}_1$ must be in valley $K$, in which case $\vec{q}_2$ is in valley $K'$.

Using approximate expressions valid near the $\vec{K}$ and $\vec{K}'$ points we thus have  
\begin{align}
V_{+\alpha;+\beta}^{ABBA}
\big(
\delta\vec{k}_1, \vec{q}_2 \mid -\delta\vec{k}_2, \vec{p}_1
\big)
=
U \Big(&
v_{+a}(\delta\vec{k}_1)
u_{\alpha a}(\vec{q}_2)
u_{+a}^*(-\delta\vec{k}_2)
v_{\beta a}^*(\vec{p}_1)
\nonumber \\
&+
v_{+b}(\delta\vec{k}_1)
u_{\alpha b}(\vec{q}_2)
u_{+b}^*(-\delta\vec{k}_2)
v_{\beta b}^*(\vec{p}_1)
\Big)
\end{align}
and
\begin{align}
V_{\beta+;\alpha+}^{ABBA}
\big(
\vec{p}_1, -\delta\vec{k}_1 \mid \vec{q}_2, \delta\vec{k}_2
\big)
=
U \Big(&
v_{\beta a}(\vec{p}_1)
u_{+a}(-\delta\vec{k}_1)
u_{\alpha a}^*(\vec{q}_2)
v_{+a}^*(\delta\vec{k}_2)
\nonumber \\
&+
v_{\beta b}(\vec{p}_1)
u_{+b}(-\delta\vec{k}_1)
u_{\alpha b}^*(\vec{q}_2)
v_{+b}^*(\delta\vec{k}_2)
\Big)
\end{align}

The function $F^{ABBA}$ is now expressed as follows:
\begin{equation}
F_{\alpha\beta}^{ABBA}(\theta_{\vec{k}_1}, \theta_{\vec{k}_2}; \theta_{\vec{p}_1}, \theta_{\vec{q}_2}) = V_{+\alpha;+\beta}^{ABBA}(\delta \vec{k}_1, \vec{q}_2 \mid -\delta \vec{k}_2, \vec{p}_1) V_{\beta+;\alpha+}^{ABBA}(\vec{p}_1, -\delta \vec{k}_1 \mid \vec{q}_2, \delta \vec{k}_2)
\label{app:F_ABBA}
\end{equation}
By substituting the expressions from equations \eqref{app:us}, we can simplify it to:
\begin{equation}
F_{\alpha\beta}^{ABBA}(\theta_{\vec{k}_1}, \theta_{\vec{k}_2}; \theta_{\vec{p}_1}, \theta_{\vec{q}_2}) = \frac{1}{2} U^2 \left( 1 - \alpha \beta \cos(\theta_{\vec{k}_1} + \theta_{\vec{k}_2} - \theta_{\vec{q}_2} - \theta_{\vec{p}_1}) \right)
\end{equation}

We are again interested in the dominant contribution $\bar{\Gamma}$ which scales as $\Lambda$ and arises from interband contributions which are summed over the entire region $k_F < q < \Lambda$.
In the limit of $k_F \ll \Lambda$, we finally find
\begin{align}
\bar{\Gamma}^{ABBA}_{\up\down\down\up}(\vec{k}_1, \vec{k}_2 ) &= -\int_{k_F}^{\Lambda} dq \ q \int d\theta \frac{1}{(2\pi)^2 \hbar v_F}  \frac{-2}{2|\vec{q}|} \frac{U^2}{2} \left( 1 + \cos(2\theta) \cos(\theta_{\vec{k}_1} + \theta_{\vec{k}_2}) + \sin(2\theta) \sin(\theta_{\vec{k}_1} + \theta_{\vec{k}_2}) \right) \\
&\to \frac{U^2}{2} \frac{2\pi}{(2\pi)^2 \hbar v_F} \Lambda
\end{align}

\subsection{Linearized gap equation}
\begin{table}[b!]
\centering
\renewcommand{\arraystretch}{1.3}
\begin{tabular}{|c|c|c|}
\hline
\textbf{Irrep} & \textbf{Planar Inversion} & \textbf{Order parameter} \\
\hline
$A_1$ $(s)$ & Even & $\psi(\nu,\theta)=1$ \\
\hline
$A_2$ & Even & Nodal \\
\hline
$E_2$ $(d \pm id)$ & Even & $\psi(\nu,\theta) = \nu e^{i \theta}$ \\
\hline
$B_1$ $(f)$ & Odd & $\psi(\nu,\theta) = \nu$ \\
\hline
$B_2$ & Odd & Nodal \\
\hline
$E_1$ $(p\pm ip)$& Odd & $\psi(\nu,\theta) = e^{i \theta}$ \\
\hline
\end{tabular}
\caption{Table of irreducible representations for the superconducting order parameter $\psi(\bk) = \psi(\nu,\theta)$. In this context, \( \theta \) is the polar angle of $\bk$ measured from the center of its valley, and $\nu $ is the valley index.}
\label{tab:my_label}
\end{table}

We now combine the different contributions to the pairing kernel and project it to the Fermi surface, which is composed of two circular pockets centered at $K$ and $K'$.
The solutions to the linearized gap equation are defined on the Fermi surface and thus denoted by $\psi(\bk) = \psi(\nu,\theta)$ with $\bk = \nu \vec{K} + k_F (\cos(\theta),\sin(\theta))$. They can be classified into valley-even and valley-odd, corresponding to a gap without or with a sign change between the two different valleys, respectively. In other words, $\psi(\nu,\theta) = \pm \psi(-\nu,\theta) $ with the plus sign corresponding to valley-even and the minus sign corresponding to valley-odd.

We can finally write down the projected pairing kernel $\bar\Gamma$ into the valley-even and valley-odd sectors:
\bea
\bar\Gamma_{\pm}(\theta_{\vec{k}_1},\theta_{\vec{k}_2}) \equiv \frac{1}{2}(\bar{\Gamma}^{ABBA}_{\up\down\down\up} \pm \bar{\Gamma}^{ABAB}_{\up\down\down\up}) =   U^2\Lambda \frac{1}{4\pi\hbar v_F}\big(\frac{1}{2}\pm (1+\cos(\theta_{\vec{k}_1}-\theta_{\vec{k}_2}))\big)\,.
\label{eq_app:gamma_pm_graphene}
\eea
Here, \( \bar{\Gamma} \) emphasizes that we are considering only the dominant contribution to the kernel, i.e. the one scaling like $\Lambda$. More explicitly, we have:
\bea
\bar\Gamma_{+}(\theta_{\vec{k}_1},\theta_{\vec{k}_2}) &=& U^2\Lambda \frac{1}{4\pi\hbar v_F}\left(\frac{3}{2}+\cos(\theta_{\vec{k}_1}-\theta_{\vec{k}_2})\right)\\
 \bar\Gamma_{-}(\theta_{\vec{k}_1},\theta_{\vec{k}_2}) &=& - U^2\Lambda \frac{1}{4\pi\hbar v_F}\left(\frac{1}{2}+\cos(\theta_{\vec{k}_1}-\theta_{\vec{k}_2})\right)
\eea
Using rotational invariance, this kernel is simply diagonalized by a Fourier series.
We find attractive (i.e. negative) eigenvalues for the pairing kernel in the valley-odd sector $\bar\Gamma_-$. We find an instability towards two superconducting orders with the same eigenvalue, in the $d+id$ and the $f$-wave channels. The gap functions are given by:
\bea
\psi_{d+id}(\nu,\theta) &=& \nu e^{i \theta} \\
\psi_{f}(\nu,\theta) &=& \nu
\eea
These belong to the $E_2$ and $B_1$ irreducible representations, see Table~\ref{tab:my_label}. The $d+id$ state is parity-even and thus spin singlet, and the $f$ state is parity-odd and thus spin-triplet.
As mentioned, these two channels are degenerate, with effective interaction
\bea
V_\text{eff}^{d+id} = V_\text{eff}^{f}  = - U^2\Lambda\frac{1}{8\pi \hbar v_F}.
\eea
Remarkably, $V_\text{eff}$ does not vanish in the limit of $k_F \to 0$, which shows that two Dirac cones of opposite chirality have a surprisingly robust tendency toward superconductivity in the low density limit.
As explained in the main text, this is due to the importance of interband excitations, which can be integrated over the whole range of momentum up to the cutoff $\Lambda$ in contrast to intraband contributions which scale like $k_F$.

 Since the pairing kernel is dominated by virtual interband electron-hole excitations with momenta of the order of the cutoff, this means the details of the UV regularization of the Dirac cone could actually matter, even in the limit of $k_F \ll \Lambda$, which is somewhat unexpected.
 In fact, a numerical calculation for the full tight-binding model of graphene, as opposed to the idealized Dirac dispersion with a cutoff $\Lambda$ studied in this appendix, shows that the degeneracy between $d+id$ and $f$ is broken in that case, and the $d+id$ state is favored (see next Appendix for more details). 
Nevertheless, the fact that both $d+id$ and $f$ orders have an effective interaction which goes to a finite constant as $k_F \to 0$ is still true in the graphene model and is thus a universal property of two Dirac cones of opposite chirality.

\section{Numerical calculation of Kohn-Luttinger superconductivity on the honeycomb lattice}

In this section we obtain numerical results for Kohn-Luttinger superconductivity for the honeycomb lattice with nearest-neighbor hopping and on-site repulsion (which we sometimes refer to as the graphene model for short). We focus on the small $k_F$ regime in order to compare against our analytical prediction from the previous Appendix.
We note that this model was already studied in Refs.~\cite{Chubukov_2014,Kagan_2014,Wolf_2018}, but we are reproducing these results here in a convenient way for comparison with our analytical results.
As a reminder, our analytic prediction for the case of two ideal Dirac cones of opposite chirality (see previous appendix) is that $V_\text{eff}$ should be finite in the limit of small $k_F$ for both $d\pm i d$ and $f$-wave orders. In this limit, pairing is driven by interband excitations with momenta up to the UV cutoff $\Lambda$. For this reason, the details of the UV regularization of the model could be important in deciding between the two orders ($d+id$ and $f$).
In order to settle this issue, we performed a numerical investigation across a range of carrier densities in graphene.

We use the full tight-binding dispersion of the honeycomb lattice (see~\ref{grapheneappendixdispersion}) and calculate and diagonalize numerically the pairing kernel, Eq.~\ref{app:Gamma_2}.
The integration over internal momentum, done over the entire Brillouin zone, is performed using a dynamic grid generation method. Initially, uniform sample points are generated with constant spacing to establish a basic grid for the integral calculation. The integral is then evaluated using these uniform points. If the error of the computed integral exceeds the predefined tolerance, additional sample points are generated on top of the basic grid using the Monte Carlo-Hastings algorithm. This iterative process continues until the error of the integral falls below the specified tolerance level. In our calculations, we set the relative error to \(5 \times 10^{-3}\).

\begin{figure}[t!]
    \centering
    \includegraphics[scale=0.3]{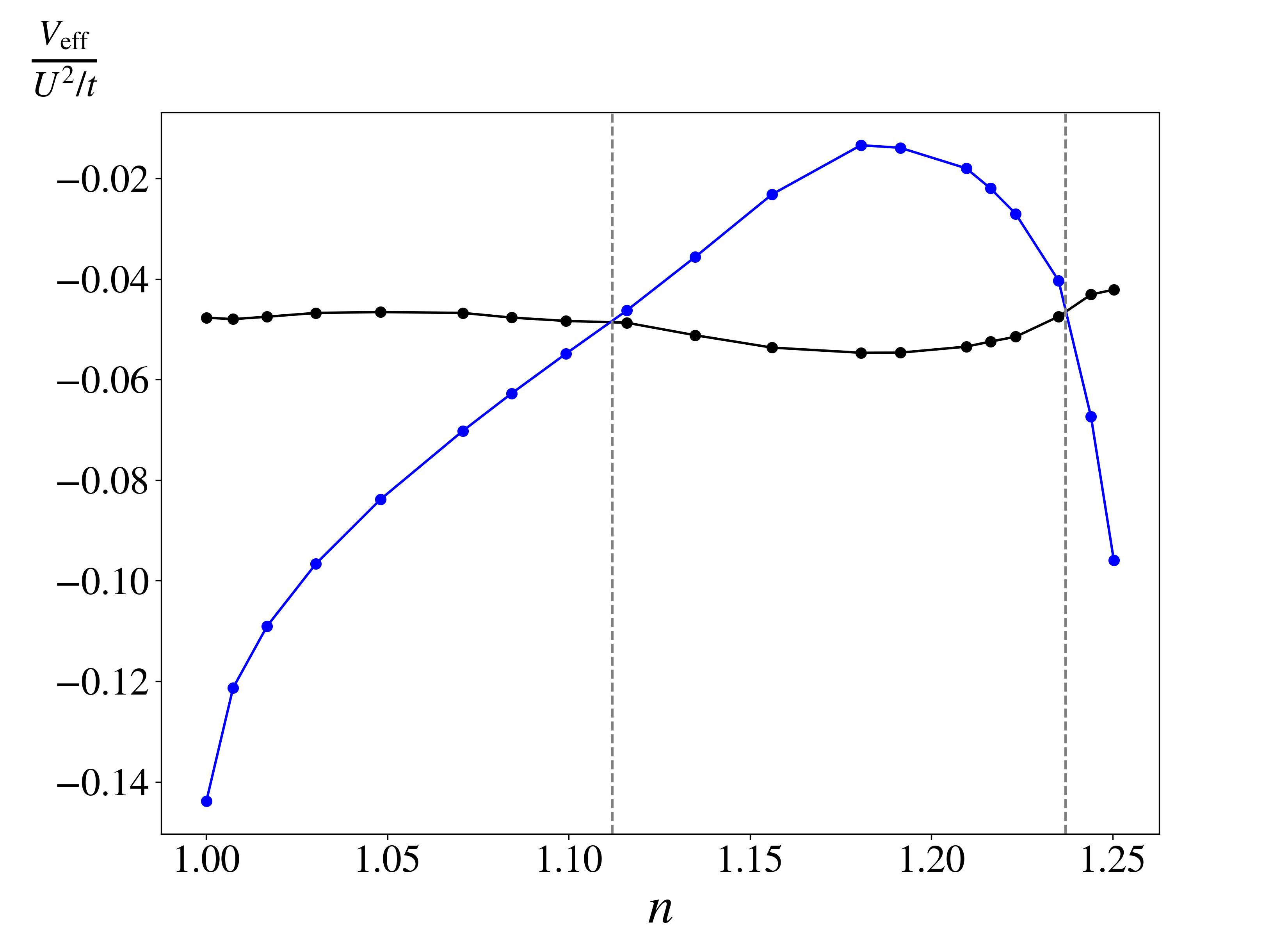}
    \caption{Effective interaction in the Cooper Channel versus density $n$, where $n=1$ corresponds to the charge neutrality point. 
    The blue curve represents the $d\pm id$ channel (corresponding to the $E_2$ irreducible representation in Table~\ref{tab:my_label}), while the black curve corresponds to the $f$ channel (associated with the $B_1$ irreducible representation in Table~\ref{tab:my_label}). The most important feature for us is the fact that both $V_\text{eff}$ go to a finite value at $n=1$. 
    In addition, two level crossings are observed at $n \approx 1.1$ and $n \approx 1.235$, indicated by dotted gray lines, in agreement with earlier work~\cite{Wolf_2018, Kagan_2014}. A Van Hove singularity occurs at $n = 1.25$.}
    \label{fig_app:effective_potential}
\end{figure}

Our results for the effective interactions $V_\text{eff}$ versus density $n$ are shown in Fig.~\ref{fig_app:effective_potential}.
We confirm our analytical prediction that the $d+id$ and $f$ wave channels are dominant at low carrier density (i.e. close to charge neutrality at $n=1$), and that $V_\text{eff}$ remains finite for each of them in the limit of $n \to 1$ (i.e. $k_F \to 0$) thanks to interband effects.
There is however a difference with our analytical prediction: the $d+id$ and $f$ eigenvalues are not degenerate, with the $d+id$ eigenvalue about three times larger for $k_F \to 0$.
This is not too surprising since, as we have just discussed, the UV details of the model matter in this case, and the analytical calculations were obtained assuming a perfectly linear Dirac dispersion up to a cutoff $\Lambda$, instead of the hexagonal Brillouin zone used here.

At larger carrier densities, we recover two eigenvalue crossings between $d+id$ and $f$ at \(n \approx 1.1\) and \(n \approx 1.235\) (see Fig.~\ref{fig_app:effective_potential}), consistent with the findings reported in \cite{Wolf_2018, Kagan_2014}.

\section{BdG Hamiltonian and Chern number}
\label{App:Chern}
In this appendix, we calculate the Chern number for the superconducting phase realized in Section \ref{sec:TPT} of the main text, with single-particle Hamiltonian
\bea
H = v_F \hat{z} \cdot (\vec{k} \times \boldsymbol{\sigma}) + (m + B k^2) \sigma^z
\eea
which means $[d_x,d_y,d_z] = [-v_F k_y, v_F k_x, m+Bk^2]$. This Hamiltonian describes the transition at $m=0$ between a trivial insulator and a Chern insulator with Chern number $C = - \sgn(B)$.
We work at positive chemical potential $\mu > 0$ and the Fermi surface is a circle with radius $k_F$.

The superconducting term reads
\bea
H_{SC} = \Delta \sum_\bk f^*_\bk c^\dagger_{\bk,+} c^\dagger_{-\bk,+} + \text{h.c.}
\eea
with
\bea
c^\dagger_{\bk,+} = u_{+,\uparrow}(\bk) c^\dagger_{k,\uparrow} + u_{+,\downarrow}(\bk) c^\dagger_{k,\downarrow}
\eea
where $\Delta$ the scale of the superconducting gap (which we take small compared to $\mu$ since we are interested in the weak coupling limit),
and $f_\bk$ is the leading eigenvector of the pairing kernel. By solving the linearized gap equation numerically, we find that the favored gap function depends on the sign of $B$ and is given by $f_\bk \sim e^{i\theta_\bk (1- 2\sgn(B))}$.

The Bogolyubov-de Gennes (BdG) Hamiltonian reads $H = \Psi_{\bk}^\dagger H(\bk) \Psi_\bk$ with
\bea
H(\bk) = \begin{bmatrix}
  d_z-\mu & d_x - i d_y & \Delta f^*_\bk u_{+,\uparrow}(\bk) u_{+,\uparrow}(-\bk) & \Delta f^*_\bk u_{+,\uparrow}(\bk) u_{+,\downarrow}(-\bk) \\  
  d_x + i d_y & -d_z-\mu & \Delta f^*_\bk u_{+,\downarrow}(\bk) u_{+,\uparrow}(-\bk) & \Delta f^*_\bk u_{+,\downarrow}(\bk) u_{+,\downarrow}(-\bk) \\  
  \Delta f_\bk u_{+,\uparrow}^*(\bk) u_{+,\uparrow}^*(-\bk) & \Delta f_\bk u_{+,\downarrow}^*(\bk) u_{+,\uparrow}^*(-\bk) & -(d_z-\mu) & d_x + i d_y\\  
  \Delta f_\bk u_{+,\uparrow}^*(\bk) u_{+,\downarrow}^*(-\bk) & \Delta f_\bk u_{+,\downarrow}^*(\bk) u_{+,\downarrow}^*(-\bk) & d_x - i d_y & -(-d_z-\mu)
\end{bmatrix}
\eea
where $\Psi_{\bk}^\dagger = [c^\dagger_{\bk,\uparrow} c^\dagger_{\bk,\downarrow} c_{-\bk,\uparrow} c_{-\bk,\downarrow}]$.
We are interested in the BdG Chern number of band $m$ defined as
\bea
\mathcal{N}_m = \int \frac{d^2k}{(2\pi)^2} \Omega_{k,m} 
\eea
with $\Omega_{k,m} = \nabla \times \vec{A}_m$ the Berry curvature, and the convention that 
\bea
\vec{A}_m = i \bra{\psi_k,m} \nabla_k \ket{\psi_k,m}
\eea
with $\ket{\psi_k,m}$ the $m$-th eigenvector of the BdG Hamiltonian.
The integral is done numerically following Ref.~\cite{Fukui}. We are interested in the sum of the BdG Chern numbers for the two occupied BdG bands, which we call simply $\mathcal{N}$.
We find in the superconducting phase that $\mathcal{N} = \sgn(B)$.
Since the BdG Chern number in the adjacent non-trivial Chern insulating phase is $\mathcal{N} = 2 C = - 2 \sgn(B)$, this means the superconductor has the \emph{opposite} chirality to the insulating phase it borders.

\section{Numerical results showing the dominance of interband contributions for the model of Section \ref{sec:TPT}}
\label{BtermApp}

\begin{figure}[h!]
    \centering
    \includegraphics[width=0.4\linewidth]{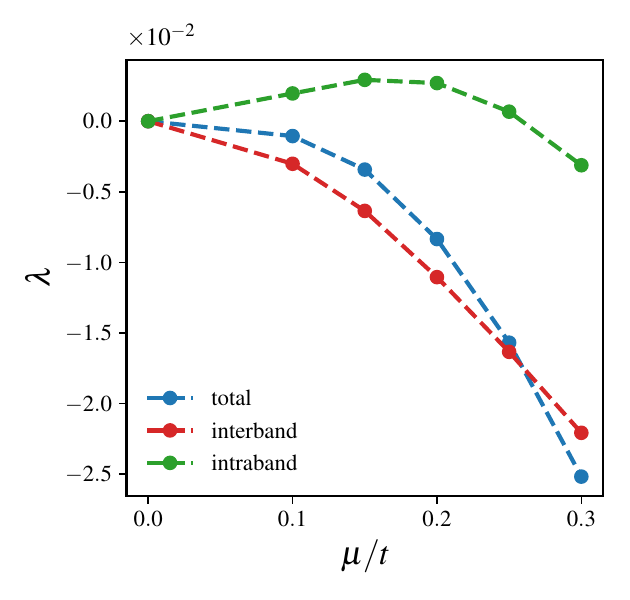}
    \caption{Decomposition into intraband and interband contributions for the pairing eigenvalue in the model of Section \ref{sec:TPT}. Parameters: $v_F / (t a / \hbar)=0.5 $, $B /(t a^2) = -0.25 $, $\Lambda a = \pi $. }
    \label{fig_app:interband_contribution}
\end{figure}


In this section, we present more numerical results about the solution of the linearized gap equation for the model
realized in Section \ref{sec:TPT} of the main text, with single-particle Hamiltonian
\bea
H = v_F \hat{z} \cdot (\vec{k} \times \boldsymbol{\sigma}) + (m + B k^2) \sigma^z.
\eea
Solving the linearized gap equation involves diagonalizing the pairing kernel
\begin{align}
\Gamma^{(2)}(\vec{k}_1,\vec{k}_2) = \sum_{\alpha\beta} \Gamma_{\alpha\beta}^{(2)}(\vec{k}_1,\vec{k}_2) = -\frac{1}{N} \sum_{\alpha,\beta} \sum_\vec{p} \chi_{\alpha\beta}(\vec{p}+\bq, \vec{p})\, F_{\alpha\beta}(\vec{k}_1, \vec{k}_2; \vec{p}+\bq, \vec{p}),
\end{align}
where \( \vec{q} = \vec{k}_1 + \vec{k}_2 \), and \( F_{\alpha\beta}(\vec{k}_1, \vec{k}_2; \vec{p}+\bq, \vec{p}) \) encapsulates the form-factor, defined explicitly in Appendix~\ref{App1}.

Since the model has rotational invariance, the Fermi surface is a circle and the eigenvectors simply correspond to Fermi surface harmonics $\Delta_\bk \sim e^{i l \theta_\bk }$.
For $B<0$, the favored gap has $l=-1$, i.e. $\Delta_\bk \sim e^{- i \theta_k}$.
We thus separately analyze in Fig.~\ref{fig_app:interband_contribution}, for the $l=-1$ sector, the interband component \( \Gamma_{\pm} = \Gamma_{+-} +\Gamma_{-+} \), the intraband component \( \Gamma_{++} \), and the total interaction \( \Gamma = \Gamma_{\pm} + \Gamma_{++} \). 
We find that, for small $\mu/t$, the interband contribution to $\lambda$ is attractive (i.e. negative) and dominates over the repulsive intraband contribution. Interband excitations are thus responsible for the superconducting instability in this model.

\section{Connection with Qi-Wu-Zhang model}
\label{QWZ}

In this Appendix, we show how the Hamiltonian of Section \ref{sec:TPT} gives a low-energy description of the Qi-Wu-Zhang model~\cite{Qi2006}, which realizes a single Dirac cone on a square lattice and realizes one half of the mercury-telluride two-dimensional topological insulator\cite{bernevig2006quantum}:
\bea
H_\text{QWZ} = -v \sin(k_x) \sigma^y + v \sin(k_y) \sigma^x + B \sigma^z (  (1-\cos(k_x)) +  (1-\cos(k_y)).
\eea
Without the $B$ term, this model would have 4 Dirac cones, at $(0,0)$, $(\pi,0)$, $(0,\pi)$, and $(\pi,\pi)$.
The $B$ term creates a gap at all Dirac cones except the one at $(0,0)$.

Close to $k=0$, the leading-order terms in $k$ are
\bea
H_\text{QWZ} \simeq - v k_x  \sigma^y + v k_y \sigma^x + B \sigma^z \frac12  ( k_x^2 + k_y^2)
\eea
We thus recover the form of the $B k^2 \sigma^z$ term we considered in Section \ref{sec:TPT}.
Additionally, by tuning separately the constant and $\cos$ terms in the $\sigma^z$ prefactor in $H_\text{QWZ}$, one can also generate a mass term $m \sigma^z$.

\section{Symmetry analysis of superconductivity for topological surface states}
\label{appsym}

In this appendix, we perform a symmetry analysis for superconducting order parameters appearing at the surface of a topological insulator. We consider the $C_{3v}$ point group, and time-reversal symmetry.

\subsection{Symmetry analysis for the $C_{3v}$ warping model}
The case of $C_{3v}$ warping discussed in Section \ref{C3}, with single-particle Hamiltonian
\bea
H = \hat{z} \cdot (\vec{k} \times \boldsymbol{\sigma}) + \eta (k_+^3 + k_-^3) \sigma^z,
\eea
has a $C_{3v}$ symmetry with a $C_3$ rotation and a mirror symmetry $x \to -x$.
First, we discuss how $c^\dagger_{k,\alpha}$ transforms under rotation in our gauge choice.
A rotation $R$ by an angle $\theta$ around the $z$ axis, with $\theta$ a multiple of $2\pi/3$, acts as $R_\theta[c^\dagger_{k,\alpha}] = e^{-i \theta  / 2} c^\dagger_{Rk,\alpha}$. (We keep the band index $\alpha$ implicitly equal to $+1$ --- corresponding to the top band--- for the rest of the discussion, but the same symmetry applies to the bottom band).
This can be shown starting from:
\bea
\ket{k,+} = \cos(\phi/2) \ket{k,\uparrow} + \sin(\phi/2)  e^{+i \Theta}\ket{k,\downarrow}
\eea
with $\cos(\phi) = d_z / |d|$ and $e^{i \Theta } = d_x + i d_y $.
If $R_\theta$ is a symmetry, then $d(R_\theta k) = R_\theta d(k)$, and thus $\Theta_{R_\theta k} = \Theta_k + \theta$ and $\phi_{R_\theta k} = \phi_k$.
Using $R_\theta \equiv e^{-i \sigma^z \theta \frac12} $ to find the action of rotation on spin space, we find:
\bea
R_\theta \ket{k} = \cos(\phi/2)  e^{-i  \theta \frac12} \ket{R_\theta k,\uparrow} + e^{+i \Theta_k} \sin(\phi/2) e^{i  \theta \frac12} \ket{R_\theta k,\downarrow}
\eea
whereas
\bea
\ket{R_\theta k} = \cos(\phi/2)  \ket{R_\theta k,\uparrow} + \sin(\phi/2) e^{+i \Theta_{R_\theta k}} \ket{R_\theta k,\downarrow}
\eea
Using $\Theta_{R_\theta k} = \Theta_k + \theta$, we find
\bea
R_\theta[c^\dagger_{k}] = e^{-i \theta /2} c^{\dagger}_{Rk}.
\eea

Now we ask how the superconducting term transforms under $R \equiv R_{2\pi/3}$.
Let the superconducting term be
\bea
H_{SC} = \sum_{k} f_k c_{k,+} c_{-k,+} 
\eea
with $f_k$ proportional to the relevant eigenvector obtained from diagonalizing the pairing kernel.
The superconducting term transforms as
\bea
R[H_{SC}] &= \sum_{k} e^{i (2\pi/3) } f_k c_{Rk,+} c_{R(-k),+}  \\
&= \sum_{k'} e^{i (2\pi/3) } f_{R^{-1}k'} c_{k',+} c_{-k',+} 
\eea
Defining $\varphi$ as the phase factor acquired by the function $f_k$ under $2\pi/3$ rotation, i.e. $f_{R^{-1}k'} \equiv e^{-i \varphi} f_{k'}$, we have
\bea
R[H_{SC}] = e^{i (2\pi/3) }  e^{-i \varphi} H_{SC}
\eea
We can thus identify $e^{i (2\pi/3) }  e^{-i \varphi}$ as the gauge-invariant quantum number $r$ of the superconducting phase under $2\pi/3$ rotations, and $r$ must thus correspond to one of the tabulated values for a given irreducible representation of $C_{3v}$.
Irreducible representations of $C_{3v}$ are $A_1$ (with $r=1$ and thus $\varphi=2\pi/3$), $A_2$ (with $r=1$ and thus $\varphi=2\pi/3$), and $E$ with $r=e^{ \pm i (2\pi/3)}$ and thus $\varphi=0$ or $\varphi = 4\pi/3$.
The two degenerate states we found in the main text correspond to $E$, with $(d+id)(p+ip) \sim (A e^{ i 2 \theta} + B e^{-i 4 \theta}) e^{i\theta}$ which has $\varphi=0$, and $(d-id)(p+ip) \sim (A e^{ -i 2 \theta} + B e^{+i 4 \theta})e^{i\theta}$ which has $\varphi = 4\pi/3$.
We found that $|B|$ is slightly larger than $|A|$, thus conferring a winding of $-3$ and $+5$, respectively.

\subsection{Time-reversal symmetry}
Using $T[c_{k,\sigma}] = \sigma c_{-k,-\sigma}$ in the spin basis, one finds $T[c_{k,\alpha}] = - e^{i \Theta_k} c_{-k,\alpha}$ with $\alpha$ the band index. This gives
\bea
T[H_{SC}] = \sum_{k} f_k^* e^{2 i\Theta_k} c_{k,+} c_{-k,+}
\eea
and thus $T[f_k] = f_k^* e^{2 i\Theta_k}$.
This  means that, in order to preserve TRS, a gap should have $f_k = f_k^* e^{2 i\Theta_k}$, which is satisfied by either $\mathrm{Arg}[\Delta_k] = \Theta_k$ or $\mathrm{Arg}[\Delta_k] = \Theta_k + \pi$. In other words, a time-reversal-symmetric gap is of the form $f_k \propto e^{i \Theta_k} g(k)$ with $g(k) \in \mathbb{R}$.
This applies to the superconducting gap originally proposed by Fu-Kane in the presence of attractive interaction, which in our notation reads $f_k \sim e^{i \Theta_k}$. It also applies to the gap we found in the quasi-1D model in Section \ref{1D}. By contrast, the $(d \pm id )\times (p+ip)$ gap we found Section \ref{C3} is obviously not time-reversal symmetric.



\section{Explicit form factors for the quasi-1D TISS model}
\label{app:quasi1D_formfactors}

In this appendix we evaluate explicitly the diagrammatic form factors
$F_{b,c,d,e}$ for the quasi-1D TISS model in Eq.~\ref{1DLattice}, assuming a contact
interaction $V_0(\mathbf q)=U$ and restricting to electrons in the conduction
band. In this limit the Bloch spinor is fully polarized along $\pm\hat y$
depending on the sign of $k_x$:
\begin{equation}
|u(\mathbf k)\rangle=
\begin{cases}
|+y\rangle,& k_x>0,\\
|-y\rangle,& k_x<0,
\end{cases}
\qquad
\eta(\mathbf k)\equiv \mathrm{sgn}(k_x)\in\{\pm1\},
\label{eq:eta_def}
\end{equation}
where $| \pm y\rangle$ are eigenstates of $\sigma^y$ with eigenvalues $\pm1$.
(Equivalently, one may take the lattice-periodic definition
$\eta(\mathbf k)=\mathrm{sgn}[\sin(k_x/2)]$; the algebra below is unchanged.)
The overlap between Bloch spinors is then given by
\begin{equation}
\langle u(\mathbf k)\,|\,u(\mathbf p)\rangle
=\delta_{\eta(\mathbf k),\,\eta(\mathbf p)},
\qquad
\eta(-\mathbf k)=-\eta(\mathbf k).
\label{eq:overlap_projector}
\end{equation}

\subsection{Vertex in the band basis}
Using the band-basis vertex in Eq.~\ref{vertexfactor}, the contact interaction gives
\begin{align}
V(\mathbf k;\mathbf p\mid \mathbf q;\mathbf s)
&=U\,\langle u(\mathbf s)|u(\mathbf k)\rangle\,\langle u(\mathbf q)|u(\mathbf p)\rangle
\nonumber\\
&=U\,\delta_{\eta(\mathbf s),\eta(\mathbf k)}\;\delta_{\eta(\mathbf q),\eta(\mathbf p)}.
\label{eq:vertex_projector}
\end{align}
In the second-order diagrams we also define, as in Appendix~A,
\begin{equation}
\mathbf q_1 \equiv \mathbf p+\mathbf k_1+\mathbf k_2.
\label{eq:q1_def}
\end{equation}

\subsection{Form factors for diagrams (b)--(e)}
We now insert Eq.~\eqref{eq:vertex_projector} into the diagram definitions
\ref{eq_app:Fb}--\ref{eq_app:Fe}. For brevity we suppress the band labels since we only consider intraband interactions.

\paragraph{Diagram (b).}
Using
\begin{equation}
F_b=
V(-\mathbf k_1;\mathbf q_1\mid \mathbf k_2;\mathbf p)\;
V(\mathbf p;\mathbf k_1\mid \mathbf q_1;-\mathbf k_2),
\end{equation}
we obtain
\begin{equation}
F_b(\mathbf k_1,\mathbf k_2;\mathbf p)
=U^2\;
\delta_{\eta(\mathbf k_1),\,\eta(\mathbf k_2)}\;
\delta_{\eta(\mathbf p),\,-\eta(\mathbf k_1)}\;
\delta_{\eta(\mathbf q_1),\,\eta(\mathbf k_1)}.
\label{eq:Fb_result}
\end{equation}

\paragraph{Diagram (c).}
Using
\begin{equation}
F_c=
V(-\mathbf k_1;\mathbf q_1\mid \mathbf p;\mathbf k_2)\;
V(\mathbf k_1;\mathbf p\mid \mathbf q_1;-\mathbf k_2),
\end{equation}
we obtain
\begin{equation}
F_c(\mathbf k_1,\mathbf k_2;\mathbf p)
=U^2\;
\delta_{\eta(\mathbf k_2),\,-\eta(\mathbf k_1)}\;
\delta_{\eta(\mathbf p),\,\eta(\mathbf q_1)}.
\label{eq:Fc_result}
\end{equation}

\paragraph{Diagram (d).}
Using
\begin{equation}
F_d=
-\,V(-\mathbf k_1;\mathbf q_1\mid \mathbf p;\mathbf k_2)\;
V(\mathbf k_1;\mathbf p\mid -\mathbf k_2;\mathbf q_1),
\end{equation}
we obtain
\begin{equation}
F_d(\mathbf k_1,\mathbf k_2;\mathbf p)
=-U^2\;
\delta_{\eta(\mathbf k_2),\,-\eta(\mathbf k_1)}\;
\delta_{\eta(\mathbf p),\,\eta(\mathbf k_1)}\;
\delta_{\eta(\mathbf q_1),\,\eta(\mathbf k_1)}.
\label{eq:Fd_result}
\end{equation}

\paragraph{Diagram (e).}
Using
\begin{equation}
F_e=
-\,V(\mathbf q_1;-\mathbf k_1\mid \mathbf p;\mathbf k_2)\;
V(\mathbf p;\mathbf k_1\mid -\mathbf k_2;\mathbf q_1),
\end{equation}
we obtain
\begin{equation}
F_e(\mathbf k_1,\mathbf k_2;\mathbf p)
=-U^2\;
\delta_{\eta(\mathbf k_2),\,-\eta(\mathbf k_1)}\;
\delta_{\eta(\mathbf p),\,-\eta(\mathbf k_1)}\;
\delta_{\eta(\mathbf q_1),\,-\eta(\mathbf k_1)}.
\label{eq:Fe_result}
\end{equation}

\subsection{Branch selection rules and $(c,d,e)$ cancellation}
The results above immediately imply two useful facts.

\paragraph{Same-branch external legs.}
If $\mathbf k_1$ and $\mathbf k_2$ lie on the same Fermi-surface branch,
$\eta(\mathbf k_1)=\eta(\mathbf k_2)$, then Eqs.~\eqref{eq:Fc_result}--%
\eqref{eq:Fe_result} give
\begin{equation}
F_c=F_d=F_e=0
\qquad
\text{(for $\eta(\mathbf k_1)=\eta(\mathbf k_2)$)},
\end{equation}
since each contains the factor $\delta_{\eta(\mathbf k_2),-\eta(\mathbf k_1)}$.

\paragraph{Opposite-branch external legs.}
If $\mathbf k_1$ and $\mathbf k_2$ lie on opposite branches,
$\eta(\mathbf k_2)=-\eta(\mathbf k_1)$, then $\delta_{\eta(\mathbf p),\eta(\mathbf q_1)}$
in Eq.~\eqref{eq:Fc_result} forces $\eta(\mathbf p)=\eta(\mathbf q_1)$.
In that case either $\eta(\mathbf p)=\eta(\mathbf q_1)=\eta(\mathbf k_1)$, for which
$F_d=-F_c$ and $F_e=0$, or $\eta(\mathbf p)=\eta(\mathbf q_1)=-\eta(\mathbf k_1)$, for which
$F_e=-F_c$ and $F_d=0$. Therefore, pointwise in $\mathbf p$,
\begin{equation}
F_c+F_d+F_e=0
\qquad
\text{(for $\eta(\mathbf k_2)=-\eta(\mathbf k_1)$)}.
\end{equation}

Combining these two cases, one finds that $F_c+F_d+F_e=0$ for any $\bk_1, \bk_2$.


\end{document}